\documentclass[11pt,a4paper]{article}

\newcommand{\myzeta}{\zeta}

\ifdefined\pdfoutput
  \pdfoutput=1
\fi

\PassOptionsToPackage{pagebackref}{hyperref}
\usepackage{jheppub}

\usepackage{amsfonts,amsthm,upgreek,bm}
\usepackage{units}
\usepackage{float}
\usepackage[export]{adjustbox}
\usepackage[shortlabels]{enumitem}
\usepackage[mathscr]{euscript}
\usepackage[normalem]{ulem}
\usepackage{tikz}
\usepackage{tikzsymbols}
\usepackage{pgfplots}
\usepackage{ytableau}
\usepackage{slashed}
\usepackage{mathtools}
\usetikzlibrary{decorations.pathmorphing}
\usetikzlibrary{arrows.meta}
\renewcommand{\tilde}{\widetilde}

\renewcommand{\d}{\mathrm{d}}

\newcommand{\p}{\widehat{p}}
\renewcommand{\tilde}{\widetilde}

\renewcommand{\Im}{{\rm Im} \,}

\newcommand{\Tr}{\operatorname{Tr}}

\newcommand{\be}{\begin{equation}}
\newcommand{\ee}{\end{equation}}
\newcommand{\ie}{\begin{equation}\begin{aligned}}
\newcommand{\fe}{\end{aligned}\end{equation}}
\newcommand{\ieg}{\begin{equation}\begin{gathered}}
\newcommand{\feg}{\end{gathered}\end{equation}}

\newcommand{\bZ}{{\mathbb Z}}

\newcommand{\bH}{{\mathbb H}}

\newcommand{\cN}{{\mathcal N}}

\newcommand{\cZ}{{\mathcal Z}}

\newcommand{\adsS}{\mathrm{AdS}_3 \times \mathrm{S}^3}
\newcommand{\adsM}{\mathrm{AdS}_3 \times \mathrm{S}^3 \times \mathcal{M}_4}

\newcommand{\adsK}{\mathrm{AdS}_3 \times \mathrm{S}^3 \times \mathrm{K3}}

\newcommand{\symK}{\textrm{Sym}^N(\mathrm{K3})}
\newcommand{\symM}{\textrm{Sym}^N(\mathcal{M}_4)}

\newcommand{\orbifold}{(\mathrm{AdS}_3 \times \mathrm{S}^3)/\mathbb{Z}_K}
\newcommand{\orbifoldM}{(\mathrm{AdS}_3 \times \mathrm{S}^3)/\mathbb{Z}_K \times \mathcal{M}_4}

\newcommand{\orbifoldK}{(\mathrm{AdS}_3 \times \mathrm{S}^3)/\mathbb{Z}_K \times \mathrm{K3}}

\newcommand{\Kthree}{\mathrm{K3}}
\newcommand{\Tfour}{\mathbb{T}^4}
\newcommand{\Mfour}{\mathcal{M}_4}

\newcommand{\Res}{\mathop{\mathrm{Res}}}
\newcommand{\qb}{\bar{q}}
\newcommand{\yb}{\bar{y}}
\newcommand{\Lb}{\bar{L}}
\newcommand{\Jb}{\bar{J}}

\newcommand{\PE}{\mathrm{PE}}

\newcommand{\Zhat}{\widehat{Z}}
\newcommand{\Sb}{\bar{S}}

\usepackage{amsmath}
\usepackage{tikz}
\usepackage{float}
\usetikzlibrary{arrows.meta}

\definecolor{PaperInk}{HTML}{263746}
\definecolor{PaperPole}{HTML}{C44E35}
\definecolor{PaperDomain}{HTML}{EEF3F7}
\definecolor{PaperGuide}{HTML}{A4ADB5}

\tikzset{
  paper figure/.style={
    font=\fontsize{10}{12}\selectfont,
    text=PaperInk
  },
  paper axis/.style={
    draw=PaperInk,
    line width=0.6pt,
    -{Latex[length=1.8mm,width=1.2mm]}
  },
  paper boundary/.style={
    draw=PaperInk,
    line width=1pt
  },
  paper pole/.style={
    draw=PaperPole,
    line width=1pt
  },
  paper continuation/.style={
    draw=PaperPole,
    line width=0.65pt,
    dash pattern=on 2.2pt off 2.2pt
  },
  paper guide/.style={
    draw=PaperGuide,
    line width=0.45pt,
    dash pattern=on 0.7pt off 1.8pt
  },
  paper intersection/.style={
    draw=PaperPole,
    fill=white,
    line width=0.9pt
  },
  paper key/.style={
    anchor=west,
    inner sep=0pt,
    align=left
  }
}

\title{Polar states in AdS$_3$ holography}

\author[a]{Ji Hoon Lee,}
\author[b,c]{Wei Li}

\affiliation[a]{Institut f\"ur Theoretische Physik, ETH Zurich, CH-8093 Z\"urich, Switzerland}
\affiliation[b]{Institute of Theoretical Physics, Chinese Academy of Sciences, 100190 Beijing, China}
\affiliation[c]{Peng Huanwu Center for Fundamental Theory, 230026 Hefei, Anhui, China}

\emailAdd{jileej@phys.ethz.ch}
\emailAdd{weili@mail.itp.ac.cn}

\abstract{Modularity determines the elliptic genus of $\symK$ at finite $N$ from the indexed degeneracies of polar states, which are states that lie below the black hole threshold in AdS$_3$/CFT$_2$. We propose that each polar degeneracy is reproduced exactly by one-loop contributions from a finite set of asymmetric $\orbifoldK$ backgrounds in type IIB string theory. We compute the classical charges of these backgrounds using supergravity and their one-loop $\frac{1}{4}$-BPS partition functions using the tensionless worldsheet with a Gauss constraint. We match these results with the residues of the grand-canonical elliptic genus at its linear poles, which determine the polar degeneracies. 
Using the chamber structure defined by these poles and labeled by Farey pairs, we give for each polar charge a contour prescription that specifies the complete finite sum of bulk orbifold contributions to its degeneracy.
Finally we interpret the charge-dependent contour choice in terms of the phase structure of the microcanonical charge lattice. 
}

\preprint{}

\begin{document}
\setcounter{tocdepth}{2}
\maketitle
\setcounter{tocdepth}{3}

\newpage

\section{Introduction, proposal, and summary}
\label{sec: intro}

Type IIB string theory on $\adsK$ with $Q_1$ and $Q_5$ units of flux is dual to a two-dimensional $\cN=(4,4)$ superconformal field theory with central charge $\mathsf{c}=6N$, whose moduli space contains the symmetric orbifold $\symK$ with $N = Q_1 Q_5 + 1$ \cite{Maldacena:1997re}. The elliptic genus of $\symK$ counts $\frac{1}{4}$-BPS states with signs and is invariant under exactly marginal deformations. At large $N$, this elliptic genus was matched with the Kaluza-Klein spectrum of supergravity on $\adsK$ \cite{deBoer:1998kjm,deBoer:1998us}.

At any finite $N \sim 1/G_N^{(3)}$, the elliptic genus is constrained exactly by modular covariance where $N = \mathsf{c}/6$ is its index as a weight-$0$ index-$N$ weak Jacobi form, which can be determined by a finite number of polar coefficients at a given value of $N$ \cite{Eichler:1985}. Therefore, the full elliptic genus of $\symK$ at finite $N$, including the index degeneracies of states above the black hole threshold, is fixed by the polar degeneracies as an input \cite{Dijkgraaf:2000fq,Manschot:2007ha,Gaberdiel:2008xb,Benjamin:2015hsa}.

Let us make this precise. The elliptic genus of $\symK$ is
\ie \label{eq: elliptic genus}
Z^{\rm RR}_N(q,y) &= \Tr_{\rm RR} \left[ (-1)^F q^{L_0 - \frac{\mathsf{c}}{24}} \qb^{\Lb_0 - \frac{\mathsf{c}}{24}} y^{2 J_0} \right] \\
&= \sum_{m\geq0,\,\ell} d(m,\ell)\, q^m y^\ell ,
\fe
where $q = e^{2 \pi i \tau}$, $y = e^{2 \pi i z}$, $(-1)^F = e^{2 \pi i (J_0 - \Jb_0)}$, and $J_0$, $\Jb_0$ are the Cartan generators of the left- and right-moving $SU(2)$ R-symmetries. We will use $(\tau,z)$ and $(q,y)$ interchangeably. The elliptic genus $Z^{\rm RR}_N$ is a weak Jacobi form of weight $0$ and index $N$ \cite{Eichler:1985}:
\ie \label{eq: Jacobi intro}
Z^{\rm RR}_N \left( \frac{a\tau+b}{c \tau + d} , \frac{z}{c \tau + d} \right) &= e^{2 \pi i N \frac{c z^2}{c \tau + d}} \, Z^{\rm RR}_N(\tau,z), \\
Z^{\rm RR}_N(\tau,z + \lambda \tau + \mu) &= e^{-2\pi i N (\lambda^2 \tau + 2 \lambda z)} \, Z^{\rm RR}_N(\tau,z),
\fe
where
\be
\begin{pmatrix} a & b \\ c & d \end{pmatrix} \in SL(2,\bZ), \qquad \lambda,\mu \in \bZ.
\ee
These properties imply that $d(m,\ell)$ depends only on $4Nm - \ell^2$ and on $\ell$ mod $2N$.

We call a coefficient $d(m,\ell)$ \textit{polar} in the Ramond sector if
\be
4Nm - \ell^2 < 0,
\ee
and we define the \textit{polar part} of the $\symK$ elliptic genus as
\be \label{eq: polar part intro}
Z^{\rm RR}_N \big|_{\rm polar} = \sum_{4Nm - \ell^2 < 0} d(m,\ell)\, q^m y^\ell.
\ee
There are finitely many independent polar coefficients modulo spectral flow, and the polar coefficients determine the elliptic genus completely \cite{Dijkgraaf:2000fq,Manschot:2007ha,Gaberdiel:2008xb,Benjamin:2015hsa}.\footnote{Schematically, we have $Z^{\rm RR}_N = \sum_{\gamma \in \Gamma_\infty \backslash SL(2,\bZ)} \big( Z^{\rm RR}_N|_{\rm polar} \big)\big|_\gamma$ where the sum over images requires a regularization \cite{Dijkgraaf:2000fq,Manschot:2007ha}.} In the bulk, the Fareytail expansion uses the polar part as an input to build the non-polar degeneracies as $SL(2,\bZ)$ images of the polar part, where the non-trivial images are interpreted in terms of BTZ black holes \cite{Dijkgraaf:2000fq,Manschot:2007ha,Kraus:2006nb}. The polar degeneracies themselves, however, are not computed by the Fareytail expansion and their AdS$_3$ bulk origin was left as an open question in \cite{Dijkgraaf:2000fq}.

\begin{figure}[t]
\centering
\includegraphics[width=8cm]{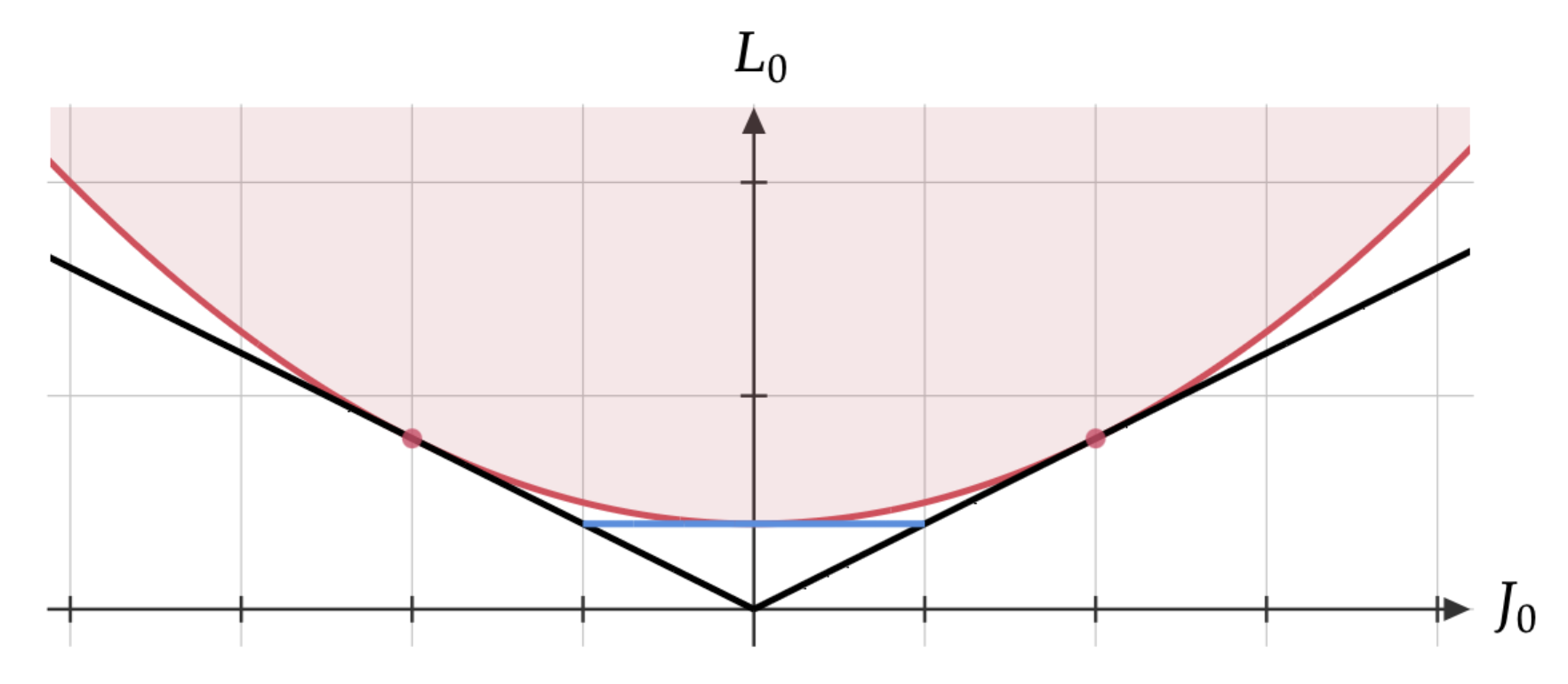}
\caption{Left-moving microcanonical diagram in the NS sector at large $N$. The polar states lie below the black hole threshold parabola $4NL_0 - (2J_0)^2 = N^2$ in red. The black lines correspond to the unitarity bound $L_0 \geq |J_0|$, and it touches the parabola at $|J_0| = \frac{N}{2}$. The blue segment corresponds to the de Boer bound \eqref{eq: de Boer bound intro} $L_0 \leq \frac{N+1}{4}$ at large $N$.}
\label{fig: polar intro}
\end{figure}

The NS-NS elliptic genus
\be \label{eq: NS elliptic genus intro}
Z^{\rm NS}_N(q,y) = q^{\frac{N}{2}} y^N Z^{\rm RR}_N(q, q^{\frac{1}{2}} y),
\ee
is obtained from the RR elliptic genus \eqref{eq: elliptic genus} via one unit of spectral flow in our normalization. The NS polarity condition is then
\be \label{eq: polar NS intro}
4 N L_0 - (2 J_0)^2 < N^2 ,
\ee
so the NS polar states are those below the black hole threshold \cite{Cvetic:1998xh} in Figure \ref{fig: polar intro}.

Let us highlight a long-standing puzzle concerning the polar states. 
In \cite{deBoer:1998us}, it was found that the large $N$ NS elliptic genus $Z^{\rm NS}_N(q,y)$ of $\symK$ agrees with the index computed from the Kaluza-Klein spectrum on $\adsK$ for
\be \label{eq: de Boer bound intro}
L_0 \leq \frac{N+1}{4}
\ee
and that the two differ beyond this so-called de Boer bound. (The comparison in \cite{deBoer:1998us} requires one to apply an ``exclusion principle'' on the Kaluza-Klein spectrum, under which the supergravity elliptic genus becomes finite and $N$-dependent. We will return to this point in Section \ref{subsec: degenerate saddles intro}.) That is, the polar states above this bound are accounted neither by Kaluza-Klein states nor by black holes. What, then, is the AdS$_3$ bulk origin of the polar degeneracies at finite $N$?

In \cite{Lee:2025veh}, we addressed the analogous question for the chiral-chiral primary states of $\symM$ at finite $N$, where $\Mfour = \Tfour$ or $\Kthree$. There, we proposed that the finite $N$ chiral-chiral primary spectrum is reproduced exactly by a sum over one-loop half-BPS partition functions of $\orbifoldM$ orbifolds and their spectral flows. These bulk partition functions were identified with residues of the grand-canonical partition function with respect to $N$. In this work, we propose an analogous answer for the polar part of the $\frac{1}{4}$-BPS elliptic genus of $\symK$.

\subsection{Preliminaries}

The elliptic genera of $\symK$ for all $N$ are generated by the Dijkgraaf-Moore-Verlinde-Verlinde (DMVV) formula \cite{Dijkgraaf:1996xw}
\be \label{eq: DMVV}
\cZ_{\rm RR}^{\textrm{GCE}}(\zeta,q,y) = \sum_{N=0}^\infty \zeta^N Z^{\rm RR}_{N}(q,y) = \prod_{n>0,\,m\geq0,\,\ell} \frac{1}{(1-\zeta^n q^m y^\ell)^{c(nm,\ell)}} ,
\ee
where $\zeta = e^{2\pi i \sigma}$ and $c(m,\ell)$ are the coefficients of the $\Kthree$ elliptic genus\footnote{We define $\eta(\tau) = q^{\frac{1}{24}} \prod_{n=1}^\infty (1-q^n)$ and
\ieg
\theta_1(\tau,z) = -i\,q^{\frac{1}{8}} y^{\frac{1}{2}} \prod_{n=1}^\infty (1-q^n)(1-yq^n)(1-y^{-1}q^{n-1}), \ \ \theta_3(\tau,z) = \prod_{n=1}^\infty (1-q^n)(1+yq^{n-\frac{1}{2}})(1+y^{-1}q^{n-\frac{1}{2}}), \\
\theta_2(\tau,z) = q^{\frac{1}{8}} y^{\frac{1}{2}} \prod_{n=1}^\infty (1-q^n)(1+yq^n)(1+y^{-1}q^{n-1}), \ \ \theta_4(\tau,z) = \prod_{n=1}^\infty (1-q^n)(1-yq^{n-\frac{1}{2}})(1-y^{-1}q^{n-\frac{1}{2}}).
\feg}
\ie \label{eq: K3 elliptic genus}
\phi_{\Kthree}(q,y) &= \sum_{m\geq0,\,\ell} c(m,\ell) \, q^m y^\ell \\
&= 8 \left[ \left( \frac{\theta_2(\tau,z)}{\theta_2(\tau,0)} \right)^2 + \left( \frac{\theta_3(\tau,z)}{\theta_3(\tau,0)} \right)^2 + \left( \frac{\theta_4(\tau,z)}{\theta_4(\tau,0)} \right)^2 \right] .
\fe
We refer to $\cZ_{\rm RR}^{\textrm{GCE}}$ as the grand-canonical elliptic genus. The fugacity $\zeta$ is conjugate to the number of copies $N = \mathsf{c}/6 \sim 1/G_N^{(3)}$.

The finite $N$ elliptic genus can be recovered from $\cZ_{\rm RR}^{\textrm{GCE}}$ via the contour integral
\be \label{eq: contour intro}
Z^{\rm RR}_N(q,y) = \oint_{\zeta=0} \frac{d\zeta}{2\pi i} \, \zeta^{-N-1} \, \cZ_{\rm RR}^{\textrm{GCE}}(\zeta,q,y).
\ee
taken around $\zeta = 0$. As in \cite{Lee:2025veh}, our strategy will be to deform this contour away from $\zeta = 0$ and to interpret the non-trivial singularities of $\cZ_{\rm RR}^{\textrm{GCE}}$ in $\zeta$ as contributions of bulk saddles \cite{DeLange:2018wbz,Eberhardt:2021jvj}. To do so, we need to understand the singularity structure of $\cZ_{\rm RR}^{\textrm{GCE}}$ on the $\zeta$-plane.

The grand-canonical elliptic genus is related to the Igusa cusp form $\Phi_{10}$ by \cite{Dijkgraaf:1996it}
\be \label{eq: Phi10 intro}
\cZ_{\rm RR}^{\textrm{GCE}}(\zeta,q,y) = \frac{- \zeta \, \eta(\tau)^{18} \theta_1(\tau,z)^2}{\Phi_{10}(\sigma,\tau,z)}.
\ee
The chemical potentials take values in the Siegel upper half-plane $\bH_2$:
\be \label{eq: Siegel intro}
(\Im \sigma) (\Im \tau) - (\Im z)^2 > 0, \quad \Im \sigma > 0, \quad \Im \tau > 0.
\ee
Since the numerator of \eqref{eq: Phi10 intro} depends on $\zeta$ only by an overall factor, all non-trivial singularities of $\cZ_{\rm RR}^{\textrm{GCE}}$ in the $\sigma$-plane are inherited from $\Phi_{10}^{-1}$.\footnote{The series expressions $\Phi_{10}^{-1} = \zeta^{-1}q^{-1}y^{\pm 1}+ \cdots$ and $\cZ_{\rm RR}^{\textrm{GCE}} = 1 + \cdots$ result in an offset in the definitions of polarity between the two contexts. The polar data of $\cZ_{\rm RR}^{\textrm{GCE}}$ is strictly contained in that of $\Phi_{10}^{-1}$.} The Fourier coefficients of $\Phi_{10}^{-1}$ exhibit wall-crossing behavior \cite{Sen:2007vb,Dabholkar:2007vk,Cheng:2007ch,Sen:2011mh,DabholkarMurthyZagier2012} that is not present in $\cZ_{\rm RR}^{\textrm{GCE}}$ due to the fact that the latter is fixed by $\symK$ to be in the chamber
\be
0 < |\Im z| \ll \Im \tau \ll \Im \sigma.
\ee
However, one encounters singularities in the $\sigma$-plane as one deforms the contour away from this chamber while keeping $\tau,z$ fixed.

The poles of $\Phi_{10}^{-1}$ occur along divisors in $\bH_2$ \cite{Borcherds1995,Kawai:1995hy,Harvey:1995fq} and are classified into two types: linear poles and quadratic poles. It was shown in \cite{Sen:2011mh,Belin:2018oza,Belin:2019jqz,Chowdhury:2019mnb} that the residues at the linear poles determine the polar data of $\Phi_{10}^{-1}$ entirely. The linear poles that descend to the singularities of $\cZ_{\rm RR}^{\textrm{GCE}}$ on the $\sigma$-plane (as double poles) lie along the divisors
\be \label{eq: linear poles}
K \sigma + M \tau + S z - R = 0
\ee
where the integers $(K, M, S, R)$ satisfy\footnote{These restrictions can be imposed using the $Sp(4,\bZ)$ symmetry of $\Phi_{10}^{-1}$. 
We fix the overall sign by choosing $K>0$. 
The $K=0$ divisors are canceled by the numerator of \eqref{eq: Phi10 intro}. 
The remaining follow from the constraints $KM=\frac{1}{4}(S^2-1)$ and $\sigma \sim \sigma+1$.}
\ieg \label{eq: linear poles conditions}
K \in \bZ_{>0}, \quad M \in \bZ_{\geq 0}, \quad S \in 2\bZ+1, \quad R=0,1,\cdots, K-1, \\
K M = \frac{1}{4}(S^2 - 1).
\feg
On the $\zeta$-plane, these poles are double poles at
\be \label{eq: linear poles zeta intro}
\zeta = e^{2\pi i \frac{R}{K}} \, q^{-\frac{M}{K}} y^{-\frac{S}{K}}.
\ee
The residues at these poles will determine the polar degeneracies of $Z^{\rm RR}_N$.

The linear poles \eqref{eq: linear poles} differ in an essential way from the singularities of the grand-canonical partition function considered for chiral-chiral primaries in \cite{Lee:2025veh}: the linear poles are double poles in $\zeta$. This reflects the fact that the elliptic genus does not distinguish states with differing right-moving R-charges $\Jb_0$. This means that, if we refine $\cZ_{\rm RR}^{\textrm{GCE}}$ by an extra fugacity for $\Jb_0$ (at the cost of at least initially computing a non-protected quantity), we can resolve each double pole into a pair of simple poles. 

The refinement we consider is the Hodge elliptic genus (at the Kummer orbifold point) \cite{Kachru:2016igs,Benjamin:2016pil}, i.e.\ the $\frac{1}{4}$-BPS partition function
\be \label{eq: hodge genus}
Z^{\rm RR}_N(q,y,\bar{y}) = \lim_{\qb \to 0} \Tr_{\rm RR} \left[ (-1)^F q^{L_0 - \frac{\mathsf{c}}{24}} \qb^{\Lb_0 - \frac{\mathsf{c}}{24}} y^{2J_0} \bar{y}^{2\Jb_0} \right],
\ee
where we introduce an additional fugacity $\bar{y}$ for the right-moving R-charge $\Jb_0$. This reduces to the elliptic genus $Z^{\rm RR}_N(q,y)$ in \eqref{eq: elliptic genus} at $\bar{y}=1$:
\be
Z^{\rm RR}_N(q,y,1) = Z^{\rm RR}_N(q,y).
\ee
Although the Hodge elliptic genus is \emph{not} protected for $\bar{y} \neq 1$, $\bar{y}$ will serve as a regulator that lifts the right-moving degeneracy between pairs of bulk saddles. The NS Hodge elliptic genus is
\be
Z^{\rm NS}_N(q,y,\bar{y}) = q^{\frac{N}{2}} y^N \bar{y}^N Z^{\rm RR}_N(q, q^{\frac{1}{2}} y, \bar{y}),
\ee
and this also reduces to the NS elliptic genus \eqref{eq: NS elliptic genus intro} at $\bar{y}=1$:
\be
Z^{\rm NS}_N(q,y,1) = Z^{\rm NS}_N(q,y).
\ee

The same refinements apply to the grand-canonical elliptic genus. At the symmetric orbifold point, the Hodge refinement of \eqref{eq: DMVV} is
\ie \label{eq: DMVV hodge intro}
\cZ_{\rm RR}^{\textrm{GCE}}(\zeta,q,y,\bar{y}) &= \sum_{N=0}^\infty \zeta^N Z^{\rm RR}_N(q,y,\bar{y}) \\
&= \prod_{n>0,\,m\geq0,\,\ell,\,\bar{\ell}} \frac{1}{(1-\zeta^n q^m y^\ell \bar{y}^{\bar{\ell}})^{c(nm,\ell,\bar{\ell})}},
\fe
where $c(m,\ell,\bar{\ell})$ are the coefficients of the RR Hodge genus of the seed $\Kthree$\footnote{The function $\theta_1$ of \cite{Kachru:2016igs} is $i$ times ours.}
\be \label{eq: K3 hodge intro}
Z_{\Kthree}(q,y,\bar{y}) = 8\Bigg[ \left(\frac{\theta_1(\tau,z)}{\theta_1^*(\tau,0)} \bar{y}_- \right)^2 + \left(\frac{\theta_2(\tau,z)}{\theta_2(\tau,0)} \bar{y}_+ \right)^2 + \left(\frac{\theta_3(\tau,z)}{\theta_3(\tau,0)} \right)^2 + \left(\frac{\theta_4(\tau,z)}{\theta_4(\tau,0)} \right)^2 \Bigg],
\ee
where $\bar{y}_+ = \frac{1}{2}(\bar{y}^{\frac{1}{2}} + \bar{y}^{-\frac{1}{2}})$, $\bar{y}_- = \frac{1}{2i}(\bar{y}^{\frac{1}{2}} - \bar{y}^{-\frac{1}{2}})$, and $\theta_1^*(\tau,0) = 2\eta(\tau)^3$. In the above, we have $\bar{\ell} \in \{-1,0,1\}$ because the Hodge genus projects to the right-moving Ramond ground states. The grand-canonical NS-NS Hodge genus is
\ie \label{eq: GC NS intro}
\cZ_{\rm NS}^{\textrm{GCE}}(\zeta,q,y,\bar{y}) &= \sum_{N=0}^\infty \zeta^N Z^{\rm NS}_N(q,y,\bar{y}) \\
&= \cZ_{\rm RR}^{\textrm{GCE}}(\zeta q^{\frac{1}{2}} y \bar{y}, \, q, \, q^{\frac{1}{2}} y, \, \bar{y}).
\fe
Both the RR and NS-NS grand-canonical Hodge genera reduce to their unrefined, protected versions at $\bar{y}=1$.

In the following, we will find that this auxiliary refinement is natural from the bulk perspective, where the residue at each simple pole corresponds to a single classically $\frac{1}{4}$-BPS geometry. Doing so will also clarify an old puzzle about the $N$-dependence of the large $N$ elliptic genus in Section \ref{subsec: degenerate saddles intro}.

\subsection{Polar states in AdS$_3$ holography}
\label{subsec: polar states intro}

In this subsection, we identify the residues of the grand-canonical Hodge elliptic genera with the one-loop $\frac{1}{4}$-BPS partition functions of the IIB theory on asymmetric orbifolds of $\adsK$ at the tensionless point. While these partition functions depend individually on $\bar{y}$, the sum over each pair $\Sb = \mp 1$ has a smooth limit $\bar{y} \to 1$ that recovers the elliptic genus. We then state our main proposal and explain how it should be read microcanonically.

In $\cZ_{\rm RR}^{\textrm{GCE}}(\zeta,q,y,\bar{y})$, the fugacity $\bar{y}$ resolves each double linear pole \eqref{eq: linear poles zeta intro} into a pair of simple poles
\be \label{eq: simple poles intro}
\zeta = e^{2\pi i \frac{R}{K}} \, q^{-\frac{M}{K}} y^{-\frac{S}{K}} \bar{y}^{-\frac{\Sb}{K}},
\ee
with $\Sb = \mp 1$, that coincide at $\bar{y}=1$. 
(The polar coefficients of the seed are $c(m,\ell,\pm1)=1$ and $c(m,\ell,0)=0$ for $4m-\ell^2=-1$.) 
In $\cZ_{\rm NS}^{\textrm{GCE}}(\zeta,q,y,\bar{y})$, these poles flow to
\be \label{eq: simple poles NS intro}
\zeta = e^{2\pi i \frac{R}{K}} \, q^{-\frac{1}{2}(1+\frac{S}{K})-\frac{1}{4K^2}(S^2-1)} y^{-(1+\frac{S}{K})} \bar{y}^{-(1+\frac{\Sb}{K})}
\ee
with $\Sb = \mp 1$ and where we used $M = \frac{1}{4K}(S^2-1)$. 
In particular, $\zeta^{-N}$ evaluated at these poles is the classical prefactor in \eqref{eq: NS2 hodge}. 
For fixed $(K,S,\Sb)$, we sum the contributions of the simple poles \eqref{eq: simple poles NS intro} over $R=0,1,\cdots,K-1$ in the contour integral that extracts the fixed $N$ NS Hodge elliptic genus. 
We identify this sum with the one-loop $\frac{1}{4}$-BPS partition function $\Zhat_{K}^{(S,\Sb)}$ of the IIB theory on a single asymmetric orbifold $\orbifoldK$ background in the tensionless limit.\footnote{This is entirely analogous to the identification of \cite{Lee:2025veh} in the simpler context of chiral-chiral primaries.} Here, $S$ and $\Sb$ specify the twists in the $SU(2)_{\textrm{L}}$ and $SU(2)_{\textrm{R}}$ isometries of $S^3$ accompanying the AdS$_3$ quotient as defined in Section~\ref{sec: classical}. The two orbifold backgrounds with fixed $(K,S)$ and $\Sb=\mp1$ are related by $2/K$ units of right-moving spectral flow.
 
Explicitly, the one-loop NS-NS Hodge elliptic genus on the $\orbifoldK$ background in the tensionless limit is
\ie \label{eq: NS2 hodge}
\Zhat_{N,(K,S,\Sb)}^{\textrm{NS}}(q,y,\bar{y}) = \ &q^{\frac{N}{2}(1+\frac{S}{K})+\frac{N}{4K^2}(S^2 -1)} y^{N(1+\frac{S}{K})} \bar{y}^{N(1+\frac{\Sb}{K})} \\
&\times \frac{1}{K} \sum_{R=0}^{K-1} e^{-2\pi i N \frac{R}{K}}\PE\left[ f_{(K,S,\Sb,R)}^{\textrm{NS}}(q,y,\bar{y}) \right].
\fe
The single-string spectrum
\be \label{eq: NS2 single}
f_{(K,S,\Sb,R)}^{\textrm{NS}}(q,y,\bar{y}) = \sum_{n=1}^\infty \Big[ e^{2\pi i n \frac{R}{K}} q^{-\frac{n}{K}(M+\frac{S}{2})} y^{-\frac{S n}{K}} \bar{y}^{-\frac{\Sb n}{K}} \cdot \frac{1}{n} \sum_{r=0}^{n-1} Z_{\Kthree}(e^{2\pi i \frac{r}{n}} q^{\frac{1}{n}},q^{\frac{1}{2}} y,\bar{y}) \Big] -1
\ee
is expressed in terms of the RR Hodge elliptic genus \eqref{eq: K3 hodge intro} of $\Kthree$. The plethystic exponential is defined as $\PE[f(\mathfrak{q})] = \exp \! \left( \sum_{n=1}^\infty \frac{1}{n} f(\mathfrak{q}^n) \right)$ where $\mathfrak{q}$ here denotes all fugacities collectively.

Each component of \eqref{eq: NS2 hodge} admits a direct physical interpretation. The prefactors with powers proportional to $N$ encode the classical conformal dimensions and R-charges of the $\orbifoldK$ geometry that we derive from supergravity in Section \ref{sec: classical}. The plethystic exponential builds the one-loop multi-string spectrum from the single-string spectrum $f_{(K,S,\Sb,R)}^{\textrm{NS}}(q,y,\bar{y})$, which we derive via the tensionless worldsheet in Section \ref{sec:worldsheet}. The sum over the $K$-th roots-of-unity $R = 0, 1, \cdots, K-1$ imposes a Gauss constraint on the spacetime spectrum. The orbifold labels $(K,S,\Sb)$ are subject to the constraints in \eqref{eq: orbifold constraint}. These constraints will be shown to arise from physical consistency conditions of asymmetric orbifold backgrounds in Sections \ref{sec: classical} and \ref{sec:worldsheet}.

As can be seen in \eqref{eq: NS2 hodge}, the classical actions of each pair $\Sb=\mp1$ of geometries become degenerate at $\bar{y} = 1$. In fact, it can be found that the one-loop spectrum of individual $\Zhat_{(K,S,\Sb)}^{\textrm{NS}}(q,y,\bar{y})$ diverges as $\bar{y} \to 1$. Nevertheless, the combined contribution 
\be \label{eq: residue pair intro}
\Zhat_{(K,S)}^{\textrm{NS}}(q,y) \equiv \lim_{\bar{y} \to 1} \sum_{\Sb = \pm 1} \Zhat_{(K,S,\Sb)}^{\textrm{NS}}(q,y,\bar{y})
\ee
can be shown to be finite, and we hypothesize that the combination \eqref{eq: residue pair intro} is a protected quantity. Each linear pole \eqref{eq: linear poles} thus corresponds to a pair of $\orbifoldK$ backgrounds whose classical actions become degenerate in the elliptic genus limit $\bar{y}=1$.

We propose that the polar part of the elliptic genus of $\symK$ at finite $N$ is given exactly by a sum over one-loop supersymmetric partition functions $\Zhat_{(K,S,\Sb)}$ of IIB string theory on asymmetric $\orbifoldK$ orbifold backgrounds:
\be \label{eq: proposal}
Z_N(q,y) \Big\vert_{\rm polar} = \sum_{K,S,\Sb} \Zhat_{(K,S,\Sb)}(q,y,\bar{y}) \bigg\vert_{\substack{\bar{y}=1 \\ {\rm polar}}},
\ee
where the limit $\bar{y} \to 1$ is taken after summing over $\Sb$, and this formula applies for both NS and Ramond sectors. 
The sum runs over
\ieg \label{eq: orbifold constraint}
K \in \bZ_{>0}, \quad S \in 2\bZ+1, \quad \Sb = \mp 1, \\
M = \frac{1}{4 K}(S^2 - 1) \in \bZ_{\geq 0}.
\feg
While in our work we compute $\Zhat_{(K,S,\Sb)}(q,y,\bar{y})$ in the tensionless limit, our proposal \eqref{eq: proposal}, if true, should not depend on the position in the moduli space in which it is computed.

The sum \eqref{eq: proposal} is best understood as a microcanonical statement (and necessarily so, due to the microcanonical restriction to the polar coefficients). For the microcanonical analysis, it is convenient to work in the Ramond sector as in Sections~\ref{sec:ExpansionCE} and \ref{sec:ExpansionMCE}.

Consider a polar charge $Q = (N,m,\ell)$ in the Ramond sector. To find the degeneracy $d(Q) = d(m,\ell)$, we first deform the contour in \eqref{eq: contour intro} away from $\zeta = 0$ at a fixed ratio $\frac{\Im z}{\Im \tau}$ and then extract the coefficient of $q^m y^\ell$ at that ratio via their contour integrals. At any given stage of the deformation, $d(Q)$ would then be the contribution of the deformed contour plus the sum of the contributions $\Res_{(K,S)}(Q)$ of the linear poles picked up until that point. The contribution $\Res_{(K,S)}(Q)$ of each linear pole is obtained as follows. We take the residue of $\cZ_{\rm RR}^{\textrm{GCE}}(\zeta,q,y)$ at the double pole in $\sigma$ labeled by $(K,S)$ and sum over the $K$-th roots-of-unity $R$. The coefficient $\Res_{(K,S)}(Q)$ of $q^m y^\ell$ can be extracted by either performing the contour integrals over $q$ and $y$ while keeping the relevant $\frac{\Im z}{\Im \tau}$ fixed or by expanding $\cZ_{\rm RR}^{\textrm{GCE}}(\zeta,q,y)$ in the region set by the ratio $\frac{\Im z}{\Im \tau}$ as is done in Section~\ref{sec:ExpansionCE}.

The set of poles that we pick up as we deform the $\sigma$ contour depends on the ratio $\frac{\Im z}{\Im \tau}$, and this set jumps if we change the ratio (Section~\ref{sec:ExpansionCE}). This is a Stokes phenomenon, similar to the one found for chiral primaries in \cite{Lee:2025veh}. By contrast, the degeneracy $d(Q)$ given by the DMVV formula should not depend on the ratio, so we are therefore free to choose the ratio depending on $Q$. We say that the deformation terminates at a given ratio $\frac{\Im z}{\Im \tau}$ if the deformed contour eventually stops contributing to $d(Q)$. For such a ratio, we find that only a finite number of poles contribute. The remaining poles do not contribute to $d(Q)$ for one of two reasons: (1) The poles lies outside $\bH_2$ at a given ratio or (2) the pole is picked up at that ratio but the corresponding geometries contribute only to higher charges than $Q$.

Microcanonically, our proposal \eqref{eq: proposal} thus states that
\be \label{eq: MCE intro}
d(Q) = \sum_{(K,S)} \mathrm{Res}_{(K,S)}(Q),
\ee
where the sum runs over the finitely many poles crossed at a ratio $\frac{\Im z}{\Im \tau}$ for which the deformation terminates. Since spectral flow relates the polar coefficients of the RR and NS-NS sectors, this means that each polar coefficient receives contributions from only a finite number of orbifold geometries.

Such a ratio is not unique in general, but by the residue theorem each of these sets of poles reproduces $d(Q)$. Our proposal \eqref{eq: MCE intro} therefore holds for any such ratio, but the set of geometries contributing to a given polar coefficient depends on this choice. That is, the sum over the orbifold labels in \eqref{eq: proposal} should be understood microcanonically as in \eqref{eq: MCE intro}, supplemented with a ratio $\frac{\Im z}{\Im \tau}$ for which the deformation terminates. In Section~\ref{sec:ExpansionMCE}, we will fix this ratio for each polar charge by imposing an additional condition.\footnote{We leave a bulk derivation of this charge-dependent contour prescription for future work, perhaps along the lines of the dyon-counting analysis of \cite{Cheng:2007ch}.}

\subsection{Degenerate saddles and the large $N$ elliptic genus}
\label{subsec: degenerate saddles intro}

Let us now see how the degenerate pairs of saddles clarify an old puzzle about the $N$-dependence of the large $N$ elliptic genus.

In the NS sector, the Fourier coefficients of the elliptic genus of $\symK$ depend non-trivially on $N$, even at low conformal weights. The Kaluza-Klein spectrum on a single $\adsK$ background, on the other hand, does not depend on $N$. In \cite{deBoer:1998us}, the two were matched at large $N$ by imposing an additional exclusion principle on the Kaluza-Klein spectrum under which the supergravity elliptic genus becomes explicitly $N$-dependent.

Our proposal \eqref{eq: proposal} suggests a different bulk origin for the $N$-dependence of the large $N$ elliptic genus. At $K=1$, we find that one should account for the Kaluza-Klein spectrum of not only the vacuum $\adsK$ with $(S,\Sb) = (-1,-1)$ but also its pair $(S,\Sb) = (-1,+1)$ related by two units of right-moving spectral flow:
\be \label{eq: K1 partner intro}
\Zhat^{\textrm{NS}}_{N,(1,-1,+1)}(q,y,\bar{y}) = \bar{y}^{2N} \, \Zhat^{\textrm{NS}}_{N,(1,-1,-1)}(q,y,\bar{y}^{-1})
\ee
which follows from \eqref{eq: NS2 hodge}. 
The one-loop partition functions $\Zhat_{1}^{(-1,\Sb)}(q,y,\bar{y})$ of the two geometries depend on $N$ only via their classical weights $1$ and $\bar{y}^{2N}$. Nevertheless, their combination \eqref{eq: residue pair intro} recovers precisely the non-trivial $N$-dependence of the elliptic genus:
\ie \label{eq: K1 pair intro}
\Zhat^{\textrm{NS}}_{N,(1,-1)}(q,y) &= (N+1) + q^{1/2} \left[ (22N-2) y^{-1} + (22N-2) y \right] + q \bigg[ (277N-323)y^{-2} \\
&+ (464N-592) + (277N-323) y^2 \bigg] + q^{3/2} \bigg[ (2576 N - 5752) y^{-3} \\
&+ (5652 N - 13716) y^{-1} + (5652 N - 13716) y + (2576 N - 5752) y^{3} \bigg] + O(q^2).
\fe
This agrees with the large $N$ NS elliptic genus of $\symK$ up to the de Boer bound \eqref{eq: de Boer bound intro}.

The mechanism for the $N$-dependence can already be seen in the leading term $N+1$ of \eqref{eq: K1 pair intro}. On the NS vacuum $(S,\Sb) = (-1,-1)$, the right-moving R-current at large $N$ generates a tower of right-moving chiral primary states with $\bar{L}_0 = \bar{J}_0 = 0,1,2,\cdots$. This would contribute to $\Zhat_{N,(1,-1,-1)}^{\textrm{NS}}$ as
\be
\frac{1}{1-\bar{y}^2}.
\ee
On the partner $(S,\Sb)=(-1,+1)$, the image of this tower under \eqref{eq: K1 partner intro} has $\bar{L}_0=\bar{J}_0=N,N-1,\cdots$ and contributes to $\Zhat_{N,(1,-1,+1)}^{\textrm{NS}}$ as
\be
\frac{\bar{y}^{2N}}{1-\bar{y}^{-2}} = \frac{- \bar{y}^{2N+2}}{1-\bar{y}^2}.
\ee
Together, they give
\be
\frac{1-\bar{y}^{2N+2}}{1-\bar{y}^2} \ \xrightarrow{\bar{y} \to 1} \ N+1.
\ee
This counts the $N+1$ right-moving chiral primaries with $\bar{L}_0 = \bar{J}_0 = 0,1,\cdots,N$ in the left-moving vacuum of $\symK$ at finite $N$. That is, the partner saddle truncates the tower on the vacuum at $\bar{L}_0 = \bar{J}_0 = N$ as required at finite $N$.\footnote{As in \cite{Lee:2025veh}, we expect the minus sign for $|\bar{y}|<1$ to be produced, in a one-loop gravitational path integral computation, via the need to Wick rotate the integration contours for an even number of bosonic negative modes. This has a direct analogy in the giant graviton expansion \cite{Imamura:2021ytr,Gaiotto:2021xce}, where the states prepared by path integrals with rotated contours \cite{Lee:2024hef} are interpreted as bulk duals of finite $N$ trace relations \cite{Lee:2023iil}.} This provides a bulk explanation of the $N$-dependence that was previously attributed to an additional exclusion principle \cite{Maldacena:1998bw}.

More generally, the same mechanism applies to every pair $\Sb = \mp 1$ of asymmetric orbifolds $\orbifoldK$ with given $(K,S)$: the sum over the pair of geometries related by $\frac{2}{K}$ units  of right-moving spectral flows generates a non-trivial $N$-dependence in their contribution to the elliptic genus.

\subsection*{Outline of the paper}

In Section~\ref{sec:ExpansionCE}, we derive a residue expansion for the polar part of the finite $N$ elliptic genus. In Section~\ref{sec: classical}, we study the asymmetric $\orbifoldK$ orbifold backgrounds and their classical charges. In Section~\ref{sec:worldsheet}, we compute the one-loop Hodge elliptic genus on $\orbifoldK$ using the tensionless worldsheet and derive the one-loop terms in the residues. In Section~\ref{sec:ExpansionMCE}, we express the polar degeneracies as a finite sum of residues, and in Section~\ref{sec:Phase} we use the result of Section~\ref{sec:ExpansionMCE} to propose a microcanonical phase diagram of the charge lattice. 
We conclude in Section~\ref{sec: discussion} with some open questions.

\section{Finite $N$ expansion in canonical ensemble}
\label{sec:ExpansionCE}

The canonical ensemble elliptic genus (in the Ramond sector) can be defined as the coefficient of $\myzeta^N$ in the $\myzeta$-series expansion of $\mathcal{Z}^{\textrm{GCE}}_{\mathrm{RR}}$:
\begin{equation}
Z^{\textrm{RR}}_{N}(q,y):=[\myzeta^N]\mathcal{Z}^{\textrm{GCE}}_{\mathrm{RR}}(\myzeta,q,y)\,,
\end{equation}
and it is a weight-$0$ index-$N$ weak Jacobi form.
It can also be evaluated via the contour integral
\begin{equation}\label{eq:ContourIntegralzeta0}
Z^{\textrm{RR}}_{N}(q,y)=\oint_{\myzeta=0}\frac{d\myzeta}{2\pi i}\,\myzeta^{-N-1}\, \mathcal{Z}^{\mathrm{GCE}}_{\mathrm{RR}}(\myzeta,q,y)\,.
\end{equation}
We will deform the contour to derive an alternative expression for $Z^{\textrm{RR}}_{N}$, which we will then interpret as the bulk evaluation of $Z^{\textrm{RR}}_{N}$.
Let us first review the chamber structure of $\mathcal{Z}^{\mathrm{GCE}}_{\mathrm{RR}}(\myzeta,q,y)$.

\subsection{Review: Linear poles, Farey pairs, and binary tree structure of CE chambers}

\subsubsection{Siegel upper half-plane}

Recall that the grand canonical ensemble elliptic genus (in the Ramond sector) is related to the Igusa cusp form $\Phi_{10}$ via \cite{Dijkgraaf:1996it,Dijkgraaf:1996xw}
\begin{equation}\label{eq:ZGCERRviaPhi10}
\mathcal{Z}^{\textrm{GCE}}_{\textrm{RR}}(\myzeta,q,y)
=\frac{-\myzeta\,\eta(\tau)^{18}\theta_1(\tau,z)^2}{\Phi_{10}(\sigma,\tau,z)}
\,,
\end{equation}
where $\Phi_{10}$ is the weight-$10$ genus-two Siegel modular form \cite{Siegel1939,Igusa1962} and $\eta(\tau)^{18}\theta_1(\tau,z)^2$ in the numerator is a weight-$10$ index-$1$ Jacobi form \cite{BruinierEtAl2008,Eichler:1985}. 

Due to the presence of the numerator, $\mathcal{Z}^{\textrm{GCE}}_{\textrm{RR}}$ is not a Siegel modular form. 
However, it is still convenient to package the chemical potentials into the genus-2 period matrix
\begin{equation}\label{eq:OmegaDef}
\Omega
=
\begin{pmatrix}
\sigma&z\\
z&\tau
\end{pmatrix}\,,
\end{equation}
which lives in the Siegel upper half-plane  \begin{equation}\label{eq:H2Def}
\mathbb{H}_2
=
\left\{
\Omega \in \mathrm{Mat}_{2\times 2}(\mathbb{C})
\,\middle|\,
\Omega = \Omega^{T}\,,
\mathrm{Det}[\mathrm{Im}\,\Omega] > 0
\,,
\mathrm{Tr}[\mathrm{Im}\,\Omega] > 0
\right\}\,.
\end{equation}
The magnitudes of the fugacities only depend on the position in its imaginary cone \cite{Cheng:2007ch}:
\begin{equation}\label{eq:H2ImagiaryCone}
\{
  (\mathrm{Im}\sigma,\mathrm{Im}\tau,\mathrm{Im} z)
\, | \,
\mathrm{Im}\tau >0\,,
\mathrm{Im}\sigma >0\,,
\mathrm{Im}\tau\,\mathrm{Im}\sigma-
\left(\mathrm{Im}z\right)^2>0
\}\,.
\end{equation}
For later convenience, as in \cite{Sen:2011mh}, we also projectivize the imaginary cone onto a 2D plane by defining
\begin{equation}\label{eq:normalized-coordinates}
X\equiv\frac{\mathrm{Im}z}{\mathrm{Im}\tau}\,, \qquad Y\equiv\frac{\mathrm{Im}\sigma}{\mathrm{Im}\tau}\,.
\end{equation}
In terms of the $(X,Y)$ coordinates, 
the boundary of the imaginary cone is a parabola
\begin{equation}\label{eq:BoundaryParabola}
Y=X^2\,.
\end{equation}

\subsubsection{Linear poles and Farey pairs}

In the DMVV formula \eqref{eq: DMVV} for 
$\mathcal{Z}^{\textrm{GCE}}_{\textrm{RR}}(\myzeta,q,y)$, there are factors 
$
1-\zeta^n q^m y^{\ell}  
$
in the denominator, which correspond to the \textit{linear poles} of $\mathcal{Z}^{\textrm{GCE}}_{\textrm{RR}}(\myzeta,q,y)$. 
Writing
\begin{equation}
1=\left|\zeta^{K}q^{M}y^{S}\right|=e^{-2\pi\,\mathrm{Im}\tau\,(KY+M+SX)}\,,
\end{equation}
we see that these linear poles correspond to the lines in the $(X,Y)$ plane, defined by
\begin{equation}\label{eq:LinearPole}
K Y+M+S X=0\,,
\end{equation}
with 
\begin{equation}
K\geq 1\,, \qquad 
S\in 2 \mathbb{Z}+1\,, \qquad
M\equiv \frac{S^2-1}{4K}\in \mathbb{Z}_{\geq 0}\,.
\end{equation}
These linear pole lines intersect with the boundary parabola $Y=X^2$ at two points
\begin{equation}\label{eq:IntersectionPoints}
[X_\textrm{L}, X_\textrm{R}]=[\frac{-S -1}{2K}\,,\,\frac{-S+ 1}{2K}]\,.
\end{equation}

To proceed, we introduce the \textit{Farey pair}, which is defined as \cite{Apostol1976}
\begin{equation}
  \left[
\frac{\alpha}{\gamma}\,,\,
\frac{\beta}{\delta}
\right]\,,
\qquad \textrm{satisfying} \quad 
\alpha,\beta\in\mathbb{Z}\,,\quad \gamma,\delta\in\mathbb{Z}_{>0}\,,
\quad 
\frac{\alpha}{\gamma}<\frac{\beta}{\delta}\,,
\quad 
\alpha\delta-\beta\gamma=-1\,.
\end{equation}
(From now on, we will always use $[
\tfrac{\alpha}{\gamma},
\tfrac{\beta}{\delta}]$ to denote Farey pairs.)
The crucial fact is that the two intersection points \eqref{eq:IntersectionPoints} (after reducing the fractions) always form a Farey pair \cite{Sen:2007vb,Mukherjee:2007nc,Cheng:2007ch}
\begin{equation}\label{eq:IntersectionFP}
[\frac{-S -1}{2K},\frac{-S+ 1}{2K}]
=\left[
\frac{\alpha}{\gamma}\,,\,
\frac{\beta}{\delta}
\right]\,,
\qquad
\frac{\alpha}{\gamma}<\frac{\beta}{\delta}\,,
\qquad
\alpha\delta-\beta\gamma=-1\,.
\end{equation}
Conversely, to each Farey pair $[
\tfrac{\alpha}{\gamma},
\tfrac{\beta}{\delta}]$, we can associate a linear pole labeled by $(K,M,S)$ via
\begin{equation}\label{eq:FPtoKMS}
K=\gamma \delta 
\,, \quad 
M=\alpha \beta
\,, \quad  
S=-(\alpha\delta+\beta\gamma)\,. 
\end{equation}
Note that since $(K,M,S)$ satisfy the constraint $4KM-S^2=-1$, we can also use the pair $(K,S)$ to label the linear pole.
To summarize, there is a one-to-one correspondence
\begin{equation}\label{eq:KStoFP}
\textrm{linear pole}\quad (K,S)
\quad \longleftrightarrow \quad 
\textrm{Farey pair} \quad [
\frac{\alpha}{\gamma},
\frac{\beta}{\delta}]\,,
\end{equation}
via the maps \eqref{eq:IntersectionFP} and \eqref{eq:FPtoKMS}.

\subsubsection{Binary tree structure of CE chambers}

The lines \eqref{eq:LinearPole} that correspond to linear poles of $\mathcal{Z}^{\textrm{GCE}}_{\textrm{RR}}(\myzeta,q,y)$ divide the $(X,Y)$ plane above the boundary parabola $Y=X^2$ into triangular chambers. 
As we will review now, these triangles form a binary tree structure \cite{Stern:1858,Brocot:1861,Cheng:2008fc}.

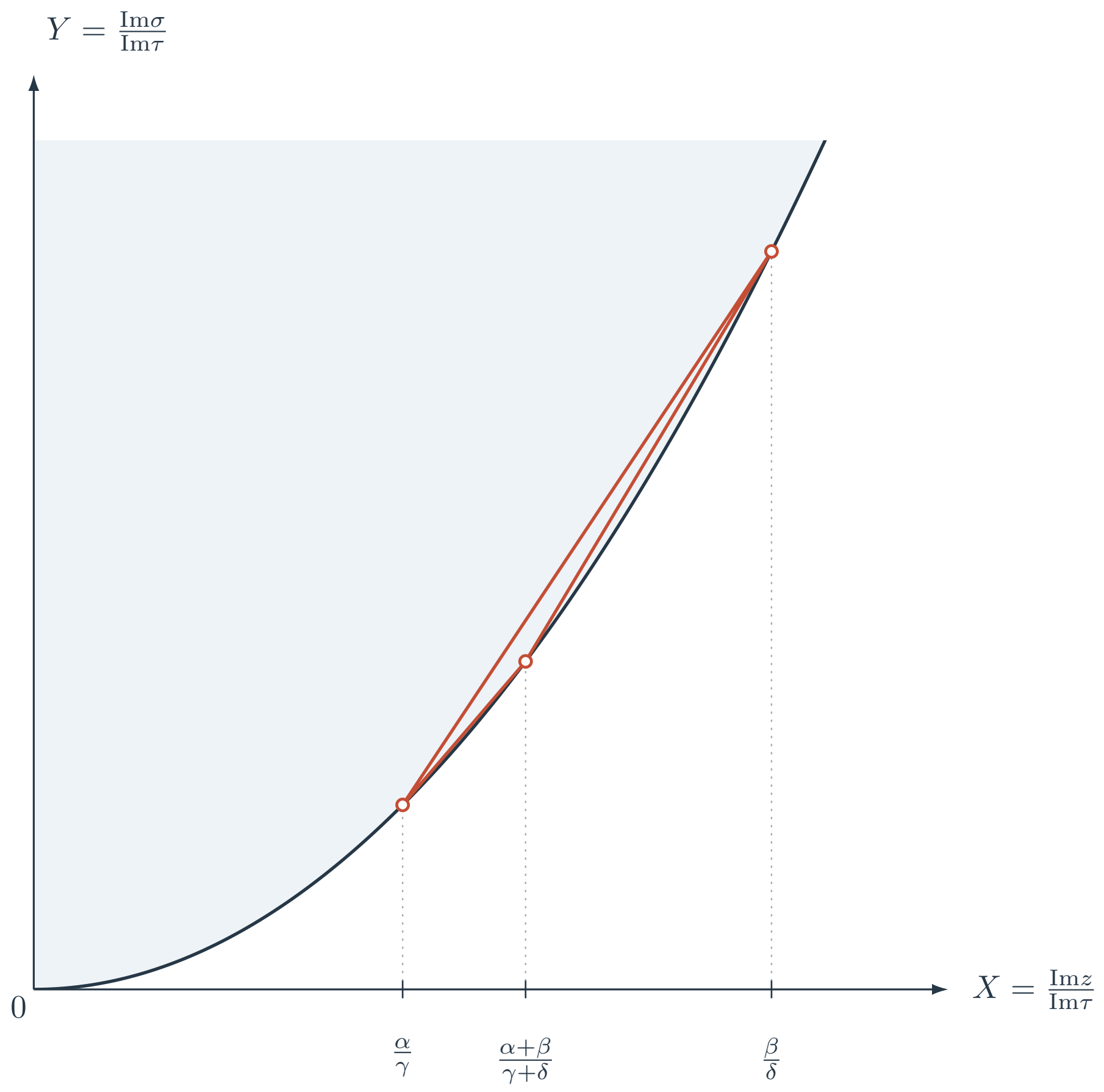
\begin{figure}[H]
\centering
\begin{tikzpicture}[paper figure,x=8cm,y=8cm]

  \begin{scope}
    \clip (0,0) rectangle (1.15,1.15);

    \path[fill=PaperDomain]
      (0,0)
      -- plot[domain=0:1.162,samples=180] (\x,{\x*\x})
      -- (1.162,1.35) -- (0,1.35) -- cycle;

    \draw[paper guide] (0.5,0) -- (0.5,0.25);
    \draw[paper guide] ({2/3},0) -- ({2/3},{4/9});
    \draw[paper guide] (1,0) -- (1,1);

    \draw[paper boundary]
      plot[domain=0:1.162,samples=180] (\x,{\x*\x});

    \draw[paper pole]
      (0.5,0.25) -- (1,1) -- ({2/3},{4/9}) -- cycle;
  \end{scope}

  \draw[paper axis] (0,0) -- (1.24,0);
  \draw[paper axis] (0,0) -- (0,1.24);

  \node[anchor=west] at (1.255,0)
    {$X=\frac{\textrm{Im}z}{\textrm{Im}\tau}$};
  \node[anchor=south west] at (0,1.255)
    {$Y=\frac{\textrm{Im}\sigma}{\textrm{Im}\tau}$};
  \node[anchor=north east,inner sep=2pt] at (0,0)
    {$0$};

  \draw[paper intersection]
    (0.5,0.25) circle[radius=1.8pt];
  \draw[paper intersection]
    ({2/3},{4/9}) circle[radius=1.8pt];
  \draw[paper intersection]
    (1,1) circle[radius=1.8pt];

  \draw[draw=PaperInk,line width=0.5pt]
    (0.5,0.012) -- (0.5,-0.012);
  \draw[draw=PaperInk,line width=0.5pt]
    ({2/3},0.012) -- ({2/3},-0.012);
  \draw[draw=PaperInk,line width=0.5pt]
    (1,0.012) -- (1,-0.012);

  \node[anchor=north,inner sep=0pt] at (0.5,-0.065)
    {$\frac{\alpha}{\gamma}$};
  \node[anchor=north,inner sep=0pt] at ({2/3},-0.065)
    {$\frac{\alpha+\beta}{\gamma+\delta}$};
  \node[anchor=north,inner sep=0pt] at (1,-0.065)
    {$\frac{\beta}{\delta}$};

\end{tikzpicture}
\caption{A linear pole intersect with the boundary parabola $Y=X^2$ at two points that correspond to a Farey pair $\Gamma=[
\tfrac{\alpha}{\gamma},
\tfrac{\beta}{\delta}]$. The mediant $\frac{\alpha+\beta}{\gamma+\delta}$ gives rise to two child Farey pairs $\Gamma_{\textrm{L}}=[\tfrac{\alpha}{\gamma},
\tfrac{\alpha+\beta}{\gamma+\delta}]$
 and $\Gamma_{\textrm{R}}=[\tfrac{\alpha+\beta}{\gamma+\delta}
, \tfrac{\beta}{\delta}]$, which correspond to two (child) pole lines.  These three lines  form a triangle $\Delta_{\Gamma}$.}
\label{fig:pole-intersections}
\end{figure}

For a given Farey pair $\Gamma=[
\tfrac{\alpha}{\gamma},
\tfrac{\beta}{\delta}]$, define its \textit{mediant}: 
\begin{equation}\label{eq:mediantDef}
X_{\textrm{M}}=\frac{\alpha+\beta}{\gamma+\delta}\,.
\end{equation}
It is important to note that the mediant produces two new Farey pairs: 
\begin{equation}
\Gamma_{\textrm{L}}=\left[\frac{\alpha}{\gamma}\,,\,
\frac{\alpha+\beta}{\gamma+\delta}\right]
\qquad \textrm{and} \qquad 
\Gamma_{\textrm{R}}=\left[
\frac{\alpha+\beta}{\gamma+\delta}
\,,\, \frac{\beta}{\delta} 
\right]\,.
\end{equation}
The three lines that correspond to $\Gamma, \Gamma_\textrm{L},\Gamma_{\textrm{R}}$ (via the one-to-one correspondence \eqref{eq:FPtoKMS}) form a triangle, which we denote by
\begin{equation}
\Delta_{\Gamma}:=
(X_{\textrm{L}}\, X_{\textrm{M}}\, X_{\textrm{R}})
=\left(\frac{\alpha}{\gamma}\,,
\frac{\alpha+\beta}{\gamma+\delta}\,,
\frac{\beta}{\delta}\right) \,,
\end{equation}
where $X_{\textrm{L/M/R}}$ are the $X$ coordinates of the three corners of the triangle, see Figure~\ref{fig:pole-intersections}.

One can define a binary tree (the so-called Stern-Brocot tree) as follows \cite{Stern:1858,Brocot:1861}. 
Each node of the tree is a Farey pair $\Gamma=[
\tfrac{\alpha}{\gamma},
\tfrac{\beta}{\delta}]$, and each such node has two children, given by $\Gamma_\textrm{L}$ and $\Gamma_{\textrm{R}}$.
Then via the map
\begin{equation}\label{eq:GammatoDelta}
\Gamma=\left[
\frac{\alpha}{\gamma},
\frac{\beta}{\delta}
\right]
\quad \longrightarrow \quad 
\Delta_{\Gamma}:=\left(
\frac{\alpha}{\gamma}
\,,\,
\frac{\alpha+\beta}{\gamma+\delta}
\,,\,
\frac{\beta}{\delta}\right)\,,
\end{equation}
the triangles $\Delta_{\Gamma}$ inherit the binary tree structure of these $\Gamma$'s.
Finally, via the one-to-one correspondence \eqref{eq:KStoFP}, the linear pole lines produce a triangulation\footnote{Note that two pole lines cannot intersect \textit{above} the boundary parabola.} of the $(X,Y)$ plane (above the parabola $Y=X^2$) where the triangles form a binary tree \cite{Cheng:2008fc}. 
This binary tree structure will be crucial for the analysis of the contour deformation. 

It will be enough to focus on the chamber structure of the fundamental
domain of the DMVV chamber under even integer spectral flows.
Recall that the 2D $\mathcal{N}=2$ superconformal algebra has an outer-automorphism called spectral flow \cite{Schwimmer:1986mf}, which acts on the zero modes by
\begin{equation}\label{eq:sf-zero-modes}
L_0 \longmapsto L_0 + w J_0 + \frac{\mathsf{c}}{24} w^2
\qquad
\text{and}
\qquad
J_0 \longmapsto J_0 + \frac{\mathsf{c}}{12} w\,,
\end{equation}
where
$w\in2\mathbb{Z}+1$ takes R$\leftrightarrow $NS, and $w\in 2\mathbb{Z}$ stays inside the R sector. 
We will first study the chamber structure in the Ramond sector, and only translate to the NS sector at the end of Section~\ref{sec:ExpansionCE}, therefore for now we can focus on $w=2n$, with $n\in \mathbb{Z}$.

Substituting 
\begin{equation}
  \mathsf{c}=6N\,,\qquad m=L_0-\frac{\mathsf{c}}{24}\,,\qquad \ell=2J_0\,,
\end{equation}
we obtain the spectral flow on the charges
\begin{equation}\label{eq:DMVVSpectralFlowCharges}
(N,m,\ell)\longmapsto
(N,m+n\ell+n^2N,\ell+2nN)\,.
\end{equation}
This spectral flow can also be realized by fixing $(N,m,\ell)$ but changing the chemical potentials:
\begin{equation}\label{eq:DMVVSpectralFlow}
(\sigma,z,\tau)
\longmapsto
(\sigma+2nz+n^2\tau,z+n\tau,\tau)\,.
\end{equation}
In terms of the projective coordinates $(X,Y)$, the spectral flow is 
\begin{equation}\label{eq:DMVVSpectralFlowXY}
(X,Y)\longmapsto(X+n,Y+2nX+n^2)\,,
\end{equation}
with $n\in\mathbb{Z}$.

Under this spectral flow, the fundamental domain of the DMVV chamber can be chosen as
\begin{equation}\label{eq:DMVVFundamentalRegion}
\Delta_0
:=\Delta_{\textrm{DMVV, fund.}}
=\left(\frac{0}{1}\,,\,\frac{1}{1}\,,\,\frac{1}{0}\right)\,,
\end{equation}
whose interior is
\begin{equation}\label{eq:DMVVFundamentalInterior}
0<X<1\,,\qquad Y>X\,.
\end{equation}
The full DMVV chamber is recovered by taking all even integer spectral-flow
images of $\Delta_0$. 
Accordingly, the representative charge range is $-2N \leq \ell \leq 0$.
We can further impose the charge conjugation symmetry, to restrict to 
$-N \leq \ell \leq 0$.

\begin{figure}[H]
\centering
\definecolor{PaperBlue}{HTML}{356B8C}
\definecolor{PaperGreen}{HTML}{4A8B49}
\definecolor{PaperYellow}{HTML}{E4B82A}

\begin{tikzpicture}[paper figure,x=8cm,y=8cm]
  \begin{scope}
    \clip (0,0) rectangle (1,1.15);

    \path[fill=PaperDomain]
      (0,0) -- (1,1) -- (1,1.15) -- (0,1.15) -- cycle;

    \path[fill=PaperPole!25]
      (0,0) -- ({1/2},{1/4}) -- (1,1) -- cycle;

    \path[fill=PaperBlue!28]
      (0,0) -- ({1/3},{1/9}) -- ({1/2},{1/4}) -- cycle;
    \path[fill=PaperBlue!28]
      ({1/2},{1/4}) -- ({2/3},{4/9}) -- (1,1) -- cycle;

    \path[fill=PaperGreen!30]
      (0,0) -- ({1/4},{1/16}) -- ({1/3},{1/9}) -- cycle;
    \path[fill=PaperGreen!30]
      ({1/3},{1/9}) -- ({2/5},{4/25}) -- ({1/2},{1/4}) -- cycle;
    \path[fill=PaperGreen!30]
      ({1/2},{1/4}) -- ({3/5},{9/25}) -- ({2/3},{4/9}) -- cycle;
    \path[fill=PaperGreen!30]
      ({2/3},{4/9}) -- ({3/4},{9/16}) -- (1,1) -- cycle;

    \path[fill=PaperYellow!45]
      (0,0) -- ({1/5},{1/25}) -- ({1/4},{1/16}) -- cycle;
    \path[fill=PaperYellow!45]
      ({1/4},{1/16}) -- ({2/7},{4/49}) -- ({1/3},{1/9}) -- cycle;
    \path[fill=PaperYellow!45]
      ({1/3},{1/9}) -- ({3/8},{9/64}) -- ({2/5},{4/25}) -- cycle;
    \path[fill=PaperYellow!45]
      ({2/5},{4/25}) -- ({3/7},{9/49}) -- ({1/2},{1/4}) -- cycle;
    \path[fill=PaperYellow!45]
      ({1/2},{1/4}) -- ({4/7},{16/49}) -- ({3/5},{9/25}) -- cycle;
    \path[fill=PaperYellow!45]
      ({3/5},{9/25}) -- ({5/8},{25/64}) -- ({2/3},{4/9}) -- cycle;
    \path[fill=PaperYellow!45]
      ({2/3},{4/9}) -- ({5/7},{25/49}) -- ({3/4},{9/16}) -- cycle;
    \path[fill=PaperYellow!45]
      ({3/4},{9/16}) -- ({4/5},{16/25}) -- (1,1) -- cycle;

    \begin{scope}[paper pole,line width=0.35pt]
      \draw (0,0) -- ({1/5},{1/25});
      \draw (0,0) -- ({1/4},{1/16});
      \draw (0,0) -- ({1/3},{1/9});
      \draw (0,0) -- ({1/2},{1/4});
      \draw (0,0) -- (1,1);

      \draw ({1/5},{1/25}) -- ({1/4},{1/16});
      \draw ({1/4},{1/16}) -- ({2/7},{4/49});
      \draw ({1/4},{1/16}) -- ({1/3},{1/9});
      \draw ({2/7},{4/49}) -- ({1/3},{1/9});

      \draw ({1/3},{1/9}) -- ({3/8},{9/64});
      \draw ({1/3},{1/9}) -- ({2/5},{4/25});
      \draw ({1/3},{1/9}) -- ({1/2},{1/4});
      \draw ({3/8},{9/64}) -- ({2/5},{4/25});
      \draw ({2/5},{4/25}) -- ({3/7},{9/49});
      \draw ({2/5},{4/25}) -- ({1/2},{1/4});
      \draw ({3/7},{9/49}) -- ({1/2},{1/4});

      \draw ({1/2},{1/4}) -- ({4/7},{16/49});
      \draw ({1/2},{1/4}) -- ({3/5},{9/25});
      \draw ({1/2},{1/4}) -- ({2/3},{4/9});
      \draw ({1/2},{1/4}) -- (1,1);
      \draw ({4/7},{16/49}) -- ({3/5},{9/25});
      \draw ({3/5},{9/25}) -- ({5/8},{25/64});
      \draw ({3/5},{9/25}) -- ({2/3},{4/9});
      \draw ({5/8},{25/64}) -- ({2/3},{4/9});

      \draw ({2/3},{4/9}) -- ({5/7},{25/49});
      \draw ({2/3},{4/9}) -- ({3/4},{9/16});
      \draw ({2/3},{4/9}) -- (1,1);
      \draw ({5/7},{25/49}) -- ({3/4},{9/16});
      \draw ({3/4},{9/16}) -- ({4/5},{16/25});
      \draw ({3/4},{9/16}) -- (1,1);
      \draw ({4/5},{16/25}) -- (1,1);
    \end{scope}

    \draw[paper guide] ({1/4},0) -- ({1/4},{1/16});
    \draw[paper guide] ({1/3},0) -- ({1/3},{1/9});
    \draw[paper guide] ({2/5},0) -- ({2/5},{4/25});
    \draw[paper guide] ({1/2},0) -- ({1/2},{1/4});
    \draw[paper guide] ({3/5},0) -- ({3/5},{9/25});
    \draw[paper guide] ({2/3},0) -- ({2/3},{4/9});
    \draw[paper guide] ({3/4},0) -- ({3/4},{9/16});

    \draw[paper boundary]
      plot[domain=0:1,samples=180] (\x,{\x*\x});
  \end{scope}

  \draw[paper guide] (1,0) -- (1,1.24);

  \draw[paper axis] (0,0) -- (1.24,0);
  \draw[paper axis] (0,0) -- (0,1.24);

  \node[anchor=west] at (1.255,0)
    {$X=\frac{\textrm{Im}z}{\textrm{Im}\tau}$};
  \node[anchor=south west] at (0,1.255)
    {$Y=\frac{\textrm{Im}\sigma}{\textrm{Im}\tau}$};
  \node[anchor=north east,inner sep=2pt] at (0,0)
    {$0$};

  \foreach \xx/\lab in {
    0.25/\frac{1}{4},
    0.333333333/\frac{1}{3},
    0.4/\frac{2}{5},
    0.5/\frac{1}{2},
    0.6/\frac{3}{5},
    0.666666667/\frac{2}{3},
    0.75/\frac{3}{4},
    1/1
  }{
    \draw[draw=PaperInk,line width=0.5pt]
      (\xx,0.012) -- (\xx,-0.012);
    \node[anchor=north,inner sep=0pt] at (\xx,-0.065)
      {$\lab$};
  }

\end{tikzpicture}
\caption{Binary tree structure of the triangular chambers.}
\label{fig:fundamental-strip-triangulation}
\end{figure}
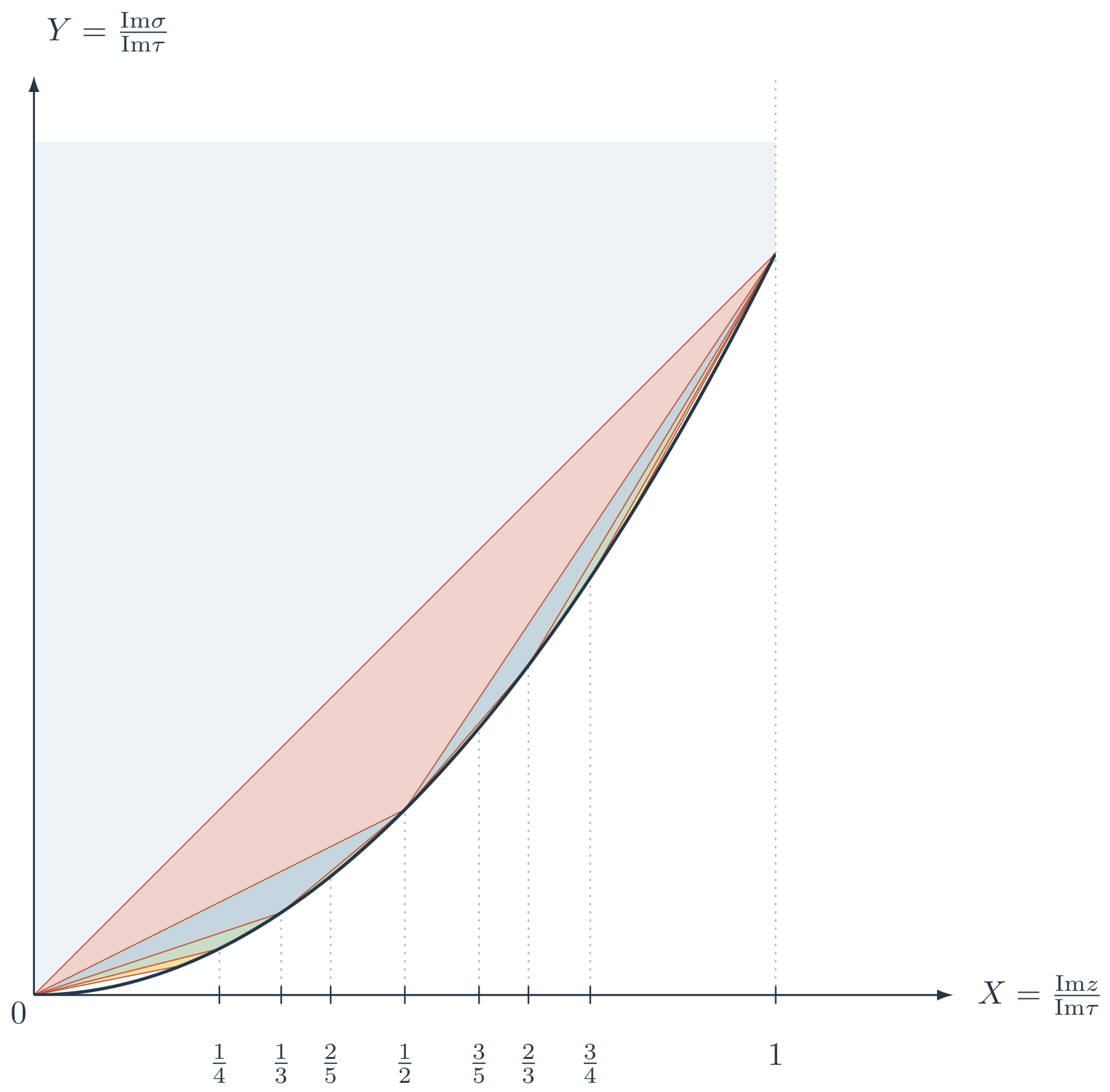

Below $\Delta_{0}$, the first few levels of the binary tree of Farey pairs are
\begin{equation}\label{eq:FareyTreeFourLevels}
\begin{tikzpicture}[
  x=1.65cm,y=1cm,
  baseline=(current bounding box.center),
  every node/.style={inner sep=2pt,outer sep=1pt}
]
  \node (root) at (0,0)
    {$\bigl[\tfrac{0}{1}\,,\,\tfrac{1}{1}\bigr]$};

  \node (L) at (-2,-1)
    {$\bigl[\tfrac{0}{1}\,,\,\tfrac{1}{2}\bigr]$};
  \node (R) at (2,-1)
    {$\bigl[\tfrac{1}{2}\,,\,\tfrac{1}{1}\bigr]$};

  \node (LL) at (-3,-2)
    {$\bigl[\tfrac{0}{1}\,,\,\tfrac{1}{3}\bigr]$};
  \node (LR) at (-1,-2)
    {$\bigl[\tfrac{1}{3}\,,\,\tfrac{1}{2}\bigr]$};
  \node (RL) at (1,-2)
    {$\bigl[\tfrac{1}{2}\,,\,\tfrac{2}{3}\bigr]$};
  \node (RR) at (3,-2)
    {$\bigl[\tfrac{2}{3}\,,\,\tfrac{1}{1}\bigr]$};

  \node (LLL) at (-3.5,-3)
    {$\bigl[\tfrac{0}{1}\,,\,\tfrac{1}{4}\bigr]$};
  \node (LLR) at (-2.5,-3)
    {$\bigl[\tfrac{1}{4}\,,\,\tfrac{1}{3}\bigr]$};
  \node (LRL) at (-1.5,-3)
    {$\bigl[\tfrac{1}{3}\,,\,\tfrac{2}{5}\bigr]$};
  \node (LRR) at (-0.5,-3)
    {$\bigl[\tfrac{2}{5}\,,\,\tfrac{1}{2}\bigr]$};
  \node (RLL) at (0.5,-3)
    {$\bigl[\tfrac{1}{2}\,,\,\tfrac{3}{5}\bigr]$};
  \node (RLR) at (1.5,-3)
    {$\bigl[\tfrac{3}{5}\,,\,\tfrac{2}{3}\bigr]$};
  \node (RRL) at (2.5,-3)
    {$\bigl[\tfrac{2}{3}\,,\,\tfrac{3}{4}\bigr]$};
  \node (RRR) at (3.5,-3)
    {$\bigl[\tfrac{3}{4}\,,\,\tfrac{1}{1}\bigr]$};

  \path[every edge/.style={
    draw,-{Latex[length=1.3mm]},line width=0.4pt
  }]
    (root.south) edge (L.north) edge (R.north)
    (L.south) edge (LL.north) edge (LR.north)
    (R.south) edge (RL.north) edge (RR.north)
    (LL.south) edge (LLL.north) edge (LLR.north)
    (LR.south) edge (LRL.north) edge (LRR.north)
    (RL.south) edge (RLL.north) edge (RLR.north)
    (RR.south) edge (RRL.north) edge (RRR.north);

  \foreach \xx in {-3.5,-2.5,-1.5,-0.5,0.5,1.5,2.5,3.5}{
    \node at (\xx,-3.55) {$\vdots$};
  }
\end{tikzpicture}
\,.
\end{equation}
The corresponding pole lines and triangular chambers are shown in Figure~\ref{fig:fundamental-strip-triangulation}.
Finally, the trees in the other strips can be obtained by even integer spectral flow \eqref{eq:DMVVSpectralFlowXY}.

\subsection{Contour deformation}
We have already summarized the pole structure of $\mathcal{Z}^{\mathrm{GCE}}$.
Now, before we perform the contour deformation, let us first study the residues of $\mathcal{Z}^{\mathrm{GCE}}$ on the linear poles.
For the linear pole $\Gamma$ labeled by $(K,M,S)$, i.e.
\begin{equation}
\Gamma:   \qquad \myzeta^{K}q^{M}y^{S}=1 \,,
\end{equation}
there are $K$ solutions for $\myzeta$:
\begin{equation}
\myzeta^{\textrm{RR}}_{(K,S),a}=\omega_K^a\left(q^My^S\right)^{-\frac{1}{K}}\,,\qquad \omega_K=e^{\frac{2\pi i}{K}}\,,\qquad a=0,\ldots,K-1\,.
\end{equation}

In the contour integral \eqref{eq:ContourIntegralzeta0}, we will deform the contour by increasing the modulus $|\myzeta|$ (from $|\myzeta|\rightarrow 0$), namely by decreasing $Y=\frac{\Im \sigma}{\Im \tau}$ (from $Y\rightarrow \infty$), while keeping 
\begin{equation}
X=\frac{\Im z}{\Im \tau}
\end{equation}
fixed. 
Since all these roots have the same modulus, when the circular $\myzeta$-contour crosses the pole-wall $\Gamma$, it crosses all $K$ roots simultaneously.
Therefore, it is natural to sum the residues of all the $K$ roots and define \begin{equation}\label{eq:ZhatDef}
\widehat{Z}^{\textrm{RR}}_{N,\Gamma}(q,y)=-\sum_{a=0}^{K-1}\mathrm{Res}_{\myzeta=\myzeta^{\textrm{RR}}_{(K,S),a}}\,
\left[\myzeta^{-N-1}\mathcal{Z}^{\mathrm{GCE}}_{\textrm{RR}}(\myzeta,q,y) \right]\,.
\end{equation}

Now the question is: when we deform the contour in \eqref{eq:ContourIntegralzeta0}, which $\Gamma$ should contribute to the left hand side $Z^{\textrm{RR}}_{N}(q,y)$?
As we will show now, the answer depends on the value of $X$.

\subsection{Aside: Branch$(X)$}

First of all,  $X$ defines a unique branch of the Stern-Brocot tree \cite{Stern:1858,Brocot:1861}, as follows. 
Recall that each segment of the Stern-Brocot tree is given by a Farey pair $[\frac{\alpha}{\gamma},
\frac{\beta}{\delta}]$.
For a given $X$, the associated Branch$(X)$ is defined as the branch whose segments all enclose $X$:
\begin{equation}\label{eq:BranchXDef}
\mathrm{Branch}(X)
:=\left\{I=\left[\frac{\alpha}{\gamma},\frac{\beta}{\delta}\right]\;:\;\frac{\alpha}{\gamma}<X<\frac{\beta}{\delta}\right\}
\end{equation} 

We can also translate this into an equivalent, but more useful, definition.
Define the height function of $I$
\begin{equation}\label{eq:HeightDef}
h_I(X):=Y_I(X)-X^2\,,
\end{equation}
which measures the vertical distance from the linear pole wall
\begin{equation}\label{eq:YIXDef}
Y_I(X):=-\frac{M+SX}{K}\,,
\end{equation} 
to the boundary parabola $Y=X^2$, at a given $X$. 
Using the map \eqref{eq:FPtoKMS}, it can be written as\footnote{See \cite{Mukherjee:2007nc} for a similar object in the context of dyon counting, in terms of Ford circles.} 
\begin{equation}
h_{I}(X)=\left(X-\frac{\alpha}{\gamma}\right)\left(\frac{\beta}{\delta}-X\right)\,.
\end{equation}
In terms of the height function, the condition 
\begin{equation}
\frac{\alpha}{\gamma}<X<\frac{\beta}{\delta}
\end{equation}
can be translated into the condition on the height function:
\begin{equation}
h_I(X)>0\,. 
\end{equation}
Namely, Branch$(X)$ can also be defined as 
\begin{equation}\label{eq:BranchXDefHeightfn}
\mathrm{Branch}(X)
:=\left\{I:h_I(X)>0\right\}\,.   
\end{equation} 

Branch$(X)$ is infinitely long for irrational $X$. 
As we descend from $Y\rightarrow \infty$ down to the boundary parabola $Y=X^2$, the chambers (and boundaries between them) that are crossed can be labeled as:
\begin{equation}
\Delta_0 \xrightarrow{\Gamma_1} \Delta_1 \xrightarrow{\Gamma_2} \Delta_2
\cdots 
\Delta_{i-1} \xrightarrow{\Gamma_{i}}  \Delta_{i} \xrightarrow{\Gamma_{i+1}}\cdots
\end{equation}
Here $\Delta_0$ is the fundamental
domain of the DMVV chamber under even integer spectral flow and we label
\begin{equation}
\Gamma_i=\left[
\frac{\alpha_i}{\gamma_i},
\frac{\beta_i}{\delta_i}
\right]
\quad \textrm{and} \quad 
\Delta_{i}:=\left(
\frac{\alpha_i}{\gamma_i}
\,,\,
\frac{\alpha_i+\beta_i}{\gamma_i+\delta_i}
\,,\,
\frac{\beta_i}{\delta_i}\right)\,.
\end{equation}
On the other hand, for (non-integer) rational $X$, Branch$(X)$ terminates:
\begin{equation}
\Delta_0 \xrightarrow{\Gamma_1} \Delta_1 \xrightarrow{\Gamma_2} \Delta_2
\cdots 
\Delta_{s-1} \xrightarrow{\Gamma_{s}}  \Delta_{s}\,,
\end{equation}
with the tip of the final chamber being
\begin{equation}
X=\frac{\alpha_s+\beta_s}{\gamma_s+\delta_s}\,.
\end{equation}

\subsection{Negative modes and Stokes chamber}

In this part, we will use the following notation 
\begin{equation}
\begin{aligned}
\textrm{Charge}: &\qquad Q:=(N,m,\ell) \,,\qquad N\in\mathbb{Z}_{\geq1}\,,\quad m\in\mathbb{Z}_{\geq0}\,,\quad \ell\in\mathbb{Z}\\
\textrm{mode}: &\qquad J:=(\nu,\mu,\lambda)\,,\qquad \nu\in\mathbb{Z}_{\geq 1}\,,\quad \mu\in\mathbb{Z}_{\geq0}\,,\quad \lambda\in\mathbb{Z}\,, \\
\textrm{Pole}: &\qquad I:=(K,M,S)\,,\qquad K\in\mathbb{Z}_{\geq 1}\,,\quad M=\frac{S^2-1}{4K}\in\mathbb{Z}_{\geq0}\,,\quad S\in2\mathbb{Z}+1\,. 
\end{aligned}  
\end{equation}
(Note that we will use $I$, in addition to $\Gamma$, to label the Farey pairs and poles.)

Consider the mode $J$, which comes from  the factor 
\begin{equation}
1-\myzeta^\nu q^\mu y^\lambda
\end{equation}
in the denominator of $\mathcal{Z}^{\textrm{GCE}}_{\textrm{RR}}(\myzeta,q,y)$.
The term $\myzeta^\nu q^\mu y^\lambda$ evaluated at the pole given by $I\neq J$ has the magnitude of the mode:
\begin{equation}
|\Lambda^J_I|=\textrm{exp} \left(
-2\pi \,
\mathrm{Im} 
\tau \, \delta_I^J(X)
\right)    
\end{equation}
with 
\begin{equation}
\delta_I^J(X):=\nu (h_I(X)-h_J(X))\,,
\end{equation}
where $h_I(X)$ was defined in \eqref{eq:HeightDef} with \eqref{eq:YIXDef} for the pole $I$ and $h_J(X)$ is defined by the same formulae with $(K,M,S)$ replaced by $(\nu,\mu,\lambda)$.
Depending on their magnitudes, there are three types of modes, with the corresponding conditions on the height functions:
\begin{equation}
\begin{aligned}
\textrm{positive mode:}\quad &|\Lambda^J_I|<1 \quad \Longleftrightarrow \quad 
h_I(X) > h_J(X) \\
\textrm{zero mode:}\quad &|\Lambda^J_I|=1 \quad \Longleftrightarrow \quad 
h_I(X) = h_J(X) \\
\textrm{negative mode:}\quad &|\Lambda^J_I|>1 \quad \Longleftrightarrow \quad 
h_I(X) < h_J(X) \\
\end{aligned}
\end{equation}

As in the chiral-chiral case in \cite{Lee:2025veh}, for a given $X$, only the linear poles whose residue $\widehat{Z}_{N,I}(q,y)$ has a \textit{finite number} of non-positive modes contribute to the finite-$N$ expansion of \eqref{eq:ContourIntegralzeta0}.
The reason is that during the analytic continuation, each negative mode (denoted by $x^{-1}$ for $|x|<1$) gives a finite upward shift of the energy:
\begin{equation}\label{eq:AnalyticContinuation}
\frac{1}{1-x^{-1}}=-\frac{x}{1-x} \,,
\end{equation}
apart from the minus sign.
Therefore, the residue $\widehat{Z}_{N,I}(q,y)$ that contains infinitely many negative modes only starts to contribute at infinitely high energy, which means essentially it never contributes in the series expansion.

Therefore, the question we want to answer now is: given $X$, what is the condition for a pole $I$ such that its residue $\widehat{Z}_{N,I}(q,y)$ has only a finite number of non-positive modes.

Depending on whether $X$ is rational or not, there are two scenarios to consider:
\begin{itemize}
\item For a generic $X$ (i.e. irrational), since $J\neq I$, we have $h_I(X) \neq h_J(X)$. 
Thus we only need to consider negative modes, i.e.\ there are no zero modes.
We first observe that
\begin{equation}\label{eq:finiteNegativeModes}
\# \{J \, | \, |\Lambda_I^J(X)|> 1 \} <\infty
\qquad\Longleftrightarrow\qquad
h_I(X)>0  \,,
\end{equation}
where $h_I(X)>0$ is the height function defined in \eqref{eq:HeightDef}.
Then the condition $h_I(X)>0$ in turn means that $I\in \textrm{Branch}(X)$. 
Finally, a negative mode $J$ must satisfy
\begin{equation}
J \in \textrm{Branch}(X) 
\qquad \textrm{and}
\qquad J< I    \,,
\end{equation}
where $J< I$ means $J$ is an ancestor of $I$ in $\textrm{Branch}(X)$.

\item For a rational $X$, \eqref{eq:finiteNegativeModes} still holds for every pole $I$ with $h_I(X)\neq 0$.
The only new case is a pole $I$ with $h_I(X)=0$, namely when $X$ is an end point of $I$.
In this case, there can be infinitely many $J\neq I$ that satisfy $0=h_J(X)=h_I(X)$ (if they share end points at $X$), producing infinitely many zero modes. 
On the other hand, for a rational $X$, the condition $h_I(X)>0$ ensures that there is no zero mode. 
\end{itemize}
Summarizing these two observations, we conclude that for a given $X$, the condition for a pole $I$ to have a finite number of non-positive modes (among all modes $J$) in its residue $\textrm{Res}_{I}(q,y)$ is therefore 
\begin{equation}
h_I(X)>0\,,
\end{equation}
or equivalently
\begin{equation}
I\in \textrm{Branch}(X)\,.
\end{equation}

Therefore, we conclude that given $X$, as we deform the contour by descending from $Y\rightarrow \infty$ down to the boundary parabola $Y=X^2$, only the linear poles that correspond to segments on Branch$(X)$ contribute.
    
\subsection{Cauchy's theorem for canonical chambers}

As we deform the contour by descending from $Y\rightarrow \infty$ down to the boundary parabola $Y=X^2$, the chambers (and boundaries between them) that are crossed will be denoted by:
\begin{equation}\label{eq:ChamberTree}
\Delta_0 \xrightarrow{\Gamma_1} \Delta_1 \xrightarrow{\Gamma_2} \Delta_2
\cdots 
\Delta_{i-1} \xrightarrow{\Gamma_{i}}  \Delta_{i} \xrightarrow{\Gamma_{i+1}}\cdots
\end{equation}

Compare two adjacent chambers
\begin{equation}
\Delta_{i-1}\xrightarrow{\ \Gamma_i\ }\Delta_i\,.    
\end{equation}
Cauchy's theorem gives
\begin{equation}
Z^{\textrm{RR}}_{N,\Delta_i}(q,y)=Z^{\textrm{RR}}_{N,\Delta_{i-1}}(q,y)-
\widehat{Z}^{\textrm{RR}}_{N,\Gamma_i}(q,y)\,,
\end{equation}
where $Z^{\textrm{RR}}_{N,\Delta_i}(q,y)$ is the contour integral in the chamber $\Delta_i$: \begin{equation}\label{eq:ContourIntegralDeltai}
Z^{\textrm{RR}}_{N,\Delta_i}(q,y)=\oint_{\Delta_i}\frac{d\myzeta}{2\pi i}\,\myzeta^{-N-1}\, \mathcal{Z}^{\mathrm{GCE}}_{\mathrm{RR}}(\myzeta,q,y)\,.
\end{equation}
and $\widehat{Z}^{\textrm{RR}}_{N,\Gamma_i}(q,y)$ is (minus) the sum of residues defined in \eqref{eq:ZhatDef}. 
Then applying Cauchy's theorem iteratively, we obtain 
\begin{equation}\label{eq:FiniteNCE}
Z^{\textrm{RR}}_{N}(q,y)
=Z^{\textrm{RR}}_{N,\Delta_0}(q,y)
=Z^{\textrm{RR}}_{N,\Delta_t}(q,y)+\sum_{i=1}^{t}\widehat{Z}^{\textrm{RR}}_{N,\Gamma_i}(q,y)\,,    
\end{equation}
for any $t$, where 
\begin{equation}
\Gamma_{i\geq 1} \in \textrm{Branch}(X) \,.     
\end{equation}

We emphasize that this identity holds as a series expansion:
\begin{equation}\label{eq:FiniteNExpCE}
Z^{\textrm{RR}}_N(q,y)
=\mathcal{E}_X[Z^{\textrm{RR}}_{N,\Delta_t}(q,y)]+\sum_{i=1}^{t}\mathcal{E}_X[\widehat{Z}^{\textrm{RR}}_{N,\Gamma_i}(q,y)]
\,,
\end{equation}
where the left hand side is expanded in the DMVV chamber
\begin{equation}\label{eq:DMVV-contourexpansion}
\begin{cases}
0<|q|\ll |y|<1<|y|^{-1}\,,
& \ell<0\,,
\\
0<|q|\ll |y|^{-1}<1<|y|\,,
& \ell>0\,,
\end{cases}
\end{equation}
whereas the right hand side is expanded in the region
\begin{equation}\label{eq:region}
|q|^{X}=|y|\,.
\end{equation}
Namely, for each factor in the denominator we first evaluate its modulus using \eqref{eq:region}, and then for each negative mode we first apply the analytic continuation \eqref{eq:AnalyticContinuation} before doing the series expansion.
Note that 
$Z_N(q,y)$ on the left hand side of \eqref{eq:FiniteNExpCE}, as a series expansion in $q,y$ in the DMVV chamber, is independent of the value of $X$.
For the right hand side, $X$ enters in two ways: via the selection of $\Gamma_i\in\textrm{Branch}(X)$, and via the series expansion of $\widehat{Z}$ in the appropriate chamber.
The right hand side is also independent of $X$ because these two effects of $X$ cancel out.

In Section~\ref{sec:ExpansionMCE}, we will prove that for any polar charge $Q=(N,m,\ell)$, there exists a choice of $X$ such that along the a priori infinite branch \eqref{eq:ChamberTree}, only the sub-branch at the beginning, from $1$ to $t$
\begin{equation}\label{eq:finiteBranch}
\Delta_0 \xrightarrow{\Gamma_1} \Delta_1
\cdots \Delta_{t-2}
\xrightarrow{\Gamma_{t-1}} 
\Delta_{t-1} \xrightarrow{\Gamma_{t}}  \Delta_{t}
\end{equation}
give non-zero contributions to the degeneracy $d(Q)$:
\begin{equation}
d(Q)=\sum^t_{i=1}\textrm{Res}_{\Gamma_i}(Q)  \,,
\end{equation}
where $\textrm{Res}_{\Gamma_i}(Q)$ is the $q^my^{\ell}$ coefficient of 
\begin{equation}\label{eq:ZhatRR}
\widehat{Z}^{\textrm{RR}}_{N,\Gamma_i}(q,y)=-\sum_{a=0}^{K-1}\mathrm{Res}_{\myzeta=\myzeta^{\textrm{RR}}_{(K,S),a}}\,
\left[\myzeta^{-N-1}\mathcal{Z}^{\mathrm{GCE}}_{\textrm{RR}}(\myzeta,q,y) \right]\,,
\end{equation}
in the expansion $\mathcal{E}_X$, where we recall that $\zeta^{\textrm{RR}}_{(K,S),a}$ is the double pole at
\begin{equation}
\myzeta^{\textrm{RR}}_{(K,S),a}=\omega_K^a\left(q^My^S\right)^{-\frac{1}{K}}\,,\qquad \omega_K=e^{\frac{2\pi i}{K}}\,,\qquad a=0,\ldots,K-1\,.
\end{equation} 
Therefore, we can write 
\begin{equation}\label{eq:FiniteNCE2}
Z^{\textrm{RR}}_{N}(q,y)\big|_{\textrm{polar}}
=\sum_{i=1}^{t}\widehat{Z}^{\textrm{RR}}_{N,\Gamma_i}(q,y)\big|_{\textrm{polar}}\,,    
\end{equation}
while keeping in mind that the identity above involves the choice of $X$ for each $Q$.

\subsection{Resolution of double poles and Hodge elliptic genus}

Let us now try to understand the meaning of the sum of the residues of the double pole \eqref{eq:ZhatRR}.
Since it is difficult to prescribe state-counting interpretation to the residue of a double pole, we will first try to resolve the double poles.
Recall that these poles are double poles because the elliptic genus is blind to the states with different $\bar{J}_0$, since $\bar{y}\rightarrow 1$.
Therefore, we can simply restore the dependence on $\bar{y}$ at the intermediate stage, and consider the Hodge elliptic genus \cite{Kachru:2016igs,Benjamin:2016pil}
\begin{equation}
Z^{\textrm{RR}}_N(q,y,\bar{y})
=\mathrm{Tr}_{\mathcal H^{\mathrm{RR}}_{\forall \otimes \textrm{vac}}}
\left[
(-1)^F q^{L_0-\frac{\mathsf{c}}{24}}y^{2J_0}\bar{y}^{2\bar{J}_0}
\right]  \,,
\end{equation}
where $\mathsf{c}=6N$, and $\textrm{vac}$ denotes the right-moving Ramond ground states.
Throughout this paper, we will use the same letter/font to denote the elliptic genus and the Hodge elliptic genus, with the latter having an additional variable $\bar{y}$, and they are related by
\begin{equation}
\mathrm{HEG}\big|_{\bar{y}=1}=\textrm{EG}. 
\end{equation}

Each double pole $\zeta^{\textrm{RR}}_{(K,S),a}$ is then resolved into two simple poles:
\begin{equation}
\myzeta^{\textrm{RR}}_{(K,S,\bar{S}),a}:=\omega_K^a \, \mathfrak{p}_{(K,S,\bar{S})}(q,y,\bar{y})\,,\qquad \omega_K=e^{\frac{2\pi i}{K}}\,,\qquad a=0,\ldots,K-1\,,
\end{equation}
with 
\begin{equation}
\mathfrak{p}_{(K,S,\bar{S})}(q,y,\bar{y}):= \left(q^My^S\bar{y}^{\bar{S} }\right)^{-\frac{1}{K}}
\end{equation}
and
\begin{equation}
\bar{S}=\pm 1\,. 
\end{equation}
The residue of a double pole can then be expressed in terms of those of the corresponding simple poles:
\begin{equation}
\mathrm{Res}_{\myzeta=\myzeta^{\textrm{RR}}_{(K,S),a}}\,
\left[\myzeta^{-N-1}\mathcal{Z}^{\mathrm{GCE}}_{\textrm{RR}}(\myzeta,q,y) \right]
=\textrm{lim}_{\bar{y}\rightarrow 1}  
\sum_{\bar{S}=\pm 1}\mathrm{Res}_{\myzeta=\myzeta^{\textrm{RR}}_{(K,S,\bar{S}),a}}\,
\left[\myzeta^{-N-1}\mathcal{Z}^{\mathrm{GCE}}_{\textrm{RR}}(\myzeta,q,y,\bar{y}) \right] 
\end{equation}
Defining
\begin{equation}\label{eq:ZhatRRnew}
\widehat{Z}^{\textrm{RR}}_{N,(K,S,\bar{S})}(q,y,\bar{y})=-\sum_{a=0}^{K-1}\mathrm{Res}_{\myzeta=\myzeta^{\textrm{RR}}_{(K,S,\bar{S}),a}}\,
\left[\myzeta^{-N-1}\mathcal{Z}^{\mathrm{GCE}}_{\textrm{RR}}(\myzeta,q,y,\bar{y}) \right]\,,
\end{equation}
we have the decomposition 
\begin{equation}\label{eq:FiniteNCE3}
Z^{\textrm{RR}}_{N}(q,y)\big|_{\textrm{polar}}
=\textrm{lim}_{\bar{y}\rightarrow 1}  
\sum_{\bar{S}=\pm 1}\sum_{(K,S)}\widehat{Z}^{\textrm{RR}}_{N,(K,S,\bar{S})}(q,y,\bar{y})\big|_{\textrm{polar}}\,,    
\end{equation}
where the sum over $(K,S)$ is read off from the branch \eqref{eq:finiteBranch}.

It is straightforward to evaluate the resolved $\widehat{Z}$:
\begin{equation}
\widehat{Z}^{\textrm{RR}}_{N,(K,S,\bar{S})}(q,y,\bar{y}) =  (q^{M}y^{S}\bar{y}^{\bar{S}})^{\frac{N}{K}}
\cdot
\frac{1}{K}\sum^{K-1}_{a=0}\omega_K^{-aN}
\textrm{PE}\left[f^{\textrm{RR}}_{(K,S,\bar{S}),a}(q,y,\bar{y})\right]
\end{equation}
where 
\begin{equation}
f^{\textrm{RR}}_{(K,S,\bar{S}),a}(q,y,\bar{y})
:=  -1+\sum_{n=1}^{\infty}\omega_K^{an}\left(q^M y^S \bar{y}^{\bar{S}}\right)^{-\frac{n}{K}}
\frac{1}{n}\sum_{r=0}^{n-1}Z_{\mathrm{seed}}^{\mathrm{RR}}\!\left(e^{2\pi i\frac{r}{n}}q^{\frac{1}{n}},y,\bar y\right)\,.
\end{equation}
Here
\begin{equation}
Z^{\mathrm{RR}}_{\mathrm{seed}}(q,y,\bar{y})
=
\operatorname{Tr}_{\mathcal H^{\mathrm{RR}}_{\mathrm{seed},\forall \otimes vac}}
\left[
(-1)^F q^{L_0-\frac{c}{24}}y^{2J_0}\bar{y}^{2\bar{J}_0}
\right]
=
\sum_{m,\ell,\bar{\ell}} c(m,\ell,\bar{\ell})q^my^\ell \bar{y}^{\bar{\ell}}\,,
\end{equation}
where the central charge $c=6$ and the degenaracy $c(m,\ell)=\sum_{\bar{\ell}}c(m,\ell,\bar{\ell})$.

We factor the final result of $\widehat{Z}^{\textrm{RR}}_{N,(K,S,\bar{S})}(q,y,\bar{y}) $ into two pieces:
\begin{equation}
\begin{aligned}
\textrm{Classical}: &\qquad (q^{M}y^{S}\bar{y}^{\bar{S}})^{\frac{N}{K}}\\
\textrm{One-loop}: &\qquad \frac{1}{K}\sum^{K-1}_{a=0}\omega_K^{-aN}
\textrm{PE}\left[f^{\textrm{RR}}_{(K,S,\bar{S}),a}(q,y,\bar{y})\right]
\end{aligned}  
\end{equation}
We will derive them in Section~\ref{sec: classical} and~\ref{sec:worldsheet}, respectively.

\subsection{NS sector}
Before we end this section, let us also  write down the corresponding expressions in the NS frame.
Applying the spectral flow \eqref{eq:sf-zero-modes} with $w=1$, the translation from the RR charges to the NS ones is:
\begin{equation}\label{eq:RtoNS}
\begin{aligned}
L_{0}^{\mathrm{NS}}
&=L_{0}^{\mathrm{RR}}+J_{0}^{\mathrm{RR}}+\frac{N}{4}=(L_{0}^{\mathrm{RR}}-\frac{N}{4})+J_{0}^{\mathrm{RR}}+\frac{N}{2}\,,\\ 
J_{0}^{\mathrm{NS}}
&=J_{0}^{\mathrm{RR}}+\frac{N}{2}\,, 
  \end{aligned}
\end{equation}
and similarly for the right-movers.
The full signed partition function in the NS sector (with exponent $L_0$ instead of $L_0-\frac{\mathsf{c}}{24}$)
is related to the one in the RR sector by 
\begin{equation}
Z_{N}^{\mathrm{NS}}(q,\bar{q},y,\bar{y})
=q^{\frac{N}{2}}\bar{q}^{\frac{N}{2}}y^{N}\bar{y}^{N} Z_{N}^{\mathrm{RR}}(q,\bar{q},q^{\frac{1}{2}}y,\bar{q}^{\frac{1}{2}}\bar{y})
\end{equation}

For the Hodge elliptic genus, the right-movers are in the Ramond ground states 
$\bar{L}_{0}^{\textrm{RR}}=\frac{N}{4}$, which  map to the right-moving chiral primaries of the NS sector, with 
$\bar{L}_{0}^{\mathrm{NS}}=\bar{J}_{0}^{\mathrm{NS}}$ \cite{Lerche:1989uy}.
Therefore the right-moving version of \eqref{eq:RtoNS} becomes
\begin{equation}
\begin{aligned}
\bar{L}_{0}^{\mathrm{NS}}
&=\bar{J}_{0}^{\mathrm{RR}}+\frac{N}{2}\,,\\ 
\bar{J}_{0}^{\mathrm{NS}}
&=\bar{J}_{0}^{\mathrm{RR}}+\frac{N}{2}\,, 
  \end{aligned}
\end{equation}
The Hodge elliptic genus in the NS sector is 
\begin{equation}
Z_{N}^{\mathrm{NS}}(q,y,\bar{y})
=q^{\frac{N}{2}}y^{N}\bar{y}^{N} Z_{N}^{\mathrm{RR}}(q,q^{\frac{1}{2}}y,\bar{y})\,.
\end{equation}
And the elliptic genus in the NS sector is 
\begin{equation}
Z_{N}^{\rm NS}(q,y) = q^{\frac{N}{2}} y^{N} Z^{\textrm{RR}}_{N}(q,q^{\frac{1}{2}} y)
\end{equation}

Going to the grand canonical ensemble gives
\begin{equation}\label{eq:ZGCENS}
\mathcal{Z}_{\mathrm{NS}}^{\mathrm{GCE}}(\zeta,q,y,\bar{y})
=\mathcal{Z}_{\mathrm{RR}}^{\mathrm{GCE}}\left( q^{\frac{1}{2}}y\bar{y}  \zeta,q,q^{\frac{1}{2}}y,\bar{y}\right) 
\end{equation}
and 
\begin{equation}
\mathcal{Z}_{\mathrm{NS}}^{\mathrm{GCE}}(\zeta,q,y)=\mathcal{Z}_{\mathrm{RR}}^{\mathrm{GCE}}\left( q^{\frac{1}{2}}y  \zeta,q,q^{\frac{1}{2}}y\right)\,.
\end{equation}
The resolved simple poles of the grand-canonical Hodge elliptic genus in \eqref{eq:ZGCENS} therefore become 
\begin{equation}
\begin{aligned} \zeta^{\mathrm{NS}}_{(K,S,\bar{S}),a}(q,y,\bar{y})&=q^{-\frac{K+S}{2K}}y^{-1}\bar{y}^{-1}\zeta^{\mathrm{RR}}_{(K,S,\bar{S}),a}(q,y,\bar{y})\\
&=\omega_{K}^{a}\left(q^{M+\frac{S+K}{2}}y^{S+K}\bar{y}^{\bar{S}+K}\right)^{-\frac{1}{K}}  \,.
\end{aligned}
\end{equation}
The NS version of the residue sum over these simple poles is 
\begin{equation}\label{eq:ZhatNSfinal}
\begin{aligned}
\widehat{Z}^{\mathrm{NS}}_{N,(K,S,\bar{S})}(q,y,\bar{y})
&=q^{\frac{N}{2}}y^{N}\bar{y}^{N} \widehat{Z}^{\mathrm{RR}}_{N,(K,S,\bar{S})}(q,q^{\frac{1}{2}}y,\bar{y})\\
&=q^{\frac{N}{2}+\frac{N}{K}(M+\frac{S}{2})}y^{N(1+\frac{S}{K})}\bar{y}^{N(1+\frac{\bar{S}}{K})}\times\frac{1}{K}\sum_{a=0}^{K-1}\omega_{K}^{-aN}\mathrm{PE}\left[f^{\mathrm{NS}}_{(K,S,\bar{S}),a}(q,y,\bar{y})\right] \,,
\end{aligned}
\end{equation}
where
\begin{equation}\label{eq:fNS} 
\begin{aligned}
f^{\mathrm{NS}}_{(K,S,\bar{S}),a}(q,y,\bar{y})
&=f^{\mathrm{RR}}_{(K,S,\bar{S}),a}(q,q^{\frac{1}{2}}y,\bar{y})\\
&=-1+\sum_{n=1}^{\infty}\omega_{K}^{an}
(q^{M+\frac{S}{2}}y^{S}\bar{y}^{\bar{S}})^{-\frac{n}{K}}
\times\frac{1}{n}
\sum_{r=0}^{n-1}Z_{\mathrm{seed}}^{\mathrm{RR}}\left(e^{\frac{2\pi ir}{n}}q^{\frac{1}{n}},q^{\frac{1}{2}}y,\bar{y}\right)\,.
\end{aligned} 
\end{equation}
Thus $\widehat{Z}^{\mathrm{NS}}_{N,(K,S,\bar{S})}(q,y,\bar{y})$ can also be factorized into the classical and one-loop factors:
\begin{equation}\label{eq:ZhatNSfactors}
\begin{aligned}
\textrm{Classical}: &\qquad q^{\frac{N}{2}+\frac{N}{K}(M+\frac{S}{2})}y^{N(1+\frac{S}{K})}\bar{y}^{N(1+\frac{\bar{S}}{K})} \,,\\
\textrm{One-loop}: &\qquad 
\frac{1}{K}\sum_{a=0}^{K-1}\omega_{K}^{-aN}\mathrm{PE}\left[f^{\mathrm{NS}}_{(K,S,\bar{S}),a}(q,y,\bar{y})\right] \,.
\end{aligned}  
\end{equation}

\section{Orbifold backgrounds and their charges}
\label{sec: classical}

\subsection{Orbifold geometries}

We now examine the $\orbifoldK$ orbifold backgrounds \cite{Martinec:2001cf,Martinec:2002xq} appearing in \eqref{eq: proposal}. These are defined by the global $\adsS$ metric
\be \label{eq: orbifold metric}
\frac{1}{\sqrt{Q_1 Q_5}} ds^2 = -(r^2 + 1) dt^2 + \frac{dr^2}{r^2 + 1} + r^2 d\phi^2 
+ d\theta^2 + \cos^2\theta d\psi^2 + \sin^2\theta d\chi^2 ,
\ee
subject to the two-parameter family of $\bZ_K$ identifications
\ie \label{eq: orbifold identification}
(\phi,\psi,\chi) &\sim (\phi, \psi + 2\pi, \chi)\sim  (\phi,\psi,\chi + 2\pi) \\
&\sim \left( \phi+\frac{2\pi}{K}, \psi + \frac{2\pi \cdot \frac{1}{2}(S+\Sb)}{K} , \chi - \frac{2\pi \cdot \frac{1}{2}(S-\Sb)}{K} \right),
\fe
with anti-periodic boundary conditions for spinors around the asymptotic spatial $\phi$-circle. Why does the sum over orbifold backgrounds in \eqref{eq: proposal} consist only of those satisfying \eqref{eq: orbifold constraint}, namely those for which $S \in 2\bZ+1$, $\Sb=\mp1$, and $M = \frac{1}{4K}(S^2 - 1) \in \bZ_{\geq 0}$? We will derive the first two constraints by examining the properties of these geometries and the supergravity fields thereon. The third will be found from the worldsheet analysis in Section~\ref{sec:worldsheet}.

In the gravitational path integral, we are instructed to consider a sum over all geometries that preserve the asymptotic boundary conditions given by the dual CFT. We are interested in asymptotically-AdS$_3$ backgrounds with NS boundary conditions around the spatial circle. The metric \eqref{eq: orbifold metric} can be brought into an asymptotically-AdS$_3$ form using the coordinates
\ieg \label{eq: coordinate change}
\tilde{t} = K t, \quad \tilde{r} = \frac{1}{K} r, \quad \tilde{\phi} = K \phi, \\
\tilde{\psi} = \psi - \left( K + \tfrac{1}{2}(S+\Sb) \right) \phi + \tfrac{1}{2}(S-\Sb) t, \\
\tilde{\chi} = \chi + \tfrac{1}{2}(S-\Sb) \phi - \left( K + \tfrac{1}{2}(S+\Sb) \right) t.
\feg
The resulting $\bZ_K$-conical defect metrics
\ie \label{eq: orbifold metric 2}
\frac{ds^2}{\sqrt{Q_1 Q_5}}  = &- \left( \tilde{r}^2 + \frac{1}{K^2} \right) d\tilde{t}^2 + \frac{1}{\tilde{r}^2 + \frac{1}{K^2}} d\tilde{r}^2 + \tilde{r}^2 d\tilde{\phi}^2 \\
&+ d\tilde{\theta}^2 + \cos^2\tilde{\theta} \left[ d\tilde{\psi} + \left( 1 + \frac{S+\Sb}{2K} \right) d\tilde{\phi} - \left( \frac{S-\Sb}{2K} \right) d\tilde{t} \right]^2 \\
&+ \sin^2\tilde{\theta} \left[ d\tilde{\chi} - \left( \frac{S-\Sb}{2K} \right) d\tilde{\phi} + \left( 1 + \frac{S+\Sb}{2K}\right) d\tilde{t} \right]^2 
\fe
have the standard identifications
\be \label{eq: orbifold identification 2}
(\tilde{\phi},\tilde{\psi},\tilde{\chi}) \sim (\tilde{\phi} + 2\pi, \tilde{\psi},\tilde{\chi}) \sim (\tilde{\phi}, \tilde{\psi} + 2\pi, \tilde{\chi}) \sim (\tilde{\phi},\tilde{\psi},\tilde{\chi} + 2\pi).
\ee
Supersymmetric and non-supersymmetric fractionally spectral flowed geometries were studied in \cite{Giusto:2012yz} and \cite{Jejjala:2005yu,Chakrabarty:2015foa}, respectively.

\subsection{Gauge holonomies and asymptotic symmetries}

The truncation of 10d IIB supergravity on $\adsK$ to the massless modes on $\Kthree$ yields 6d $(2,0)$ supergravity on $\adsS$. For our purposes, we can further restrict the 6d supergravity to the sector described by $SU(1,1|2)_L \times SU(1,1|2)_R$ Chern-Simons supergravity on AdS$_3$ \cite{Knizhnik:1986wc,Bershadsky:1986ms,Sevrin:1993vv,Achucarro:1986uwr,Witten:1988hc,Howe:1995zm,deBoer:1998kjm,Henneaux:1999ib}. The reduction on $S^3$ produces the 3d metric
\be \label{eq: orbifold metric 3}
ds^2 = - \left( \tilde{r}^2 + \frac{1}{K^2} \right) d\tilde{t}^2 + \frac{1}{\tilde{r}^2 + \frac{1}{K^2}} d\tilde{r}^2 + \tilde{r}^2 d\tilde{\phi}^2
\ee
and the $SU(2)_L \times SU(2)_R$ gauge fields restricted to a surface of constant $\tilde{t}$
\be \label{eq: gauge fields}
A = \left( 1 + \frac{S}{K} \right) T^3 d\tilde{\phi}, \quad \bar{A} = \left( 1 + \frac{\Sb}{K} \right) T^3 d\tilde{\phi},
\ee
where $T^a = \frac{1}{2} \sigma^a$. The values of these gauge fields at spatial infinity can be interpreted as the spectral flow automorphism, i.e. large gauge transformation, of the $\cN = (4,4)$ superconformal algebra formed by the asymptotic symmetry generators of Chern-Simons supergravity. The left-moving generators obey the small $\cN=4$ superconformal algebra
\ie \label{eq: N4 SCA}
[L_m, L_n] &= (m-n) L_{m+n} + \tfrac{\mathsf{c}}{12}\, m(m^2-1)\, \delta_{m+n,0}, \\
[L_m, J^a_n] &= -n\, J^a_{m+n}, \\
[L_m, G^{\pm\alpha}_r] &= \left( \tfrac{m}{2} - r \right) G^{\pm\alpha}_{m+r}, \\
[J^3_m, J^3_n] &= \tfrac{k}{2}\, m\, \delta_{m+n,0}, \quad
[J^3_m, J^\pm_n] = \pm J^\pm_{m+n}, \\
[J^+_m, J^-_n] &= 2 J^3_{m+n} + k\, m\, \delta_{m+n,0}, \quad [J^\pm_m, J^\pm_n] = 0 , \\
[J^3_m, G^{\pm\alpha}_r] &= \pm \tfrac{1}{2}\, G^{\pm\alpha}_{m+r}, \\
[J^\pm_m, G^{\mp\alpha}_r] &= G^{\pm\alpha}_{m+r} , \quad [J^\pm_m, G^{\pm\alpha}_r] = 0, \\
\{ G^{+\alpha}_r, G^{-\beta}_s \} &= \epsilon^{\alpha\beta} \Big[ 2 L_{r+s} + 2(r-s)\, J^3_{r+s} + \tfrac{\mathsf{c}}{3} \left( r^2 - \tfrac{1}{4} \right) \delta_{r+s,0} \Big], \\
\{ G^{\pm\alpha}_r, G^{\pm\beta}_s \} &= \mp 2\, \epsilon^{\alpha\beta} (r-s)\, J^\pm_{r+s},
\fe
where $\mathsf{c} = 6N$,\footnote{We use the exact central charge $\mathsf{c}=6N$. For $\Kthree$, the one-loop correction $\delta c=6$ combines with the classical value $6Q_1Q_5=6(N-1)$ to give $c=6N$ \cite{ArabiArdehali:2018mil}.} $k = c/6$, $\epsilon^{+-} = -\epsilon^{-+} = 1$, the index $\pm$ on $G^{\pm\alpha}_r$ is the $\mathfrak{su}(2)_R$ doublet index and $\alpha = \pm$ is the outer $\mathfrak{su}(2)$ index. In the NS sector, $m,n \in \bZ$ and $r,s \in \bZ + \frac{1}{2}$. The right-moving generators satisfy the analog relations.

The case $K=1$ and $S=\Sb=-1$ corresponds to global AdS$_3$ with vanishing gauge fields, dual to the NS vacuum of the CFT. For general $S, \Sb$ at $K=1$, spinors coupling to the gauge fields acquire holonomy phases $e^{i \pi (1+S)\sigma^3}$ and $e^{i \pi (1+\Sb) \sigma^3}$ around the $\tilde\phi$-circle. These phases equal $+1$, and hence preserve the anti-periodicity of spinors, precisely when $S, \Sb \in 2\bZ+1$, which is the condition for NS boundary conditions.

Let us now consider the more general case $K \geq 2$ and $S,\Sb \in 2\bZ+1$. Without considering the effect of gauge holonomies, the anti-periodic spinors in the covering space \eqref{eq: orbifold metric} descend in the base space \eqref{eq: orbifold metric 3} to those that are anti-periodic as they go $K$ times around the $\tilde\phi$-circle. That is, the left-moving spinor harmonic
\be
\zeta_r =
\begin{pmatrix}
e^{+i \frac{r}{K} \tilde\phi} \\
e^{-i \frac{r}{K} \tilde\phi}
\end{pmatrix},
\quad r \in \bZ + \tfrac{1}{2}
\ee
picks up the phase $\exp(\pm2 \pi i \frac{r}{K})$ as it goes once around the $\tilde\phi$-circle. Now consider the effect of gauge holonomies
\ie
e^{i \oint A} &= e^{i \pi \left( 1+\frac{S}{K} \right)\sigma^3} \\
e^{i \oint \bar{A}} &= e^{i \pi \left( 1+\frac{\Sb}{K} \right) \sigma^3} 
\fe
on the spinors that couple to $A,\bar{A}$. We find that the gauge-covariant combination
\be
e^{ -i \int_0^{\tilde\phi} A } \zeta_r =
\begin{pmatrix}
e^{-i \frac{1}{2} \left( 1 + \frac{S-2r}{K} \right) \tilde\phi} \\
e^{+i \frac{1}{2} \left( 1 + \frac{S-2r}{K} \right) \tilde\phi}
\end{pmatrix}
\ee
is anti-periodic under $\tilde\phi \to \tilde\phi + 2\pi$ if
\be \label{eq: antiperiodic condition}
r = \tfrac{S}{2} \ {\rm mod} \ K,
\ee
and similarly on the right. The two components of $\zeta_r$ correspond to the two $\mathfrak{su}(2)_R$ components $G^\pm$ of the supercurrent, with mode labels $\pm r$ respectively. The constraint \eqref{eq: antiperiodic condition} thus implies that the supercurrents have modes $r = \pm\frac{S}{2}$ mod $K$. The bosonic $\cN = (4,4)$ generators that are periodic under $\tilde\phi \to \tilde\phi + 2 \pi$ upon accounting for holonomies can be found in a similar manner.

Explicitly, the base-space generators in \eqref{eq: N4 SCA} are built from auxiliary covering-space generators $\hat{L}_m, \hat{J}^a_m, \hat{G}^{\pm\alpha}_{\hat{r}}$, which obey \eqref{eq: N4 SCA} with $c \to \hat{c} = c/K$ and $\hat{r} \in \bZ + \frac{1}{2}$, as
\ieg \label{eq: covering map}
L_n = \frac{1}{K} \big( \hat{L}_{Kn} + \eta\, \hat{J}^3_{Kn} \big) + \frac{\hat{c}\eta^2}{24K} \delta_{n,0} + \frac{\hat{c}}{24} \left( K - \frac{1}{K} \right) \delta_{n,0}, \\
J^3_n = \hat{J}^3_{Kn} + \frac{\hat{c}}{12}\, \eta\, \delta_{n,0}, \quad J^\pm_n = \hat{J}^\pm_{Kn \pm \eta}, \quad G^{\pm\alpha}_r = \frac{1}{K^{1/2}}\, \hat{G}^{\pm\alpha}_{Kr \pm \frac{\eta}{2}},
\feg
where $\eta \equiv K + S$. The two constant terms in $L_n$ arise from spectral flow and the vacuum-energy shift required by the $K$-fold relabeling of Virasoro modes, respectively. The map \eqref{eq: covering map} is an algebra homomorphism, so the base generators obey \eqref{eq: N4 SCA} with the standard NS moding. The covering fermion modes $Kr \pm \eta/2$ are half-integers if and only if $S$ is odd, and the covering modes appearing on the right-hand side of \eqref{eq: covering map} are those in Table~\ref{tab: mode constraint}. Evaluating \eqref{eq: covering map} at $\hat{L}_0 = \hat{J}^3_0 = 0$ with $\hat{c} = 6N/K$ reproduces the classical charges \eqref{eq: classical charges asymmetric} below. The right-moving sector is identical with $S \to \Sb$.

\begin{table}[t]
\centering
\renewcommand{\arraystretch}{1.4}
\begin{tabular}{l@{\hspace{1.5em}}c}
\hline\hline
Generator & Mode restrictions \\
\hline
$\hat{L}_m$, $\hat{J}_m^3$ & $m = 0$ mod $K$ \\
$\hat{J}_m^\pm$ & $m = \pm S$ mod $K$ \\
$\hat{G}_{\hat{r}}^{\pm \alpha}$ & $\hat{r} = \pm \frac{S}{2}$ mod $K$ \\
\hline\hline
\end{tabular}
\caption{Mode restrictions on left-moving $\cN = 4$ superconformal generators in the covering space that descend to those of Chern-Simons supergravity on \eqref{eq: orbifold metric 3} with spectral flow parameter $S$. }
\label{tab: mode constraint}
\end{table}

We find that the $\cN = (4,4)$ modes satisfying the periodicity constraints in Table~\ref{tab: mode constraint} form a closed superalgebra. The mode restrictions can be interpreted as the result of a fractional spectral flow by $1+S/K$, $1+\Sb/K$ units implemented by the gauge holonomies. The $SU(1,1|2)_L \times SU(1,1|2)_R$ Chern-Simons supergravity on the base space \eqref{eq: orbifold metric 3} is thus well-defined, with fields expanded in modes satisfying the constraints of Table~\ref{tab: mode constraint}. The asymptotic symmetry algebra is correspondingly generated by the $\cN = (4,4)$ modes satisfying these constraints. These fields satisfy the correct boundary conditions to contribute to the NS Hodge elliptic genus.

\subsection{Classical charges and the elliptic genus}

We now turn to the classical energies and charges of these geometries. Using the relation \cite{Balasubramanian:2000rt,Maldacena:2000dr} between the metric and gauge fields in asymptotically-AdS$_3$ spacetimes and the conformal dimensions and charges in the dual CFT, we find that these solutions possess
\ieg \label{eq: classical charges asymmetric}
L_0 = \frac{N}{2} \left( 1 + \frac{S}{K} \right) + \frac{N}{4 K^2} (S^2 - 1) \\
\Lb_0 = \frac{N}{2} \left( 1 + \frac{\Sb}{K} \right) + \frac{N}{4 K^2} (\Sb^2 - 1) \\
J_0 = \frac{N}{2} \left( 1 + \frac{S}{K} \right), \quad \Jb_0 = \frac{N}{2} \left( 1 + \frac{\Sb}{K} \right).
\feg
The Hodge elliptic genus \eqref{eq: hodge genus} is defined in terms of a projection $\qb \to 0$ of the signed R-R partition function to the right-moving Ramond ground states. This translates in the NS-NS sector to the right chiral primary condition $\Lb_0 - \Jb_0 = 0$, which is satisfied by \eqref{eq: classical charges asymmetric} only if $\Sb = \mp 1$. With $\Sb = \mp 1$, we find that the dimensions and charges of asymmetric $\orbifoldK$ orbifolds precisely reproduce the classical weights seen in \eqref{eq: NS2 hodge}.

An important feature of the constraint $\Sb = \mp 1$ is that it yields two geometries for each choice of $K$ and $S$. While these geometries are physically distinct, their classical actions coincide in the elliptic genus limit $\bar{y} \to 1$. This two-fold degeneracy is largely responsible for the non-trivial $N$-dependence of the finite $N$ elliptic genus $Z^{\rm NS}_N$.

Furthermore, not all values of $S \in 2\bZ + 1$ yield consistent orbifold backgrounds. As will be shown in Section~\ref{sec:worldsheet}, the level-matching condition for asymmetric orbifolds \cite{Narain:1986qm,Gaberdiel:2023dxt} requires $M = \frac{1}{4K}(S^2 - 1) \in \bZ_{\geq 0}$. Together with the right chiral primary condition $\Sb = \mp 1$, this accounts for all the constraints \eqref{eq: orbifold constraint}. We thus establish a 1-to-2 correspondence between each $(K,S)$ family of linear, grand-canonical poles in \eqref{eq: linear poles}, with $R$ summed, and a pair of $\orbifoldK$ backgrounds whose classical actions become degenerate in the elliptic genus limit.

\section{One-loop string partition functions on $\orbifoldK$ orbifolds}
\label{sec:worldsheet}

In this section, we compute the one-loop supersymmetric partition function 
on asymmetric $\orbifoldK$ orbifold backgrounds, using the tensionless string worldsheet theory  of \cite{Gaberdiel:2023dxt}.
The goal is to show that the one-loop factor
\begin{equation}
\begin{aligned}
\textrm{One-loop}: &\qquad \frac{1}{K}\sum^{K-1}_{a=0}\omega_K^{-aN}
\textrm{PE}\left[f^{\textrm{RR}}_{(K,S,\bar{S}),a}(q,y,\bar{y})\right]
\end{aligned}  
\end{equation}
with 
\begin{equation}
f^{\textrm{RR}}_{(K,S,\bar{S}),a}(q,y,\bar{y})
:=  -1+\sum_{n=1}^{\infty}\omega_K^{an}\left(q^M y^S \bar{y}^{\bar{S}}\right)^{-\frac{n}{K}}
\frac{1}{n}\sum_{r=0}^{n-1}Z_{\mathrm{seed}}^{\mathrm{RR}}\!\left(e^{2\pi i\frac{r}{n}}q^{\frac{1}{n}},y,\bar y\right)\,,
\end{equation}
is the one-loop Hodge elliptic genus of IIB string on asymmetric $\orbifoldK$ orbifold backgrounds.

\subsection{Review: tensionless strings on $\orbifoldM$}

Let us first review the tensionless string worldsheet theory, on the $\orbifoldM$ orbifold background, labeled by $(S,\bar{S})$.\footnote{
See e.g.\ \cite{Martinec:2018nco,Martinec:2025xoy} for the worldsheet theory (at $k>1$) study of perturbative string spectra on non-supersymmetric and supersymmetric backgrounds in terms of the null-gauged WZW models. 
}
It reduces to the  tensionless string worldsheet theory in the $\adsM$ upon setting 
\begin{equation}
K=1
\,, \qquad 
S=\bar{S}=-1\,.  
\end{equation}
We describe the left-moving sector, the right-moving one is parallel.

The tensionless string worldsheet theory is based on the super affine algebra $\mathfrak{psu}(1,1|2)_1$ \cite{Eberhardt:2018ouy}.
It has a bosonic subalgebra $
\mathfrak{sl}(2,\mathbb{R})_{1}\oplus \mathfrak{su}(2)_{1}$ and $8$ supercharges.
Let us denote the   $\mathfrak{sl}(2,\mathbb{R})_{1}$ modes by $\mathcal{J}^{3,\pm}_m$, the $\mathfrak{su}(2)_{1}$ modes by $\mathcal{K}^{3, \pm}_m$, and the worldsheet Virasoro generator by $\mathcal{L}_m$.
The mass-shell condition is $\mathcal{L}_0=0$.
The $\mathfrak{psu}(1,1|2)_1$ algebra has a spectral flow outer-automorphism, labeled by $w\in \mathbb{Z}$ \cite{Maldacena:2000hw,Eberhardt:2018ouy}. 
Its actions on the bosonic currents are
\begin{equation}\label{eq:WSSpectralFlow}
\begin{aligned}
&\Sigma^{w}(\mathcal{L}_m)=\mathcal{L}_{m}-w(\mathcal{J}^3_m-\mathcal{K}^3_m)\,,\\
&\Sigma^{w}(\mathcal{J}^3_m)=\mathcal{J}^3_{m}+\frac{w}{2}\delta_{m,0}
\,,\qquad 
\Sigma^{w}(\mathcal{J}^{\pm}_m)=\mathcal{J}^{\pm}_{m\mp w}\,,
\\
&\Sigma^{w}(\mathcal{K}^3_m)=\mathcal{K}^3_{m}+\frac{w}{2}\delta_{m,0}
\,, \qquad 
\Sigma^{w}(\mathcal{K}^{\pm}_m)=\mathcal{K}^{\pm}_{m\pm w}
\,.
\end{aligned}
\end{equation}
Note that the actions on $\mathcal{J}^{\pm}$ and $\mathcal{K}^{\pm}$ fix the actions on the other three generators due to the commutation relations of the  bosonic subalgebra $
\mathfrak{sl}(2,\mathbb{R})_{1}\oplus \mathfrak{su}(2)_{1}$ and the Virasoro algebra. 

For the orbifold background $\orbifoldM$ labeled by $(S,\bar{S})$, the worldsheet orbifold generator is
\begin{equation}\label{eq:asymmetric-generator}
g
=\exp\left[\frac{2\pi i}{K}\left(\mathcal{J}_0^3+S\mathcal{K}_0^3\right)\right]
\otimes
\exp\left[-\frac{2\pi i}{K}\left(\bar{\mathcal{J}}_0^3+\bar{S}\bar{\mathcal{K}}_0^3\right)
\right]\,,
\end{equation}
with level matching condition\footnote{Note that $S$ is related to the parameter $s$ in \cite{Gaberdiel:2023dxt} via $S_{\textrm{here}}=-2s|_{\textrm{there}}-1$.}
\begin{equation}\label{eq:asymmetric-level}
\frac{S^2-\bar{S}^2}{4K}
\in
\mathbb{Z}\,.
\end{equation}
In order to align with the orbifold action, it is more convenient to modify the spectral flow action on the currents $\mathcal{J}^{\pm}$ and $\mathcal{K}^{\pm}$ in \eqref{eq:WSSpectralFlow} to\footnote{We are free to choose the spectral flow of the $\mathfrak{su}(2)$ factor since it is just a relabeling of the states for $\mathfrak{su}(2)$. However, a choice that is aligned with the orbifold action would make the description of the twisted sector simpler. } 
\begin{equation}\label{eq:eq:WSSpectralFlowOrbifoldJK}
\Sigma^{w}(\mathcal{J}^{\pm}_m)=\mathcal{J}^{\pm}_{m\mp w}
\,, \qquad 
\Sigma^{w}(\mathcal{K}^{\pm}_m)=\mathcal{K}^{\pm}_{m\mp S w}\,,
\end{equation}
with spectral flow parameter
\begin{equation}
w=\frac{n}{K}\,, \qquad n\in \mathbb{N}\,.
\end{equation}
This then determines the action on the remaining bosonic currents
\begin{equation}
\begin{aligned}
\Sigma^{w}(\mathcal{L}_m)&=\mathcal{L}_{m}-w(\mathcal{J}^3_m+S\mathcal{K}^3_m)+\frac{S^2-1}{4}w^2\delta_{m,0}\,,\\
\Sigma^{w}(\mathcal{J}^3_m)&=\mathcal{J}^3_{m}+\frac{w}{2}\delta_{m,0}\,,\qquad 
\Sigma^{w}(\mathcal{K}^3_m)=\mathcal{K}^3_{m}-\frac{Sw}{2}\delta_{m,0}\,.
\end{aligned}
\end{equation}

Let us now consider a (not necessarily physical) state in the unflowed sector (with $w=0$) with worldsheet charge $(\mathcal{L}_0, \mathcal{J}^3_0, \mathcal{K}^3_0)$, and consider the $w$-spectrally-flowed image of these states
and demand them to be physical.
Since $\mathfrak{psu}(1,1|2)_1$ can be written in terms of $2$ free bosons and $4$ free fermions, one would expect that the physical state condition will remove all these free field excitation modes, and thus retain only the ground states of the $\mathfrak{psu}(1,1|2)_1$ representation, together with arbitrary states in $\mathcal{M}_4$. 
Furthermore, as the analysis of \cite{Gaberdiel:2021njm} suggests, the $\mathfrak{su}(2)$ R-symmetry of the $\mathcal{M}_4$ theory gets identified with the $\mathfrak{su}(2)$ subalgebra of $\mathfrak{psu}(1,1|2)$ in this process. 
Thus the physical states in each $w$-cycle twisted sector are in one-to-one correspondence with the states in $\mathcal{M}_4$, with the identification 
 \begin{equation}
\mathcal{L}_0=h^{\mathcal{M}_4}
\qquad \textrm{and} \qquad 
\mathcal{K}^3_0= \frac{1}{2}+k^{\mathcal{M}_4} \,,
\end{equation}
where $h^{\mathcal{M}_4}$ and  $k^{\mathcal{M}_4}$ are  the total conformal dimension and the $\mathcal{K}^3_0$ charge of the excitations of the sigma model of $\mathcal{M}_4$.
The mass-shell condition $\Sigma^{(w)}(\mathcal{L}_0)=0$
then fixes the charges of the $w$-spectrally-flowed state to be
\begin{equation}\label{eq:JKworldsheetNS}
\begin{aligned}
\mathcal{J}^3_0&=\frac{h^{\mathcal{M}_4}}{w}+\frac{w-S}{2}-Sk^{\mathcal{M}_4}+\frac{S^2-1}{4}w\,,\\
\mathcal{K}^3_0&=\frac{-Sw+1}{2}+k^{\mathcal{M}_4} \,.
\end{aligned}
\end{equation}
The charges \eqref{eq:JKworldsheetNS} are to be compared to the spacetime charges in the NS sector
\begin{equation}
h_{\mathrm{NS}}
=\frac{i}{2}(\partial_t-\partial_{\phi})
=\mathcal{J}^{3}_{0},
\qquad
j_{\mathrm{NS}}
=\frac{i}{2}(\partial_{\chi}-\partial_{\psi})
=\mathcal{K}^{3}_{0}\,,
\end{equation}
where $(t,\phi,\chi,\psi)$ are from the global $\adsS$ metric \eqref{eq: orbifold metric}, subject to the orbifold identification \eqref{eq: orbifold identification}. 
It is actually more convenient to compare to the boundary CFT in the Ramond sector, using charges defined with respect to the Ramond form of the metric \eqref{eq: orbifold metric 2}:
\begin{equation}\label{eq:AdS3JKROrbifold}
\begin{aligned}
\mathcal{J}_{0,\mathrm{R}}^3
&=\frac{i}{2}(\partial_{\tilde{t}_{\textrm{R}}}-\partial_{\tilde{\phi}_{\textrm{R}}})
=\frac{1}{K}\left(\mathcal{J}_0^3+S\mathcal{K}_0^3\right)
=\frac{h^{\mathcal{M}_4}}{wK}-wM\,,\\
\mathcal{K}_{0,\mathrm{R}}^3
&=\frac{i}{2}(\partial_{\tilde{\chi
}_{\textrm{R}}}-\partial_{\tilde{\psi}_{\textrm{R}}})
=
\mathcal{K}_0^3
\qquad \qquad\quad \, =\frac{-Sw+1}{2}+k^{\mathcal{M}_4}\,,
\end{aligned}
\end{equation}
where 
\begin{equation}
\begin{aligned}
&\tilde{t}_{\textrm{R}}:=\tilde{t}=Kt
\,, \qquad  \tilde{\phi}_{\textrm{R}}:=\tilde{\phi}=K\phi
  \\
&\tilde{\chi}_{\textrm{R}}:=\tilde{\chi}+\tilde{t} 
\,, \qquad  \tilde{\psi}_{\textrm{R}}:=\tilde{\psi}+\tilde{\phi}\,.
\end{aligned}
\end{equation}
Here $w=\frac{n}{K}$ with $n\in\mathbb{N}$, and we have defined 
\begin{equation}
M:=\frac{S^2-1}{4K}  \,.
\end{equation}
The right-moving sector is similar.

Let us first check that this worldsheet computation indeed reproduces the single-particle spacetime CFT spectrum.
We can do this in the Ramond sector, and the result for the NS sector computation can be obtained by spectral flow.

It is useful to first recall how the tensionless worldsheet reproduces the spectrum of $\textrm{Sym}^{N}(\mathcal{M}_4)$.
In the seed theory (with central charge $c=\bar{c}=6$), choose the background Ramond ground state $|0_1^{++}\rangle_{\mathrm{R}}$ with
\begin{equation}\label{eq:seed-background-ground}
h^{\textrm{seed}}_{\textrm{R},0}
=\frac{c}{24}=\frac{1}{4}\,,
\qquad
j^{\textrm{seed}}_{\textrm{R},0}
=\frac{1}{2}\,.
\end{equation}
A generic state in the seed theory has conformal dimension and $\mathfrak{su}(2)_\mathrm{R}$ charge
\begin{equation}
h^{\textrm{seed}}_{\textrm{R}}=N^{\mathrm{CFT}}+\frac{1}{4}
\,, \qquad
j^{\textrm{seed}}_{\textrm{R}}=k^{\textrm{CFT}}+\frac{1}{2}\,,
\end{equation}
where $N^{\mathrm{CFT}}$ is the excitation of the seed theory, and $k^{\textrm{CFT}}$ is $\mathfrak{su}(2)_\mathrm{R}$ charge of these excitations.

In the symmetric orbifold, the single-particle spectrum comes from  a single $w$-cycle twisted sector.
We work with the convention that the ground state in this sector is $|0^{++}_w\rangle_{\mathrm{R}}$, with the remaining $N-w$ copies all in the ground state $|0_1^{--}\rangle_{\mathrm{R}}$.
The $w$ copies of $|0_1^{--}\rangle_{\mathrm{R}}$, before being replaced by $|0^{++}_w\rangle_{\mathrm{R}}$, have total charges
\begin{equation}\label{eq:background-gs}
h_{\textrm{R},0}
=w \, h^{\textrm{seed}}_{\textrm{R},0}
=\frac{w}{4}\,,
\qquad
j_{\textrm{R},0}
=-w \, j^{\textrm{seed}}_{\textrm{R},0}
=-\frac{w}{2}\,.
\end{equation}
The excited state in the $w$-twisted sector (i.e.\ above $|0^{++}_w\rangle_{\mathrm{R}}$) has charges
\begin{equation}
h_{\textrm{R}}=\frac{N^{\mathrm{CFT}}}{w}+\frac{w}{4}
\,, \qquad
j_{\textrm{R}}=k^{\textrm{CFT}}+\frac{1}{2} \,.
\end{equation}
The difference is then
\begin{equation}\label{eq:Deltahj}
\Delta h_{\textrm{R}}=h_{\textrm{R}}-h_{\textrm{R},0}=\frac{N^{\mathrm{CFT}}}{w}
\,, \qquad
\Delta j_{\textrm{R}}=j_{\textrm{R}}-j_{\textrm{R},0}=k^{\textrm{CFT}}+\frac{w+1}{2}
\end{equation}
The analysis for the right-moving sector is similar.
\eqref{eq:Deltahj} is to be compared with \eqref{eq:AdS3JKROrbifold} at $K=1,S=\bar{S}=-1$.
We see that they agree 
as long as
\begin{equation}\label{eq:WSvsST}
h^{\mathcal{M}_4}=N^{\mathrm{CFT}} 
\qquad \textrm{and}\qquad
k^{\mathcal{M}_4}=k^{\textrm{CFT}}\,.
\end{equation}
The first identity is true by definition. 
The second identity has been proven for the $\mathbb{T}^4$ case for the first few states, see \cite{Gaberdiel:2021njm}.

Let us now generalize this computation to the $\orbifoldM$.
As before, we first work in the Ramond sector.
It was proposed in \cite{Gaberdiel:2023dxt} that for the spacetime CFT that is dual to the tensionless string worldsheet theory on the $\orbifoldM$ orbifold background, the ground state $|\textrm{GS}\rangle_{\textrm{orbifold}}$ should be the (previously excited) state in the original $\textrm{Sym}^N(\mathcal{M}_4)$ where the $N$ copies are arranged into $N/K$ $K$-cycles.\footnote{The CFT description of fractional spectral flow on $K$-cycles was already studied in \cite{Giusto:2012yz,Chakrabarty:2015foa}.}
For each $K$-cycle, its ground state $|0^{++}_{K}\rangle_{\mathrm{R}}$ has charges
\begin{equation}\label{eq:K-background-ground}
h^{\textrm{single}}_{\textrm{R},0}=
\frac{K}{4}+M\,,
\qquad
j^{\textrm{single}}_{\textrm{R},0}=-\frac{S}{2}\,,
\end{equation}
which reduces to \eqref{eq:seed-background-ground} upon setting $K=1$ and $S=-1$.

The $w$-spectrally-flowed sector of the worldsheet theory in the $(S,\bar{S})$ $\orbifoldM$ background corresponds to the single particle spectrum in the $n=wK$ twisted sector.
The ground state in this sector is
$|0^{++}_{wK}\rangle_{\mathrm{R}}$, with the remaining $N-wK$ copies
arranged into $\frac{N}{K}-w$ number of $K$-cycles, each in the ground state $|0^{--}_{K}\rangle_{\mathrm{R}}$.
The $w$ copies of $K$-cycles, before being replaced by one $wK$-cycle, have total charges
\begin{equation}\label{eq:background-gs-orbifold}
h_{\textrm{R},0}
=w \, h^{\textrm{single}}_{\textrm{R},0}
=\frac{wK}{4}+w M\,,
\qquad
j_{\textrm{R},0}
=-w \, j^{\textrm{single}}_{\textrm{R},0}
=\frac{Sw}{2}\,.
\end{equation}
On the other hand, an excited state in the $n=wK$ twisted sector has charges
\begin{equation}
h_{\textrm{R}}=\frac{N^{\mathrm{CFT}}}{wK}+\frac{wK}{4}
\,, \qquad
j_{\textrm{R}}=\frac{1}{2}+k^{\textrm{CFT}} \,.
\end{equation}
Therefore, the charges of the excited state relative to the ground state $|\textrm{GS}\rangle_{\textrm{Orbifold}}$
 are the difference:
\begin{equation}\label{eq:ExcitationOrbifold}
\begin{aligned}
\Delta h_{\textrm{R}}&=h_{\textrm{R}}-h_{\textrm{R},0}=\frac{N^{\mathrm{CFT}}}{wK}-wM\,,\\
\Delta j_{\textrm{R}}&=j_{\textrm{R}}-j_{\textrm{R},0}=\frac{-Sw+1}{2}+k^{\textrm{CFT}} \,.
\end{aligned}
\end{equation}
The excitation charges \eqref{eq:ExcitationOrbifold} are to be compared with the worldsheet charges \eqref{eq:AdS3JKROrbifold}.
We see that we have the match 
\begin{equation}\label{eq:WSSTMatchOrbifold}
\Delta h_{\textrm{R}}= \mathcal{J}^3_{0,\mathrm{R}} 
\qquad\textrm{and}\qquad
\Delta j_{\textrm{R}}= \mathcal{K}^3_{0,\mathrm{R}}
\end{equation}
as long as 
\begin{equation}\label{eq:WSvsST1}
h^{\mathcal{M}_4}=N^{\mathrm{CFT}} 
\qquad \textrm{and}\qquad
k^{\mathcal{M}_4}=k^{\textrm{CFT}}\,.
\end{equation}

\subsection{Worldsheet and spacetime CFT partition function}

We have explained the matching of the full spectrum. 
We now use this to show that the residue computed in Section~\ref{sec:ExpansionCE} is indeed the one-loop Hodge elliptic genus of IIB string on the asymmetric $\orbifoldK$ orbifold.
In fact, the computation can be done at the level of the signed partition function, and we only specialize to HEG at the very last step.

Let's first compare the (signed) partition function of the $w=\frac{1}{K}$ sector of the worldsheet theory 
\begin{equation}
\mathsf{z}^{(w=\frac{1}{K})}
(q,\bar{q},y,\bar{y}):= 
\mathrm{Tr}_{\mathcal H^{(w=\frac{1}{K})}_{\mathrm{ws,phys}}}
\left[
(-1)^F
q^{\mathcal{J}^{3}_{0,\mathrm{R}}}
\bar{q}^{\bar{\mathcal{J}}^{3}_{0,\mathrm{R}}}
y^{2\mathcal{K}^{3}_{0,\mathrm{R}}}
\bar{y}^{2\bar{\mathcal{K}}^{3}_{0,\mathrm{R}}}
\right] 
\end{equation}
with the (signed) partition function of the seed spacetime CFT (sigma model of $\mathcal{M}_4$) in the RR sector 
\begin{equation}\label{eq:ZRRseed}
Z^{\mathrm{RR}}_{\mathrm{seed}}(q,\bar q,y,\bar{y})
=
\mathrm{Tr}_{\mathcal H^{\mathrm{RR}}_{\mathrm{seed}}}
\left[
(-1)^F 
q^{L_0-\frac{c}{24}}\bar q^{\bar{L}_0-\frac{\bar{c}}{24}}y^{2J_0}\bar{y}^{2\bar J_0}
\right]
=
\sum_{m,\bar m,\ell,\bar{\ell}} d(m,\bar m,\ell,\bar{\ell})q^m\bar q^{\bar m}y^\ell \bar{y}^{\bar{\ell}}
\end{equation}
with $c=\bar{c}=6$.
Using the identification  \eqref{eq:WSvsST1}, we know that 
a state in the $w=\frac{1}{K}$ sector of the worldsheet theory with charge $h^{\mathcal{M}_4} $ and $k^{\mathcal{M}_4}$
corresponds to a state in the seed spacetime CFT with charge 
\begin{equation}\label{eq:ChargeMapn=1}
\begin{aligned}
\mathcal{J}^{3}_{0,\mathrm{R}}
&=h^{\mathcal{M}_4}-\frac{M}{K}\\
&=N^{\textrm{CFT}}-\frac{M}{K}
=(L_0-\frac{c}{24})-\frac{M}{K}=m-\frac{M}{K}
 \\
2\mathcal{K}^{3}_{0,\mathrm{R}}
&=2k^{\mathcal{M}_4} +1-\frac{S}{K}=2k^{\textrm{CFT}}+1-\frac{S}{K}\\
&=2J_0-\frac{S}{K}=\ell-\frac{S}{K} \,,
\end{aligned}
\end{equation}
and similarly for the right sector.
Therefore the term $q^{\mathcal{J}^{3}_{0,\mathrm{R}}}
\bar{q}^{\bar{\mathcal{J}}^{3}_{0,\mathrm{R}}}
y^{2\mathcal{K}^{3}_{0,\mathrm{R}}}
\bar{y}^{2\bar{\mathcal{K}}^{3}_{0,\mathrm{R}}}$ in the $w=\frac{1}{K}$-spectrally flowed sector contributes
\begin{equation}
\mathfrak{p}_K(q,\bar{q},y,\bar{y})
\cdot
q^{m}\bar q^{\bar m}y^{\ell}\bar{y}^{\bar{\ell}}
\end{equation}
to $\mathsf{z}^{(w=\frac{1}{K})}
(q,\bar{q},y,\bar{y})$, where we have pulled out a prefactor that is independent of $(m,\bar{m},\ell, \bar{\ell})$:
\begin{equation}\label{eq:pKDef}
\mathfrak{p}_K(q,\bar{q},y,\bar{y})
:= ( q^{M}\, \bar{q}^{\bar{M}}\, y^{S} \,\bar{y}^{\bar{S}})^{-\frac{1}{K}}\,.
\end{equation}
This gives the identity:
\begin{equation}\label{eq:zWStoZspacetimew1K}
\mathsf{z}^{(w=\frac{1}{K})}
(q,\bar{q},y,\bar{y})
=\mathfrak{p}_K(q,\bar{q},y,\bar{y}) \cdot 
Z^{\mathrm{RR}}_{\mathrm{seed}}(q,\bar q,y,\bar{y}) \,. 
\end{equation}

In the worldsheet theory, the $w$-spectrally flowed sector corresponds to a long string of length $n=wK$. 
Therefore, the map between the charges of the worldsheet and those of the spacetime CFT \eqref{eq:ChargeMapn=1} becomes
\begin{equation}\label{eq:ChargeMapn}
\begin{aligned}
\mathcal{J}^{3}_{0,\mathrm{R}}
&=\frac{h^{\mathcal{M}_4}}{wK}-w M\\
&=\frac{N^{\textrm{CFT}}}{n}-n\cdot\frac{M}{K}
=\frac{L_0-\frac{c}{24}}{n}-n\cdot\frac{M}{K}
=\frac{m}{n}-n\cdot\frac{M}{K}
 \\
2\mathcal{K}^{3}_{0,\mathrm{R}}
&=2k^{\mathcal{M}_4} +1-Sw\\
&=2k^{\textrm{CFT}}+1-n\cdot\frac{S}{K}
=2J_0-n\cdot \frac{S}{K}
=\ell-n\cdot \frac{S}{K}\,,
\end{aligned}
\end{equation}
and similar for the right-moving sector. 
Therefore the term $q^{\mathcal{J}^{3}_{0,\mathrm{R}}}
\bar{q}^{\bar{\mathcal{J}}^{3}_{0,\mathrm{R}}}
y^{2\mathcal{K}^{3}_{0,\mathrm{R}}}
\bar{y}^{2\bar{\mathcal{K}}^{3}_{0,\mathrm{R}}}$ in the $w=\frac{n}{K}$-spectrally flowed sector now contributes
\begin{equation}
 \mathfrak{p}_K(q,\bar{q},y,\bar{y})^n \cdot
q^{m/n}\bar q^{\bar m/n}y^{\ell}\bar{y}^{\bar{\ell}}.
\end{equation}
In addition, we need to impose 
the worldsheet level matching in the $w$-spectrally flowed sector
\begin{equation}\label{eq:level-matchingSingleString}
\frac{m-\bar m}{n}\in \mathbb Z 
\,, \qquad 
n=wK
\,, \quad 
\textrm{with} \quad 
n\in\mathbb{Z}\,,
\end{equation}
which can be realized using the cyclic projection operator
\begin{equation}
\Pi_n
=
\frac{1}{n}\sum_{r=0}^{n-1}g_n^r,
\qquad
g_n=\exp\left[\frac{2\pi i}{n}\left(L_0-\bar L_0\right)\right].
\end{equation}
Therefore the (signed) partition function of the $w=\frac{n}{K}$-spectrally flowed sector is
\begin{equation}\label{eq:zWStoZspacetimewnK}
\begin{aligned}
\mathsf{z}^{(w=\frac{n}{K})}(q,\bar{q},y,\bar{y})&
= \mathfrak{p}_K(q,\bar{q},y,\bar{y})^n \cdot
\sum_{\substack{m,\bar m,\ell,\bar{\ell}\\ m-\bar m\in n\mathbb Z}}
d(m,\bar m,\ell,\bar{\ell})q^{m/n}\bar{q}^{\bar m/n}y^{\ell}\bar{y}^{\bar{\ell}}\\
&= \mathfrak{p}_K(q,\bar{q},y,\bar{y})^n\cdot
 \frac{1}{n}\sum_{r=0}^{n-1}
Z^{\mathrm{RR}}_{\mathrm{seed}}
\left(e^{2\pi i r/n}q^{1/n},e^{-2\pi i r/n}\bar q^{1/n},y,\bar{y}\right)\,.
\end{aligned}
\end{equation}
The second factor in \eqref{eq:zWStoZspacetimewnK} is actually
\begin{equation}\label{eq:chiDef}
\begin{aligned}
\chi_n^s(q,\bar{q},y,\bar{y})
&:=
\mathrm{Tr}_{(\mathcal{H}_{(n)}^{\mathrm{RR}})^{\mathbb{Z}_n}}
\left[
(-1)^F
q^{L_0-\frac{c_n}{24}}
\bar{q}^{\bar{L}_0-\frac{\bar{c}_n}{24}}
y^{2J_0}\bar{y}^{2\bar{J}_0}
\right]  \\
&=\frac{1}{n}\sum_{r=0}^{n-1}
Z^{\mathrm{RR}}_{\mathrm{seed}}
\left(e^{2\pi i r/n}q^{1/n},e^{-2\pi i r/n}\bar q^{1/n},y,\bar{y}\right) \,,
\end{aligned}
\end{equation}
where $\mathcal{H}_{(n)}^{\mathrm{RR}}$ is the Hilbert space of a single $n$ cycle and  $(\mathcal{H}_{(n)}^{\mathrm{RR}})^{\mathbb{Z}_n}$ is its cyclically invariant subspace, $c_n=\bar{c}_n=6n$.
Therefore the identity for $w=\frac{1}{K}$ \eqref{eq:zWStoZspacetimew1K} is generalized to 
\begin{equation}\label{eq:zWStoZspacetimeRRwbK}
\begin{aligned}
\mathsf{z}^{(w=\frac{n}{K})}(q,\bar{q},y,\bar{y})
&=\mathfrak{p}_K(q,\bar{q},y,\bar{y})^n \cdot\chi_n^s(q,\bar{q},y,\bar{y})\,.  
\end{aligned}
\end{equation}

Now that we have obtained the worldsheet partition function for an arbitrary spectrally flowed sector, we can assemble them to get the full worldsheet partition function (which captures the single-particle excitation on top of the orbifold background). 
Naively, one would have thought that we should first evaluate the full (signed) partition function of the tensionless worldsheet by summing over all the spectrally-flowed sectors:
\begin{equation}
\mathsf{z}^{\textrm{ws}}_{
K
}(q,\bar{q},y,\bar{y})=\sum^{\infty}_{n=1} \mathsf{z}^{(w=\frac{n}{K})}_{
}(q,\bar{q},y,\bar{y}) \,,
\end{equation}
and then obtain the full spacetime partition function by taking the PE of $\mathsf{z}^{\textrm{ws}}_{K}$. 
However, as already explained in \cite{Lee:2025veh}, we need to impose the following constraint. 

From the dictionary \eqref{eq:ChargeMapn}, the spacetime charges of a multi-string excitation, including both the classical and one-loop contribution, are
\begin{equation}
\begin{aligned}
L^{\textrm{R}}_0-\frac{N}{4} &=N\frac{M}{K}  +\sum_i (\frac{m_i}{n_i}-n_i \frac{M}{K}) 
=(N-\sum_i n_i)\frac{M}{K}  +\sum_i \frac{m_i}{n_i}
\\
J^{\textrm{R}}_0 &=\frac{1}{2}\left(N\frac{S}{K} +\sum_i (\ell_i-n_i \frac{S}{K})\right)
=\frac{1}{2}\left((N-\sum_i n_i)\frac{S}{K} +\sum_i \ell_i \right)\,,
\end{aligned}
\end{equation}
and similarly for the right-movers. 
The constraints are
\begin{equation}
L^{\textrm{R}}_0-\bar{L}^{\textrm{R}}_0\in \mathbb{Z}
\, \qquad 
\textrm{and}
\qquad
J^{\textrm{R}}_0, \bar{J}^{\textrm{R}}_0\in \frac{1}{2}\mathbb{Z}\,.
\end{equation}
Given that $S^2-4KM=1$, we have gcd$(S,K)=1$. 
And using the level-matching condition for each string \eqref{eq:level-matchingSingleString}, both constraints give
\begin{equation}
N-\sum_i n_i \in K \, \mathbb{Z}\,.
\end{equation}

As in \cite{Lee:2025veh}, this can be implemented using the identity
\begin{equation}
\frac{1}{K}\sum_{a=0}^{K-1}\omega_K^{a(\sum_i n_i-N)}=\begin{cases}1\,,&N-\sum_i n_i\in K\mathbb Z\,,\\0\,,&\text{otherwise}\,.\end{cases} 
\end{equation}
Namely, define 
\begin{equation}
f^{\mathrm{RR}}_{(K,S,\bar{S}),a}(q,\bar{q},y,\bar y)=-1+\sum_{n=1}^{\infty}\omega_K^{an}\mathsf{z}^{(w=\frac{n}{K})}(q,\bar{q},y,\bar y)\,
\end{equation}
then 
\begin{equation}
\frac{1}{K}\sum_{a=0}^{K-1}\omega_K^{-aN}\mathrm{PE}\!\left[f^{\mathrm{RR}}_{(K,S,\bar S),a}(q,\bar q,y,\bar y)\right]
\end{equation}
computes the full one-loop (signed) partition function in the $\orbifoldM$ orbifold background.

\subsection{Hodge elliptic genus}

The identity above is at the level of the full spectrum and the (signed) partition function. 
We can now specialize to the Hodge elliptic genus.
We now consider $\mathcal{M}_4$=K3.

In the signed partition function of the seed theory in the Ramond sector \eqref{eq:ZRRseed}, since $\bar{m}\geq 0$, the limit $\bar{q}\rightarrow 0$ exists and selects the right-moving Ramond ground states with $\bar{m}=0$, which gives the seed Hodge elliptic genus in the RR sector:
\begin{equation}
\begin{aligned}
Z^{\mathrm{RR}}_{\mathrm{seed}}(q,y,\bar{y})
&:=\textrm{lim}_{\bar{q}\rightarrow 0} Z^{\mathrm{RR}}_{\mathrm{seed}}(q,\bar{q},y,\bar{y})\\
&=
\operatorname{Tr}_{\mathcal H^{\mathrm{RR}}_{\mathrm{seed},\forall \otimes vac}}
\left[
(-1)^F q^{L_0-\frac{c}{24}}y^{2J_0}\bar{y}^{2\bar{J}_0}
\right]
=
\sum_{m,\ell,\bar{\ell}} c(m,\ell,\bar{\ell})q^my^\ell \bar{y}^{\bar{\ell}}  
\end{aligned}
\end{equation}
with $c(m,\ell,\bar{\ell})=\d(m,0,\ell,\bar{\ell})$.

Now consider the tensionless worldsheet theory in the orbifold backgrounds of Section~\ref{sec: classical}. 
All these backgrounds satisfy $\bar{S}=\pm1$, and hence 
\begin{equation}
\bar{M}=0\,.  
\end{equation}
Therefore, in the $w$-spectrally flowed worldsheet signed partition functions \eqref{eq:zWStoZspacetimewnK}, the right-moving charge is
\begin{equation}
\bar{\mathcal{J}}^3_{0,\textrm{R}}
=\frac{\bar{m}}{n}-n\cdot\frac{\bar{M}}{K}
=\frac{\bar{m}}{n}\,,
\end{equation}
with $\bar{m}\geq 0$.
Therefore, the limit $\bar{q}\rightarrow 0$ exists and selects the states with $\bar{m}=0$, producing the corresponding $w$-spectrally flowed worldsheet Hodge elliptic genus:
\begin{equation}\label{eq:zWStoZspacetimewnKHEG}
\begin{aligned}
\mathsf{z}^{(w=\frac{n}{K})}(q,y,\bar{y})&:=\textrm{lim}_{\bar{q}\rightarrow 0} \mathsf{z}^{(w=\frac{n}{K})}(q,\bar{q},y,\bar{y}) \\
&
= \mathfrak{p}_K(q,y,\bar{y})^n \cdot
\sum_{\substack{m,\ell,\bar{\ell}\\ m\in n\mathbb Z}}
c(m,\ell,\bar{\ell})q^{m/n}y^{\ell}\bar{y}^{\bar{\ell}}\\
&= \mathfrak{p}_K(q,y,\bar{y})^n\cdot
 \frac{1}{n}\sum_{r=0}^{n-1}
Z^{\mathrm{RR}}_{\mathrm{seed}}
\left(e^{2\pi i r/n}q^{1/n},y,\bar{y}\right)\,,
\end{aligned}
\end{equation}
where 
\begin{equation}\label{eq:pKDefHEG}
\mathfrak{p}_K(q,y,\bar{y})
:= ( q^{M}\, y^{S} \,\bar{y}^{\bar{S}})^{-\frac{1}{K}}\,.
\end{equation}
Similarly, the HEG version of \eqref{eq:chiDef} is 
\begin{equation}\label{eq:chiDefHEG}
\begin{aligned}
\chi_n^s(q,y,\bar{y})
&:=
\mathrm{Tr}_{(\mathcal{H}_{(n)}^{\mathrm{RR}})^{\mathbb{Z}_n},\forall \otimes vac}
\left[
(-1)^F
q^{L_0-\frac{c_n}{24}}
y^{2J_0}\bar{y}^{2\bar J_0}
\right]  \\
&=\frac{1}{n}\sum_{r=0}^{n-1}
Z^{\mathrm{RR}}_{\mathrm{seed}}
\left(e^{2\pi i r/n}q^{1/n},y,\bar{y}\right)\,.
\end{aligned}
\end{equation}
And the HEG specialization of \eqref{eq:zWStoZspacetimeRRwbK} is
\begin{equation}\label{eq:zWStoZspacetimeRRwbKHEG}
\begin{aligned}
\mathsf{z}^{(w=\frac{n}{K})}(q,y,\bar{y})
&=\mathfrak{p}_K(q,y,\bar{y})^n \cdot\chi_n^s(q,y,\bar{y})\,.  
\end{aligned}
\end{equation}

Finally, define 
\begin{equation}
f^{\mathrm{RR}}_{(K,S,\bar{S}),a}(q,y,\bar y)=-1+\sum_{n=1}^{\infty}\omega_K^{an}\mathsf{z}^{(w=\frac{n}{K})}(q,y,\bar y)\,
\end{equation}
then 
\begin{equation}
\frac{1}{K}\sum_{a=0}^{K-1}\omega_K^{-aN}\mathrm{PE}\!\left[f^{\mathrm{RR}}_{(K,S,\bar S),a}(q,y,\bar y)\right]
\end{equation}
computes the full one-loop HEG in the $\orbifoldM$ orbifold background.  
And $y\rightarrow q^{\frac{1}{2}}y$ gives the NS version of the one-loop HEG \eqref{eq:ZhatNSfactors} with \eqref{eq:fNS}.

\section{Finite $N$ expansion in microcanonical ensemble}
\label{sec:ExpansionMCE}

The main result of the analysis in Section~\ref{sec:ExpansionCE} is \eqref{eq:FiniteNExpCE}, which we reprint here:
\begin{equation}\label{eq:FiniteNExpCE1}
Z^{\textrm{RR}}_N(q,y)
=\mathcal{E}_X[Z^{\textrm{RR}}_{N,\Delta_t}(q,y)]+\sum_{i=1}^{t}\mathcal{E}_X[\widehat{Z}^{\textrm{RR}}_{N,\Gamma_i}(q,y)]
\,.     
\end{equation}
The expression on the r.h.s. depends explicitly on the value of $X$, giving 
infinitely many Stokes chambers. 
In this section, we are going to show that for each polar state $Q$ (with $\Delta(Q)<0$), there is a preferred choice of $X$ for which the Fourier coefficient of the remaining contour integral $Z^{\textrm{RR}}_{N,\Delta_t}$ (at charge $Q$) vanishes.
The microcanonical degeneracy is then a finite sum of residues, each of which can be reproduced by summing the HEG of a pair of orbifolds, as already shown in Section~\ref{sec: classical} and Section~\ref{sec:worldsheet}. 

\subsection{Charges and shifted charges}

To obtain a finite sum, we need to determine when the contribution from 
the remaining chambers vanish. 
The vanishing condition can be written in terms of the charge $Q$, or more conveniently, in terms of a shifted charge $Q'$ (to be defined below).

Recall that the three chemical potentials can be packaged into the genus-two period matrix $\Omega$. 
Accordingly, it is natural to assemble the charge $(N,m,\ell)$ into the associated charge matrix $Q$
\begin{equation}\label{eq:locked-period-charge}
Q
:=
\begin{pmatrix}
N&\ell/2\\
\ell/2&m
\end{pmatrix}\,,
\end{equation}
with the natural SL$(2,\mathbb{Z})$ pairing
\begin{equation}
Q\cdot \Omega =N\sigma+m\tau+\ell z\,.
\end{equation} 
Given $Q$, its degeneracy is
\begin{equation}
\begin{aligned}
d_{\textrm{D}}(Q)
&=
\oint_{\mathcal C_{\textrm{D}}}
\frac{dy}{2\pi i\,y^{\ell+1}}
\frac{dq}{2\pi i\,q^{m+1}}
\frac{d\myzeta}{2\pi i\,\myzeta^{N+1}}
\,
\mathcal{Z}^{\mathrm{GCE}}_{\textrm{RR}}(\Omega)=
\oint_{\mathcal{C}_{\textrm{D}}}
d\Omega \,  e^{-2\pi i \textrm{Tr}[Q \Omega]}
\,
\mathcal{Z}^{\mathrm{GCE}}_{\textrm{RR}}(\Omega)\\
&=[\myzeta^Nq^my^{\ell}]\mathcal{E}_{\textrm{D}}[\mathcal{Z}^{\mathrm{GCE}}_{\textrm{RR}}(\Omega)]\,,
\end{aligned}
\end{equation}
where $\mathcal{C}_{\textrm{D}}$ is the standard asymptotic contour inside the DMVV chamber: \begin{equation}\label{eq:dmvv-asymptotic}
0<\left|\mathrm{Im}z\right|
\ll
\mathrm{Im}\tau
\ll
\mathrm{Im}\sigma\,,
\end{equation}
and the corresponding expansion $\mathcal{E}_{\textrm{D}}$ is w.r.t.
\begin{equation}\label{eq:dmvv-contourexpansion}
\begin{cases}
0<|\myzeta|\ll |q|\ll |y|<1<|y|^{-1}\,,
& \ell<0\,,
\\
0<|\myzeta|\ll |q|\ll |y|^{-1}<1<|y|\,,
& \ell>0\,.
\end{cases}
\end{equation}
Below, we often first consider 
\begin{equation}\label{eq:FundamentalDomainDef}
-2N< \ell < 0
\end{equation}
and obtain the other sectors by spectral flow. 


As we will show now, it will be convenient to define a \textit{shifted charge} $Q'$, and express the degeneracies in terms of $Q'$, as follows. 
The main reason is that the series expansion of the numerator of $\mathcal{Z}^{\mathrm{GCE}}_{\textrm{RR}}$ does not start from 1, but instead:
\begin{equation}
-\myzeta\,\eta(\tau)^{18}\theta_1(\tau,z)^2=\myzeta q\left(y^{-1}-2+y\right)\prod_{n=1}^{\infty}(1-q^n)^{20}(1-q^ny)^2(1-q^ny^{-1})^2\,.    
\end{equation}
\begin{itemize}
\item For $\ell<0$, we want $|y|<1$, therefore the numerator starts with $
\myzeta qy^{-1}$.   
\item 
For $\ell>0$, we want $|y|>1$, therefore the numerator starts with 
$\myzeta qy$.
\end{itemize}
Therefore it will be more convenient to consider the rescaled $\tilde{\mathcal{Z}}^{\mathrm{GCE}}_{\textrm{RR}}(\Omega)$, defined as
\begin{equation}\label{eq:ZtildeDef}
\tilde{\mathcal{Z}}^{\mathrm{GCE}}_{\textrm{RR}}(\Omega):=
\begin{cases}
\begin{aligned}
\frac{1}{\myzeta qy} \mathcal{Z}^{\mathrm{GCE}}_{\textrm{RR}}(\Omega)&\qquad |y| > 1 \\
\frac{1}{\myzeta q}\mathcal{Z}^{\mathrm{GCE}}_{\textrm{RR}}(\Omega) &\qquad |y|=1\\
\frac{1}{\myzeta qy^{-1}}\mathcal{Z}^{\mathrm{GCE}}_{\textrm{RR}}(\Omega) &\qquad |y|<1
\end{aligned}    
\end{cases}
\end{equation}

Accordingly, the DMVV degeneracy can be rewritten in terms of the shifted charge $Q'=(N',m',\ell')$:
\begin{equation}\label{eq:dQContourIntegralQprime}
\begin{aligned}
d_{\textrm{D}}(Q)&=
\oint_{\mathcal C_{\textrm{D}}}
d\Omega \,  e^{-2\pi i \textrm{Tr}[Q' \Omega]}
\,
\tilde{\mathcal{Z}}^{\mathrm{GCE}}_{\textrm{RR}}(\Omega)\\
&=[\myzeta^{N'}q^{m'}y^{\ell'}]\mathcal{E}_{\textrm{D}}[\tilde{\mathcal{Z}}^{\mathrm{GCE}}_{\textrm{RR}}(\Omega)]
\end{aligned}
\end{equation}
where 
the shifted charge is defined as
\begin{equation}
Q':=\begin{pmatrix}
N'&\ell'/2\\
\ell'/2&m'
\end{pmatrix}  \,,
\end{equation}
with
\begin{equation}
\begin{aligned}
N':=N-1 \,,
\qquad 
m':=m-1 \,,
\qquad
\ell':=
\begin{cases}
\begin{aligned}
\ell-1 &\qquad \ell \ge 1 \\
0 &\qquad \ell=0\\
\ell+1 &\qquad \ell \le -1    
\end{aligned}    
\end{cases}
\end{aligned}   
\end{equation}
Note that the degeneracy $d(Q)$ is still a function of $Q$; the shifted charge $Q'$ is convenient for extracting the Fourier coefficient w.r.t.\  the rescaled $\tilde{\mathcal{Z}}^{\mathrm{GCE}}_{\textrm{RR}}$.

\subsection{MCE Cauchy's theorem and terminal chambers}

Similar to the canonical expansion, we now apply Cauchy's theorem for the new contour integral form \eqref{eq:dQContourIntegralQprime}.
For any given $X$, consider the first $t$ segments $\Gamma_i\in \textrm{Branch}(X)$, and the triangular chambers $\Delta_i$ defined by  $\Gamma_i$ via \eqref{eq:GammatoDelta}: 
\begin{equation}\label{eq:FiniteBrancht}
\Delta_0 \xrightarrow{\Gamma_1} \Delta_1
\cdots \Delta_{t-2}
\xrightarrow{\Gamma_{t-1}} 
\Delta_{t-1} \xrightarrow{\Gamma_{t}}  \Delta_{t}\,.
\end{equation}
Applying Cauchy's theorem iteratively gives
\begin{equation}\label{eq:dDQexpansion}
d_{\textrm{D}}(Q)=d_{\Delta_t}(Q)+\sum^t_{i=1} \textrm{Res}_{\Gamma_i}\,,
\end{equation} 
where the contour integral is defined by
\begin{equation}\label{eq:dGammaIntegralDef}
\begin{aligned}
d_{\Delta_{\Gamma}}(Q)&:= \oint_{\mathcal C_{\Gamma}}
d\Omega_{\Gamma} \,  e^{-2\pi i \textrm{Tr}[Q' \Omega_{\Gamma}]}\tilde{\mathcal{Z}}^{\mathrm{GCE}}_{\textrm{RR}}(\Omega_{\Gamma})\,,
\end{aligned}
\end{equation}
with $\Delta_{\Gamma}$ denoting the (triangular) chamber whose top line is $\Gamma$. 
The residue on the (codimension-1) pole wall $\Gamma$ is
\begin{equation}\label{eq:ResGammaDef}
\mathrm{Res}_{\Gamma}(Q):=\oint_{\mathcal T_{\Gamma}}d\Omega_{\Gamma}\,e^{-2\pi i\,\mathrm{Tr}[Q'\Omega_{\Gamma}]}\tilde{\mathcal{Z}}^{\mathrm{GCE}}_{\textrm{RR}}(\Omega_{\Gamma})\,.
\end{equation}
The residue from the pole $\Gamma_i$ is related to the contour integrals of the two neighboring chambers that are separated by the pole lines corresponding to $\Gamma_i$ via
\begin{equation}\label{eq:CauchyMCE}
\mathrm{Res}_{\Gamma_i}(Q)=d_{\Delta_{i-1}}(Q)-d_{\Delta_i}(Q)\,.   
\end{equation}
We emphasize that both the contour integral \eqref{eq:dGammaIntegralDef} and the residue \eqref{eq:ResGammaDef} are defined for $Q$ but will be directly computed in terms of the shifted charge $Q'$.

Given a polar state with $Q$, i.e.\ one satisfying $\Delta(Q)<0$, we would like to determine a contour $\mathcal{C}_{Q}$ (specified by the value of $X$) such that $d_{\textrm{D}}(Q)$ can be computed as
\begin{equation}\label{eq:dDMVVQDecomposition}
d_{\textrm{D}}(Q)=\sum^t_{i=1} \textrm{Res}_{\Gamma_i}(Q)
\end{equation}
for some value of $t$, where $\Gamma_i\in \textrm{Branch}(X)$ for some value of $X$, and $i=1,\dots,t$.
We denote such a value as $X_{\mathcal{C}}(Q)$.

However, specifying $X_{\mathcal{C}}(Q)$ is not enough, as it only gives the direction along which to deform the  contour, i.e.\ decreasing $Y$ at fixed $X=X_{\mathcal{C}}(Q)$, without specifying when to stop. 
We also need to specify the 
\textit{terminal chamber} $\Delta_t$, which satisfies\footnote{Note that it is enough to specify the terminal chamber since the set of pole walls crossed only depends on the terminal chamber $\Delta_t$.}
\begin{equation}\label{eq:dterminalcondition}
d_{\Delta_{t-1}}(Q) \neq 0 
\qquad \textrm{and} \qquad
d_{\Delta_i}(Q)=0  \textrm{ for all } i\geq t   \,.
\end{equation}
This implies
\begin{equation}\label{eq:Resterminalcondition1}
\mathrm{Res}_{\Gamma_t}(Q)\neq 0   
\end{equation}
and
\begin{equation}\label{eq:ResterminalityCondition2}
\mathrm{Res}_{\Gamma_i}(Q)=0  
\,,\qquad 
i\geq t+1 \ \text{along Branch}(X_{\mathcal{C}}(Q)) \,.
\end{equation} 
Once we have determined the terminal chamber, and denoting the pole line $\Gamma_t$ on top of $\Delta_t$ by $\Gamma_t=[\frac{\alpha}{\gamma},\frac{\beta}{\delta}]$, then we can simply choose its mediant as $X_{\mathcal{C}}(Q)$:
\begin{equation}\label{eq:XCQDef}
\boxed{
X_{\mathcal{C}}(Q)
=\frac{\alpha+\beta}{\gamma+\delta} \,. 
}
\end{equation}

\bigskip

Let us now spell out the terminal conditions \eqref{eq:dterminalcondition} more explicitly, in terms of the charge $Q$.
First, let us derive the condition for a triple contour integral $d_{\Delta_{\Gamma}}(Q)$ to vanish.
Since we know under what condition the triple contour integral in the DMVV chamber vanishes, we can use a suitable SL$(2,\mathbb{Z})$ transformation to map the chamber $\Delta_{\Gamma}$ to the DMVV chamber.\footnote{This strategy is similar to the one used by \cite{Sen:2011mh} in the context of dyon counting and by \cite{Belin:2018oza} to compute polar degeneracies in symmetric orbifolds.}

Recall that when the chamber is the DMVV chamber, or rather, the fundamental domain given by
\begin{equation}
\Delta_{0}=\left[\frac{0}{1}\,, \frac{1}{1}\,, \frac{1}{0}\right]
\end{equation}
the contour integral surrounding $\myzeta=q=y=0$, satisfying \eqref{eq:dmvv-asymptotic}, gives $d_{\textrm{D}}(Q)$.
For the integral \eqref{eq:dGammaIntegralDef} in the chamber $\Gamma$, $\Omega_{\Gamma}$ denotes the period matrix in chamber $\Delta_{\Gamma}$, where $\Gamma=[\frac{\alpha}{\gamma}\,,\, \frac{\beta}{\delta}]$, and it is related to the periodic matrix $\Omega$ in the (fundamental domain of the)  DMVV chamber via
\begin{equation}\label{eq:period-basis-transform}
\Omega\longmapsto \Omega_{\Gamma}= M_{\Gamma}\Omega M_{\Gamma}^T
\,.
\end{equation}
Here the matrix
\begin{equation}\label{eq:MGammaDef}
M_{\Gamma}=
\begin{pmatrix}
\beta&\alpha\\
\delta&\gamma
\end{pmatrix} \in \mathrm{SL}(2,\mathbb{Z}) \,,
\end{equation}
and it is embedded in the Sp$(4,\mathbb{Z})$ via \begin{equation}\label{eq:sp4-action}
\Omega \longmapsto (A\Omega+B)(C\Omega+D)^{-1}
\qquad
\textrm{for}
\quad
\begin{pmatrix}
A & B\\
C & D
\end{pmatrix}=\begin{pmatrix}
M_{\Gamma} & 0\\
0 & M_{\Gamma}^{-T}
\end{pmatrix}
\in
\mathrm{Sp}(4,\mathbb{Z})\,.
\end{equation}

Now in order to find the condition for $d_{\Delta_{\Gamma}}(Q)$ to vanish, we use the SL$(2,\mathbb{Z})$ transformation \eqref{eq:MGammaDef} to pull $\Omega_{\Gamma}$ back to the DMVV chamber:
\begin{equation}\label{eq:dDeltaQPullback}
\begin{aligned}
d_{\Delta_{\Gamma}}(Q)&
=\oint_{\mathcal{C}_0} d\Omega\, e^{-2 \pi i \textrm{Tr}[Q'_{\Gamma} \Omega]} \tilde{\mathcal{Z}}^{\mathrm{GCE}}_{\textrm{RR}}(M_{\Gamma} \Omega M_{\Gamma}^T)\,,\\
&=\left[\zeta^{N'_{\Gamma}}q^{m'_{\Gamma}}y^{\ell'_{\Gamma}}\right]\mathcal{E}_{\mathrm{D}}\!\left[\tilde{\mathcal{Z}}^{\mathrm{GCE}}_{\mathrm{RR}}\!\left(M_{\Gamma}\Omega M_{\Gamma}^{T}\right)\right]\,,
\end{aligned}
\end{equation}
where the shifted charge $Q'_{\Gamma}$ is related to the original shifted charge $Q'$ via
\begin{equation}
Q'_{\Gamma}:=M_{\Gamma}^{T}Q'M_{\Gamma}\,,
\end{equation}
and explicitly
\begin{equation}\label{eq:NmLGamma}
\begin{split}
N'_\Gamma&:=\beta^2N'+\beta\delta \ell'+\delta^2m'\,,\\
m'_\Gamma&:=\alpha^2N'+\alpha\gamma \ell'+\gamma^2m'\,,\\
\ell'_\Gamma&:=2\alpha\beta N'+(\beta\gamma+\alpha\delta)\ell'
    +2\gamma\delta m'\,.
\end{split}
\end{equation}
And it will be convenient to define
\begin{equation}
\hat{N}_{\Gamma}=N'_{\Gamma}+1
\qquad \textrm{and} \qquad 
\hat{m}_{\Gamma}=m'_{\Gamma}+1 \,. 
\end{equation}

In terms of the pullback expression \eqref{eq:dDeltaQPullback} for $d_{\Delta_{\Gamma}}(Q)$, and using the definition of $\tilde{\mathcal{Z}}^{\mathrm{GCE}}_{\textrm{RR}}$ in \eqref{eq:ZtildeDef}, one can immediately obtain the condition for the $d_{\Delta_{\Gamma}}(Q)$ to vanish: it is when the transformed shifted charge $Q'_{\Gamma}$ falls out of the charge support of the series expansion of $\tilde{\mathcal{Z}}^{\mathrm{GCE}}_{\textrm{RR}}(M_{\Gamma} \Omega M_{\Gamma}^T)$:
\begin{equation}\label{eq:dDelta0}
\left.
\begin{aligned} 
\hat{N}_{\Gamma}&<0\\ 
\text{or}\qquad \hat{m}_{\Gamma}&<0\\ \text{or}\qquad \hat{N}_{\Gamma}+\hat{m}_{\Gamma}+\ell'_{\Gamma}&<1 \end{aligned}
\right\} 
\quad \Longrightarrow \quad 
d_{\Delta_{\Gamma}}(Q)=0\,. 
\end{equation}
Equivalently,
\begin{equation}\label{eq:dDeltaneq0}
d_{\Delta_{\Gamma}}(Q) \neq 0 \quad \Longrightarrow \quad 
\hat{N}_{\Gamma}\geq 0 \,,
\quad 
\hat{m}_{\Gamma}\geq 0 \,,
\quad 
\hat{N}_{\Gamma}+\hat{m}_{\Gamma}+\ell'_{\Gamma} \geq 1\,.
\end{equation} 
The terminal condition \eqref{eq:dterminalcondition} can then be translated into \eqref{eq:dDelta0} for $\Gamma=\Gamma_{i\geq t}$ together with \eqref{eq:dDeltaneq0} for $\Gamma=\Gamma_{t-1}$.

\subsection{Solving for terminal chambers}
\label{ssec:QuantumAlgorithm}

Before we solve these terminal conditions, let us simplify them further.  
The formulae \eqref{eq:NmLGamma} suggest that we can define one function to express both $\hat{N}_{\Gamma}$ and $\hat{m}_{\Gamma}$.
We define the \textit{height function}
\begin{equation}
H_{Q'}[p,q]:=     N' \, p^2+\ell' \, pq +m'q^2  
\end{equation}
In terms of $H_{Q'}$, 
\begin{equation}
\begin{aligned}
\hat{N}_{\Gamma}&=H_{Q'}[\beta,\delta]+1\,,\\ 
\hat{m}_{\Gamma}&=H_{Q'}[\alpha,\gamma]+1\,,\\ 
\hat{N}_{\Gamma}+\hat{m}_{\Gamma}+\ell'_{\Gamma}-1&=H_{Q'}[\alpha+\beta,\gamma+\delta]+1\,. \end{aligned} 
\end{equation}

The first part of the terminal condition \eqref{eq:dterminalcondition}, i.e.\ 
\eqref{eq:dDeltaneq0} for $\Gamma=\Gamma_{t-1}$ is 
\begin{equation}\label{eq:dDeltaneq0i}
d_{\Delta_{\Gamma_{t-1}}}(Q) \neq 0 \quad \Longrightarrow \quad 
\hat{N}_{\Gamma_{t-1}}\geq 0 \,,
\quad 
\hat{m}_{\Gamma_{t-1}}\geq 0 \,,
\quad 
\hat{N}_{\Gamma_{t-1}}+\hat{m}_{\Gamma_{t-1}}+\ell'_{\Gamma_{t-1}} \geq 1\,.
\end{equation}
In order to translate these inequalities into those in terms of $\Gamma_{i}$, we 
use the following identities:
\begin{itemize}
\item If $\Gamma_i$ is the left child of $\Gamma_{i-1}$, then 
\begin{equation}\label{eq:Nmhatleft}
\begin{aligned} \hat{N}_{\Gamma_{i}}&=\hat{N}_{\Gamma_{i-1}}+\hat{m}_{\Gamma_{i-1}}+\ell'_{\Gamma_{i-1}}-1\,,\qquad \hat{m}_{\Gamma_{i}}=\hat{m}_{\Gamma_{i-1}}\,. 
\end{aligned}
\end{equation}
\item If $\Gamma_i$ is the right child of $\Gamma_{i-1}$, then 
\begin{equation}\label{eq:Nmhatright}
  \begin{aligned} \hat{N}_{\Gamma_{i}}&=\hat{N}_{\Gamma_{i-1}}\,,\qquad \hat{m}_{\Gamma_{i}}=\hat{N}_{\Gamma_{i-1}}+\hat{m}_{\Gamma_{i-1}}+\ell'_{\Gamma_{i-1}}-1 \,. 
  \end{aligned} 
\end{equation}
\end{itemize}
Either way, \eqref{eq:dDeltaneq0i} implies 
\begin{equation}\label{eq:dDeltaneq0inew}
\hat{N}_{\Gamma_{t}}\geq 0 \,,
\quad 
\hat{m}_{\Gamma_{t}}\geq 0 \,.
\end{equation}

On the other hand, \eqref{eq:dDelta0} for $\Gamma=\Gamma_t$ means that if any of the three inequalities holds, then  $d_{\Delta_{\Gamma_{t}}}(Q)= 0$.
Since the first two are already shown to fail by \eqref{eq:dDeltaneq0inew}, we should just impose the last inequality, namely
\begin{equation}
\hat{N}_{\Gamma}+\hat{m}_{\Gamma}+\ell'_{\Gamma} < 1 \,,
\end{equation}
for $\Gamma=\Gamma_{t}$.
In total, the collective conditions to 
define the terminal chamber are
\begin{equation}
\hat{N}_{\Gamma}\geq 0 \,,
\quad 
\hat{m}_{\Gamma}\geq 0 \,,
\quad 
\hat{N}_{\Gamma}+\hat{m}_{\Gamma}+\ell'_{\Gamma} < 1  \,,
\end{equation}
for $\Gamma=\Gamma_{t}$.

In terms of $H_{Q'}$, the conditions on the terminal chamber \eqref{eq:dterminalcondition}
lead us to impose the conditions on the terminal $\Gamma_t=[\frac{\alpha}{\gamma}\,,\, \frac{\beta}{\delta}]$ (linear pole-wall on top of the terminal chamber):
\begin{equation}\label{eq:terminality}
\boxed{
\begin{aligned}
H_{Q'}[\alpha, \gamma]+1 &\geq 0 
\, \qquad\\
H_{Q'}[\beta, \delta]+1 &\geq 0
\, \qquad\\
H_{Q'}[\alpha+\beta, \gamma+\delta]+1&<0   \,.   
\end{aligned}
}
\end{equation}

\bigskip

Now the task is as follows.
For a given polar charge $Q$, we would like to find an $X_*(Q)$, such that as we descend,  following Farey intervals containing $X_*(Q)$, we will eventually reach a terminal chamber whose top wall $\Gamma_t=[\frac{\alpha}{\gamma}\,,\, \frac{\beta}{\delta}]$ satisfies the set of inequalities \eqref{eq:terminality}.
Note that this $X_*(Q)$ is not to be confused with $X_{\mathcal{C}}(Q)$, which was defined in \eqref{eq:XCQDef}.

To find such an $X_*(Q)$, first rewrite the last inequality in \eqref{eq:terminality} into
\begin{equation}\label{eq:3rdInequality}
 \left|\frac{\alpha+\beta}{\gamma+\delta}+\frac{\ell'}{2N'}\right|
 <\sqrt{\frac{-\Delta(Q')}{4N'^2}-\frac{1}{N'(\gamma+\delta)^{2}}}\,,
\end{equation} 
where
\begin{equation}
\Delta(Q'):=4N'm'-\ell'^2  \,.
\end{equation}
Note that for polar charge with $N>1$, 
\begin{equation}
\Delta(Q')  <0\,,
\end{equation}
and one can show that what is inside the square root in \eqref{eq:3rdInequality} is positive for $N>1$.

On the other hand, since both $X_*(Q)$ and $\frac{\alpha+\beta}{\gamma+\delta}$ are enclosed by $\Gamma_t=[\frac{\alpha}{\gamma}\,,\, \frac{\beta}{\delta}]$, their distance is bounded by
\begin{equation}
\left|\frac{\alpha+\beta}{\gamma+\delta}-X_*(Q)\right|\leq\frac{\beta}{\delta}-\frac{\alpha}{\gamma}=\frac{1}{\gamma\delta}\,.
\end{equation}
As we descend along $X_*(Q)$, since $\gamma \delta$ increases, this bound approaches zero. 
Therefore,  choosing 
\begin{equation}\label{eq:XstarQDef}
\boxed{
X_*(Q):=-\frac{\ell'}{2N'}   \,,
}
\end{equation}
can ensure the l.h.s. of \eqref{eq:3rdInequality} approaches zero as we descend and \eqref{eq:3rdInequality} can be satisfied.

\bigskip

Now that we have determined $X_*(Q)$, we apply the following iterative procedure to determine the terminal chamber $\Gamma_t=[\frac{\alpha}{\gamma}\,,\, \frac{\beta}{\delta}]$:
\begin{enumerate}
\item Given a polar state $Q$ within the fundamental domain \eqref{eq:FundamentalDomainDef}, consider $[0,1]$ as the initial chamber. (Other polar charges are related by spectral flows). 
This is the step 1 of the iterative process. 
\item Given the chamber $\Gamma_i=[\frac{\alpha_i}{\gamma_i},\frac{\beta_i}{\delta_i}]$ for step $i$, if $H_{Q'}[\alpha_i+\beta_i, \gamma_i+\delta_i]+1<0$, then the iteration terminates, and $\Gamma_t=\Gamma_i$.
\item Otherwise, compute the mediant $\mu_i=\frac{\alpha_i+\beta_i}{\gamma_i+\delta_i}$.
And the chamber $\Gamma_{i+1}$ for step $i+1$ is determined as follows:\footnote{Note that $\mu_i=X_*(Q)$ cannot occur at this step: in that case the iteration has already terminated at step 2.}
\begin{equation}\label{app:FareyconditionQ}
\begin{cases}
\Gamma_{i+1}=\Gamma_{\mathrm{R}}=[\mu_i,\frac{\beta_i}{\delta_i}],
& \qquad \mu_i< X_*(Q),
\\[2pt]
\Gamma_{i+1}=\Gamma_{\mathrm{L}}=[\frac{\alpha_i}{\gamma_i},\mu_i],
& \qquad \mu_i> X_*(Q),
\end{cases}
\end{equation}
\item Go to step 2 again, until the iteration terminates. (This iteration is bound to terminate due to the argument above that leads to the choice \eqref{eq:XstarQDef}.)
\end{enumerate}
This gives the unique solution for the terminal chamber $\Gamma_t=[\frac{\alpha}{\gamma}\,,\, \frac{\beta}{\delta}]$ that satisfies the set of inequalities \eqref{eq:terminality} together with the constraint\footnote{Note that the set of inequalities \eqref{eq:terminality} alone does not guarantee a unique solution for all polar states. } 
\begin{equation}\label{eq:Xconstraint}
\frac{\alpha}{\gamma}\leq X_*(Q)=-\frac{\ell'}{2N'} \leq \frac{\beta}{\delta}\,,
\end{equation}
 which then uniquely specifies the set of orbifolds that contribute to $d_{\textrm{D}}(Q)$:
\begin{equation}\label{eq:dDMVVQDecomposition1}
d_{\textrm{D}}(Q)=\sum^t_{i=1} \textrm{Res}_{\Gamma_i}(Q)\,,
\end{equation}
for all polar states. 

As an illustration, Table~\ref{tab:N8-polar-states} lists all the polar states within the fundamental domain \eqref{eq:FundamentalDomainDef} at $N=8$, together with their terminal chambers, the orbifolds that contribute, and the decomposition \eqref{eq:dDMVVQDecomposition1} of their degeneracies.

\begin{table}[H]
\centering
\small
\renewcommand{\arraystretch}{1.15}
\begin{tabular*}{\linewidth}{@{}c@{\extracolsep{\fill}}ccl@{}}
\hline
$(\ell,m)$ & $\Gamma_t$ & $(K,S)$ orbifolds & $d_{\textrm{D}}(Q)=\sum^t_{i=1}\textrm{Res}_{\Gamma_i}(Q)$ \\
\hline
$(-1,0)$ & $[0,\tfrac{1}{2}]$ & $(1,-1)$, $(2,-1)$ & $6223062=6123696+99366$ \\
$(-2,0)$ & $[0,\tfrac{1}{2}]$ & $(1,-1)$, $(2,-1)$ & $1940970=1939362+1608$ \\
$(-3,0)$ & $[0,\tfrac{1}{2}]$ & $(1,-1)$, $(2,-1)$ & $467730=467724+6$ \\
$(-4,0)$ & $[0,1]$ & $(1,-1)$ & $92118$ \\
$(-5,0)$ & $[0,1]$ & $(1,-1)$ & $14856$ \\
$(-6,0)$ & $[0,1]$ & $(1,-1)$ & $1893$ \\
$(-6,1)$ & $[0,1]$ & $(1,-1)$ & $226800$ \\
$(-7,0)$ & $[0,1]$ & $(1,-1)$ & $174$ \\
$(-7,1)$ & $[0,1]$ & $(1,-1)$ & $31500$ \\
\hline
$(-8,0)$ & $[0,1]$ & $(1,-1)$ & $9$ \\
$(-8,1)$ & $[0,1]$ & $(1,-1)$ & $3120$ \\
\hline
$(-9,2)$ & $[0,1]$ & $(1,-1)$ & $31500$ \\
$(-9,1)$ & $[0,1]$ & $(1,-1)$ & $174$ \\
$(-10,3)$ & $[0,1]$ & $(1,-1)$ & $226800$ \\
$(-10,2)$ & $[0,1]$ & $(1,-1)$ & $1893$ \\
$(-11,3)$ & $[0,1]$ & $(1,-1)$ & $14856$ \\
$(-12,4)$ & $[0,1]$ & $(1,-1)$ & $92118$ \\
$(-13,5)$ & $[\tfrac{1}{2},1]$ & $(1,-1)$, $(2,-3)$ & $467730=467724+6$ \\
$(-14,6)$ & $[\tfrac{1}{2},1]$ & $(1,-1)$, $(2,-3)$ & $1940970=1939362+1608$ \\
$(-15,7)$ & $[\tfrac{1}{2},1]$ & $(1,-1)$, $(2,-3)$ & $6223062=6123696+99366$ \\
\hline
\end{tabular*}
\caption{The polar states within the fundamental domain \eqref{eq:FundamentalDomainDef} at $N=8$, labeled by $(\ell,m)$.
The residues are listed in the order of the poles $\Gamma_1,\dots,\Gamma_t$.
Note the reflection symmetry $(\ell,m)\mapsto(-\ell-2N,\,m+\ell+N)$, which  leaves the two states with $\ell=-N$ in the middle invariant.}
\label{tab:N8-polar-states}
\end{table}

\section{Phase diagram of microcanonical charge lattice}
\label{sec:Phase}

In this section, we will show that the microcanonical charge lattice can be divided into phases, where charges within the same region share the same terminal chamber, and thus their microcanonical degeneracies $d_{\textrm{D}}(Q)$ can be decomposed in terms of the set of orbifolds in \eqref{eq:dDMVVQDecomposition1}.

\subsection{Classical tessellation of  unshifted charge lattice}

We will first show that the set of the terminality conditions \eqref{eq:terminality} can be viewed as a quantum deformation of the tangent lines of the polar parabola of the charge lattice. 
These tangent lines give a triangular tessellation of the charge lattice, and by checking which triangular chamber a given charge $Q$ sits in, one can read off all the orbifolds that \textit{potentially} could contribute to the degeneracy $d_{\textrm{D}}(Q)$. 

\subsubsection{Ramond sector}

Consider the charge lattice in the Ramond sector for fixed $N$.
The polar states, namely those satisfying
$\Delta(Q)=4Nm-\ell^2<0$,    
sit below the 
polar parabola:
\begin{equation}\label{eq:MCEparabolaR}
\textrm{m}=\frac{\textrm{l}^2 }{4N}  \,, 
\end{equation}
where we are using the font $(\textrm{l},\textrm{m})$ to denote the coordinates whose integer points are $(\ell, m)$, for a fixed $N$.
As we now show, we can use those tangents of this parabola with fractional slopes to tessellate the charge lattice. 
Let's first consider the range $-2N\leq \ell\leq 0$.

An arbitrary tangent line to the parabola $\textrm{m}=\frac{\textrm{l}^2 }{4N}$ with slope $-s$, with $s\in \mathbb{R}_{\geq 0}$, is 
\begin{equation}\label{eq:tangentR}
T_{s}(\textrm{l})= \textrm{m} =-s \textrm{l}-Ns^2\,,
\end{equation}
which touches the parabola at the point $(\textrm{l}_0,\textrm{m}_0)=(-2Ns,Ns^2)$.

Now let's consider the case when the slope is a rational number $s=\frac{p}{q}$. 
We observe that such a tangent line $T_{p/q}(\textrm{l})$ is given by the zero locus of the quadratic 
\begin{equation}
\textrm{H}_{(N,\textrm{m},\textrm{l})}[p,q]:= N\, p^2+\textrm{l} \, pq +\textrm{m} \, q^2  
 =q^2(\textrm{m}-T_{p/q}(\textrm{l})) \,.
\end{equation}
Note that $\textrm{m},\textrm{l},\textrm{H}\in \mathbb{R}$.
This real quadratic specializes to the one defined on the integer lattice: 
\begin{equation}\label{eq:HQDefR}
H_{Q}[p,q]:= N\, p^2+\ell \, pq +m \, q^2   =q^2(m-T_{p/q}(\ell))\,,
\end{equation}
where charge $Q=(N,m,\ell)$.
This gives 
\begin{equation}
\begin{aligned}
&H_{Q}[p,q]>0 \qquad \Longleftrightarrow \qquad Q \textrm{ lies above } T_{p/q}  \\
&H_{Q}[p,q]=0 \qquad \Longleftrightarrow \qquad Q \textrm{ lies on } T_{p/q}  \\
&H_{Q}[p,q]<0 \qquad \Longleftrightarrow \qquad Q \textrm{ lies below } T_{p/q}  \,.
\end{aligned}
\end{equation}

\begin{figure}[H]
\centering
\definecolor{RRBlue}{HTML}{155DA8}
\definecolor{RROrange}{HTML}{C14514}
\definecolor{RRGreen}{HTML}{3F832D}
\definecolor{RRRose}{HTML}{B5386A}
\definecolor{RRGold}{HTML}{A56D07}
\definecolor{RRIndigo}{HTML}{514FA1}
\begin{tikzpicture}[paper figure,x=0.6875cm,y=0.6875cm]
  \definecolor{PaperBlue}{HTML}{356B8C}

  \coordinate (d) at ({-8/3},0);
  \coordinate (e) at ({-20/3},{4/3});
  \coordinate (f) at ({-28/3},{8/3});
  \coordinate (g) at ({-40/3},{16/3});

  \fill[PaperPole!20] (-8,0)--(-4,0)--(-12,4)--cycle;
  \fill[PaperBlue!47] (-4,0)--(d)--(e)--cycle
                      (-12,4)--(f)--(g)--cycle;

  \draw[paper axis] (-16.56,0)--(0.8,0) node[right] {$\textrm{l}$};
  \draw[paper axis] (0,0)--(0,8.56) node[above] {$\textrm{m}$};
  \draw[paper continuation] (d)--(0,0);
  \draw[PaperPole,semithick] (-8,0)--(d) (-12,4)--(-4,0);
  \draw[PaperPole,thick] (-16,8)--(-8,0);
  \draw[PaperBlue,thick] (d)--(e) (f)--(g);
  \draw[paper boundary] plot[domain=-16:0,samples=120] (\x,{\x*\x/32});

  \foreach \x/\y/\ink in {
    -11/3/black,-10/2/black,-9/1/black,-9/2/black,-8/0/black,
    -8/1/black,-7/0/black,-7/1/black,-6/0/black,-5/0/black,
    -6/1/RRBlue,-4/0/RRBlue,-12/4/RROrange,-10/3/RROrange}
    \fill[\ink] (\x,\y) circle[radius=1.95pt];

  \foreach \x/\y/\ink/\turn in {
    -3/0/RRBlue/0,-2/0/RRGreen/0,-1/0/RRGold/0,
    -13/5/RROrange/180,-14/6/RRRose/180,-15/7/RRIndigo/180}
    \fill[\ink,shift={(\x,\y)},rotate=\turn]
      (0,3.5pt)--(-3.1pt,-1.8pt)--(3.1pt,-1.8pt)--cycle;

  \foreach \p/\q/\ink in {1/3/PaperBlue,1/2/PaperPole,2/3/PaperBlue}
    \draw[\ink,fill=white,semithick] ({-16*\p/\q},{8*(\p/\q)^2})
      circle[radius=1.8pt]
      node[above right,text=\ink,yshift=11pt] {$-\tfrac{\p}{\q}$};
  \draw[PaperPole,fill=white,semithick] (-16,8) circle[radius=1.8pt]
    node[above right,text=PaperPole] {$-1$};
  \draw[PaperPole,fill=white,semithick] (0,0) circle[radius=1.8pt]
    node[above left,text=PaperPole] {$0$};

  \foreach \x/\tick in {-16/-2N,-12/-\tfrac{3N}{2},-8/-N,-4/-\tfrac{N}{2},0/0}
    \node[below] at (\x,0) {$\tick$};
  \foreach \y/\tick in {4/\tfrac{N}{2},8/N}
    \node[right] at (0,\y) {$\tick$};
\end{tikzpicture}
\caption[Classical RR chambers at $N=8$]{Tessellation of charge lattice in the RR frame at $N=8$. We show the tangent lines with slope $-1$, $0$ (the horizontal axis), $-\frac{1}{2}$, $-\frac{1}{3}$, and $-\frac{2}{3}$.
Charges in different chambers are distinguished by their \textit{colors}, with the corresponding terminal chambers being $[0,1]$,
\textcolor{RRBlue}{$[0,\tfrac{1}{2}]$}, \textcolor{RROrange}{$[\tfrac{1}{2},1]$},
\textcolor{RRGreen}{$[0,\tfrac{1}{4}]$}, \textcolor{RRRose}{$[\tfrac{3}{4},1]$},
\textcolor{RRGold}{$[0,\tfrac{1}{8}]$}, and \textcolor{RRIndigo}{$[\tfrac{7}{8},1]$}.
Different shapes correspond to their different quantum chambers.}
\label{fig:N8-RR-classical}
\end{figure}
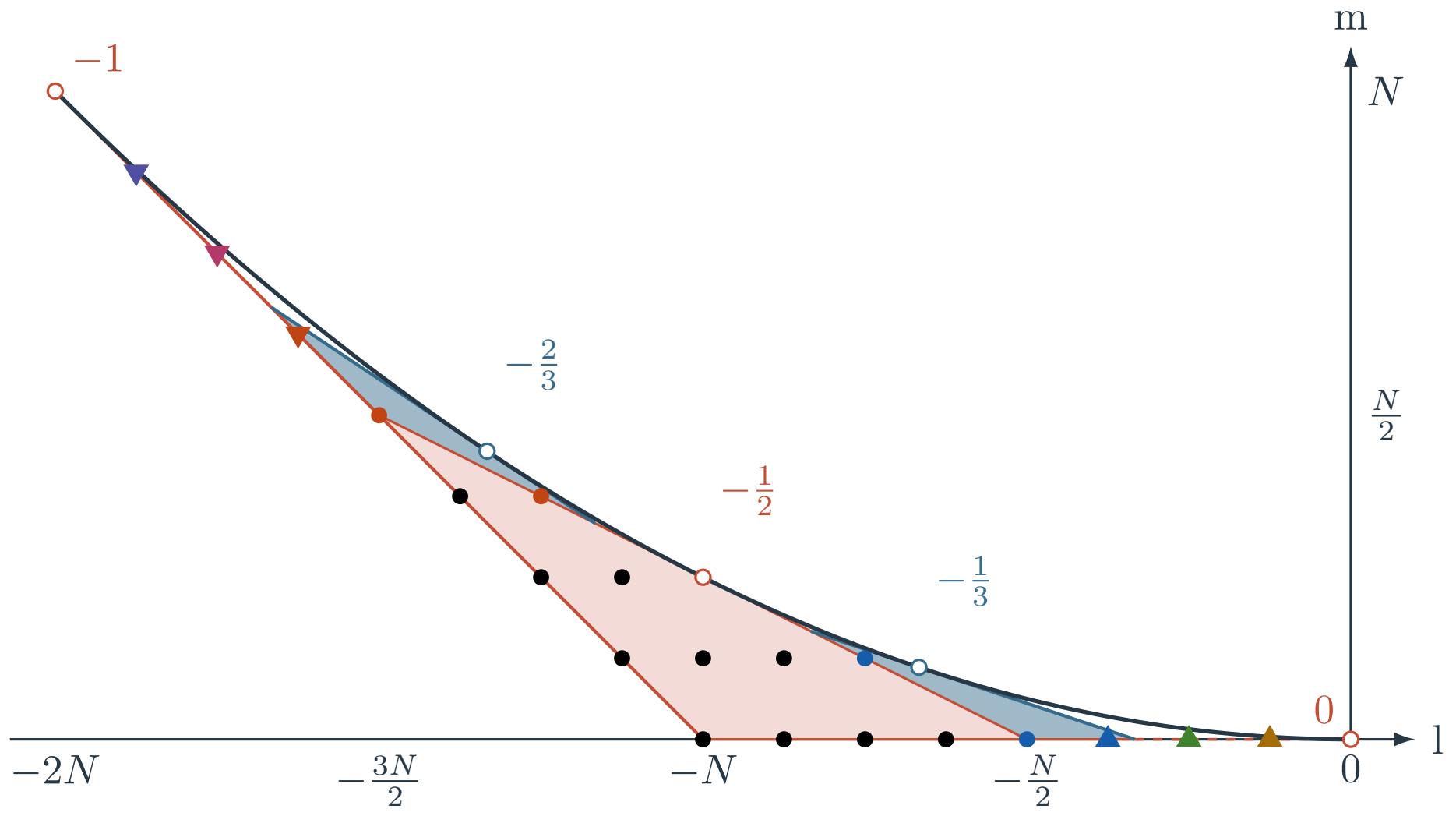

Now in the charge lattice, start with the region bounded by the polar parabola and the tangents $T_{0}$ and $T_{1}$; we will refer to this region as $[0,1]$, see Figure~\ref{fig:N8-RR-classical}.
The tangent $T_{\frac{1}{2}}$ cuts out a triangle from this region, leaving two smaller regions: the lower one bounded by the tangents $T_{0}$ and $T_{\frac{1}{2}}$ together with the parabola, and the upper one bounded by the tangents $T_{\frac{1}{2}}$ and $T_{1}$ together with the parabola. 
Therefore, it is natural to label them by the two child Farey pairs of $[0,1]$:  $[0,\tfrac{1}{2}]$
and $[\tfrac{1}{2},1]$.

Repeating this operation, with the mediant $\frac{\alpha+\beta}{\gamma+\delta}$ generating the two 
child Farey pairs for each parent $[\frac{\alpha}{\gamma},\frac{\beta}{\delta}]$, we again obtain a Stern-Brocot tree \eqref{eq:FareyTreeFourLevels}.
Since the fractions generated this way are dense in $[0,1]$, these triangles fill the original region below the parabola.
Furthermore, there is no overlap between different triangles. 
Therefore, we have obtained a tessellation (called classical tessellation) given by the triangles bounded by
\begin{equation}\label{eq:TessellationRboundary}
H_{Q}[\alpha, \gamma]=0
\,, \qquad
H_{Q}[\beta, \delta]=0
\,, \qquad
H_{Q}[\alpha+\beta, \gamma+\delta]=0 \,,
\end{equation}
with $[\frac{\alpha}{\gamma},\frac{\beta}{\delta}]$ all Farey pairs.
The triangle associated with $[\frac{\alpha}{\gamma},\frac{\beta}{\delta}]$ is the region
\begin{equation}\label{eq:TessellationR}
H_{Q}[\alpha, \gamma]\geq 0
\,, \qquad
H_{Q}[\beta, \delta]\geq 0
\,, \qquad
H_{Q}[\alpha+\beta, \gamma+\delta]< 0\,.
\end{equation}
For example, for the leading triangle: $[\frac{0}{1},\frac{1}{2},\frac{1}{1}]$, the three lines are
\begin{equation}
\begin{aligned}
\frac{0}{1}:& \qquad \textrm{m}=0 
\, \qquad\\
\frac{1}{2}:& \qquad \textrm{m}=-\frac{1}{2}\textrm{l}-\frac{1}{4}N
\, \qquad\\
\frac{1}{1}:& \qquad \textrm{m}=-\textrm{l}-N \,.
\end{aligned}
\end{equation}
See Figure~\ref{fig:N8-RR-classical} for the case of $N=8$.

This set of inequalities looks very similar to the set of terminality condition \eqref{eq:terminality}.
We will call the chambers defined by these $H_{Q}$ the \textit{classical chambers}, and those defined by $H_{Q'}+1$ \textit{quantum chambers}.

\noindent
Now we explain how to determine which classical terminal chamber a charge $Q$ sits in, which is useful since the triangles can get rather small at deeper levels.

\begin{enumerate}
\item Given a polar state $Q$ within the fundamental domain \eqref{eq:FundamentalDomainDef}, define\footnote{Note that this value of $X$ also played an important role in dyon-counting in \cite{Cheng:2007ch,Cheng:2008fc,LopesCardoso:2020pmp}.} 
\begin{equation}
x_*(Q):=-\frac{\ell}{2N}\,,
\end{equation} 
and consider $[0,1]$ as the initial chamber. (Other polar charges are related by spectral flows). 
This is the step 1 of the iterative process. 
\item Given the chamber $\Gamma_i=[\frac{\alpha_i}{\gamma_i},\frac{\beta_i}{\delta_i}]$ for step $i$, 
if $H_{Q}[\alpha_i+\beta_i, \gamma_i+\delta_i]< 0$, then the iteration terminates. 
\item Otherwise, compute the mediant $\mu_i=\frac{\alpha_i+\beta_i}{\gamma_i+\delta_i}$. 
And the chamber $\Gamma_{i+1}$ for step $i+1$ is determined as follows:
\begin{equation}\label{app:FareyconditionAppC}
\begin{cases}
\Gamma_{i+1}=\Gamma_{\mathrm{R}}=[\mu_i,\frac{\beta_i}{\delta_i}],
& \qquad \mu_i< x_*(Q)\,,
\\[2pt]
\Gamma_{i+1}=\Gamma_{\mathrm{L}}=[\frac{\alpha_i}{\gamma_i},\mu_i],
& \qquad \mu_i> x_*(Q)\,.
\end{cases}
\end{equation}
\item Go back to step 2, until the iteration terminates. (Note that this procedure will always terminate because $H_{Q}[x_*(Q),1]=\frac{\Delta(Q)}{4N}<0$ and the successive Farey pairs shrink in size and hug $x_*(Q)$ tighter and tighter.) 
\end{enumerate}
This gives the unique solution for the set of inequalities \eqref{eq:TessellationR}.
In particular the terminal chamber encloses $x_*(Q)$:
\begin{equation}\label{eq:ClassicalEnclosure}
  \frac{\alpha}{\gamma}< x_*(Q)=-\frac{\ell}{2N}< \frac{\beta}{\delta}\,.
\end{equation}

We show the charges in Figure~\ref{fig:N8-RR-classical} below, together with the terminal classical chambers and the branches that they belong to.  
\begin{table}[H]
\centering
\scriptsize
\setlength{\tabcolsep}{2pt}
\renewcommand{\arraystretch}{1.08}

\definecolor{RRBlue}{HTML}{155DA8}
\definecolor{RROrange}{HTML}{C14514}
\definecolor{RRGreen}{HTML}{3F832D}
\definecolor{RRRose}{HTML}{B5386A}
\definecolor{RRGold}{HTML}{A56D07}
\definecolor{RRIndigo}{HTML}{514FA1}

\begin{tabular*}{\linewidth}{@{}p{0.27\linewidth}@{\extracolsep{\fill}}p{0.36\linewidth}p{0.34\linewidth}@{}}
\hline
RR sector charges $(\ell,m)$ & Classical chamber and branch & $(K,S)$ orbifolds \\
\hline
\raggedright $\bullet$ $0\leq m\leq 3$,
  $-8-m\leq\ell\leq-5-2m$
  & \raggedright $[0,1]$
  & \raggedright $(1,-1)$ \tabularnewline
\hline
\raggedright \textcolor{RRBlue}{$\bullet$} $(-6,1)$, $(-4,0)$\newline
  \textcolor{RRBlue}{$\blacktriangle$} $(-3,0)$
  & \raggedright $[0,1]$ $\to\textcolor{RRBlue}{[0,\tfrac{1}{2}]}$
  & \raggedright $(1,-1)$, $(2,-1)$ \tabularnewline
\hline
\raggedright \textcolor{RROrange}{$\bullet$} $(-12,4)$, $(-10,3)$\newline
  \textcolor{RROrange}{$\blacktriangledown$} $(-13,5)$
  & \raggedright $[0,1]$ $\to\textcolor{RROrange}{[\tfrac{1}{2},1]}$
  & \raggedright $(1,-1)$, $(2,-3)$ \tabularnewline
\hline
\raggedright \textcolor{RRGreen}{$\blacktriangle$} $(-2,0)$
  & \raggedright $[0,1]$ $\to[0,\tfrac{1}{2}]$ $\to[0,\tfrac{1}{3}]$
    $\to\textcolor{RRGreen}{[0,\tfrac{1}{4}]}$
  & \raggedright $(1,-1)$, $(2,-1)$, $(3,-1)$, $(4,-1)$ \tabularnewline
\hline
\raggedright \textcolor{RRRose}{$\blacktriangledown$} $(-14,6)$
  & \raggedright $[0,1]$ $\to[\tfrac{1}{2},1]$ $\to[\tfrac{2}{3},1]$
    $\to\textcolor{RRRose}{[\tfrac{3}{4},1]}$
  & \raggedright $(1,-1)$, $(2,-3)$, $(3,-5)$, $(4,-7)$ \tabularnewline
\hline
\raggedright \textcolor{RRGold}{$\blacktriangle$} $(-1,0)$
  & \raggedright $[0,1]$ $\to[0,\tfrac{1}{2}]$ $\to[0,\tfrac{1}{3}]$
    $\to[0,\tfrac{1}{4}]$ $\to[0,\tfrac{1}{5}]$ $\to[0,\tfrac{1}{6}]$
    $\to[0,\tfrac{1}{7}]$ $\to\textcolor{RRGold}{[0,\tfrac{1}{8}]}$
  & \raggedright $(1,-1)$, $(2,-1)$, $(3,-1)$, $(4,-1)$,
    $(5,-1)$, $(6,-1)$, $(7,-1)$, $(8,-1)$ \tabularnewline
\hline
\raggedright \textcolor{RRIndigo}{$\blacktriangledown$} $(-15,7)$
  & \raggedright $[0,1]$ $\to[\tfrac{1}{2},1]$ $\to[\tfrac{2}{3},1]$
    $\to[\tfrac{3}{4},1]$ $\to[\tfrac{4}{5},1]$ $\to[\tfrac{5}{6},1]$
    $\to[\tfrac{6}{7},1]$ $\to\textcolor{RRIndigo}{[\tfrac{7}{8},1]}$
  & \raggedright $(1,-1)$, $(2,-3)$, $(3,-5)$, $(4,-7)$,
    $(5,-9)$, $(6,-11)$, $(7,-13)$, $(8,-15)$ \tabularnewline
\hline
\end{tabular*}
\caption{RR sector charges at $N=8$ in Figure~\ref{fig:N8-RR-classical}.
}
\label{tab:N8-RR-classical}
\end{table}
\noindent
The last column gives the orbifolds that correspond to the Farey pairs in the branch, using the dictionary \eqref{eq:FPtoKMS}.
Although this looks very reminiscient of the orbifolds that contribute to the residues associated to each pole lines, we emphasize that they do not all contribute, as we will show in the next subsection.
Namely, if we evaluate the residues on the corresponding poles, some of them at the end of the branch might be zero.

It is important to note that the constraint \eqref{eq:Xconstraint}, which we reprint here 
\begin{equation}\label{eq:Xconstraintnew}
\frac{\alpha}{\gamma}\leq X_*(Q) =-\frac{\ell'}{2N'} \leq \frac{\beta}{\delta}
\end{equation}
 is also satisfied for the classical terminal chamber $[\frac{\alpha}{\gamma},\frac{\beta}{\delta}]$, as we demonstrate now. 
Start from the classical enclosure \eqref{eq:ClassicalEnclosure}.
First, the distance between the classical $x_*(Q)$ and the quantum $X_*(Q)$ is bounded by 
\begin{equation}\label{eq:Shift1}
  \bigl|X_*(Q)-x_*(Q)\bigr|=\frac{|N+\ell|}{2N(N-1)}\leq\frac{1}{2N}\,,
\end{equation}
using \eqref{eq:FundamentalDomainDef}.
On the other hand, let $x$ be either $\frac{\alpha}{\gamma}$ or $\frac{\beta}{\delta}$ of the classical chamber. 
Define 
\begin{equation}\label{eq:HQxDef}
H_{Q}(x):= H_{Q}[x,1]    
\end{equation}
By \eqref{eq:TessellationR}, we have $H_{Q}(x)\geq 0$. 
Since $-\Delta(Q)>0$ for polar states, completing the square gives
\begin{equation}\label{eq:Shift2}
\bigl(x-x_*(Q)\bigr)^{2}=\frac{4N H_{Q}(x)-\Delta(Q)}{4N^{2}}\geq\frac{1}{4N^{2}}\,.
\end{equation}
\eqref{eq:Shift1} and \eqref{eq:Shift2} together mean that the shift from $x_*(Q)$ to $X_*(Q)$
cannot move it out of the Farey pair $[\frac{\alpha}{\gamma},\frac{\beta}{\delta}]$, and we thus have \eqref{eq:Xconstraintnew}.

\subsubsection{Neveu-Schwarz sector}

Since the comparison with gravity is more transparent in the NS frame,  
we now translate the tessellation into the NS frame. 

Applying the map from the Ramond to the NS frame, see \eqref{eq:RtoNS}, to $m=L_{0}^{\mathrm{RR}}-\frac{N}{4}$ and 
$\ell=2J_{0}^{\mathrm{RR}}$, we have
\begin{equation}
  J_{0}^{\mathrm{NS}}=\frac{\ell+N}{2}\,,\qquad L_{0}^{\mathrm{NS}}=m+\frac{\ell+N}{2}\,.
\end{equation}
As in the RR frame, we use the text font to denote the coordinates 
\begin{equation}
\textrm{J}_{0}^{\mathrm{NS}}=\frac{\textrm{l}+N}{2}\,,\qquad \textrm{L}_{0}^{\mathrm{NS}}=\textrm{m}+\frac{\textrm{l}+N}{2}\,,
\end{equation}
whose (half) integer points are the charges.

The polar parabola in the RR frame \eqref{eq:MCEparabolaR} maps to another parabola in the NS frame:
\begin{equation}
\textrm{L}_{0}^{\mathrm{NS}}=\frac{(\textrm{J}_{0}^{\mathrm{NS}})^{2}}{N}+\frac{N}{4}\,.
\end{equation}  
The tangent line \eqref{eq:tangentR} in the RR frame (with slope $-s$) is mapped to the tangent line in the NS frame
\begin{equation}
T_{s}^{\mathrm{NS}}(\textrm{J}_{0}^{\mathrm{NS}})=\textrm{L}_{0}^{\mathrm{NS}}=(1-2s)\textrm{J}_{0}^{\mathrm{NS}}+Ns(1-s)\,,
\end{equation}
with slope
\begin{equation}
1+2(-s)\,.   
\end{equation}
It touches the parabola at the point $\left(N\left(\frac{1}{2}-s\right),\,N\left(s^{2}-s+\frac{1}{2}\right)\right)$.

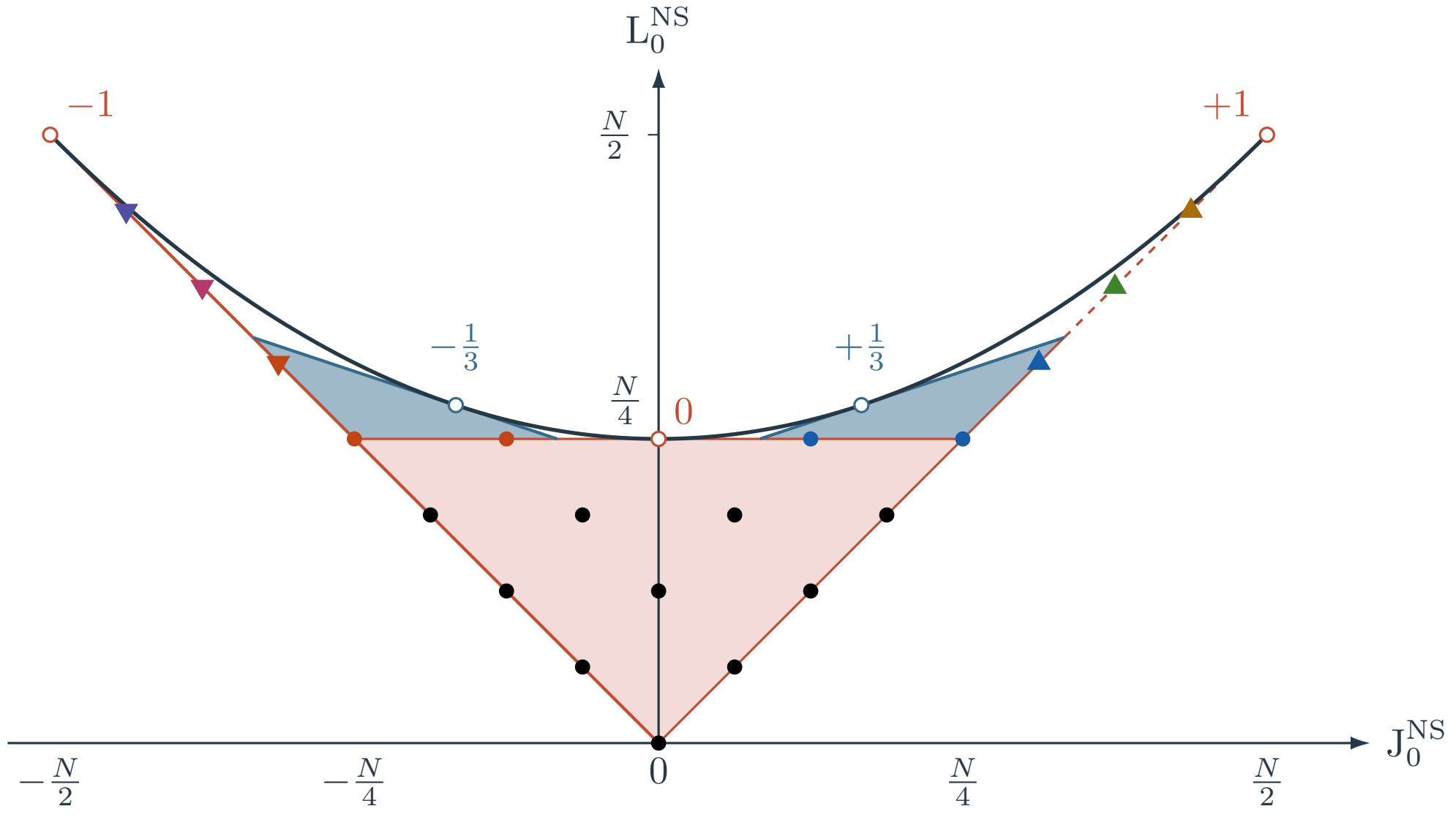
\begin{figure}[H]
\centering
\begin{tikzpicture}[paper figure,x=1.375cm,y=1.375cm]
  \definecolor{PaperBlue}{HTML}{356B8C}
  \definecolor{RRBlue}{HTML}{155DA8}
  \definecolor{RROrange}{HTML}{C14514}
  \definecolor{RRGreen}{HTML}{3F832D}
  \definecolor{RRRose}{HTML}{B5386A}
  \definecolor{RRGold}{HTML}{A56D07}
  \definecolor{RRIndigo}{HTML}{514FA1}

  \fill[PaperPole!20] (0,0)--(2,2)--(-2,2)--cycle;
  \foreach \side in {-1,1}
    \fill[PaperBlue!47] ({2*\side},2)--({2*\side/3},2)
      --({8*\side/3},{8/3})--cycle;

  \draw[paper axis] (-4.28,0)--(4.68,0)
    node[right] {$\textrm{J}_{0}^{\mathrm{NS}}$};
  \draw[paper axis] (0,0)--(0,4.44)
    node[above] {$\textrm{L}_{0}^{\mathrm{NS}}$};
  \draw[paper continuation] ({8/3},{8/3})--(4,4);
  \draw[PaperPole,semithick] (0,0)--({8/3},{8/3}) (-2,2)--(2,2);
  \draw[PaperPole,thick] (-4,4)--(0,0);
  \foreach \side in {-1,1}
    \draw[PaperBlue,thick] ({2*\side/3},2)--({8*\side/3},{8/3});
  \draw[paper boundary] plot[domain=-4:4,samples=120] (\x,{2+\x*\x/8});

  \foreach \ellRR/\mRR/\ink in {
    -11/3/black,-10/2/black,-9/1/black,-9/2/black,-8/0/black,
    -8/1/black,-7/0/black,-7/1/black,-6/0/black,-5/0/black,
    -6/1/RRBlue,-4/0/RRBlue,-12/4/RROrange,-10/3/RROrange}
    \fill[\ink] ({(\ellRR+8)/2},{\mRR+(\ellRR+8)/2})
      circle[radius=1.95pt];
  \foreach \ellRR/\mRR/\ink/\turn in {
    -3/0/RRBlue/0,-2/0/RRGreen/0,-1/0/RRGold/0,
    -13/5/RROrange/180,-14/6/RRRose/180,-15/7/RRIndigo/180}
    \fill[\ink,shift={({(\ellRR+8)/2},{\mRR+(\ellRR+8)/2})},rotate=\turn]
      (0,3.5pt)--(-3.1pt,-1.8pt)--(3.1pt,-1.8pt)--cycle;

  \draw[PaperPole,fill=white,semithick] (-4,4) circle[radius=1.8pt]
    node[above right,text=PaperPole] {$-1$};
  \draw[PaperPole,fill=white,semithick] (4,4) circle[radius=1.8pt]
    node[above left,text=PaperPole] {$+1$};
  \foreach \side/\sign in {-1/-,1/+}
    \draw[PaperBlue,fill=white,semithick] ({4*\side/3},{20/9})
      circle[radius=1.8pt] node[above,text=PaperBlue,yshift=5pt]
      {$\sign\tfrac{1}{3}$};
  \draw[PaperPole,fill=white,semithick] (0,2) circle[radius=1.8pt]
    node[above right,text=PaperPole] {$0$};

  \foreach \x/\tick in {-4/-\tfrac{N}{2},-2/-\tfrac{N}{4},0/0,
                         2/\tfrac{N}{4},4/\tfrac{N}{2}}
    \node[below] at (\x,0) {$\tick$};
  \node[above left] at (0,2) {$\tfrac{N}{4}$};
  \draw[PaperInk,thin] (0,4)--(-0.07,4) node[left] {$\tfrac{N}{2}$};
\end{tikzpicture}
\caption{Tessellation of charge lattice in the NS frame at $N=8$. We show the tangent lines with slopes $\pm 1$, $0$, and  $\pm\frac{1}{3}$.
Colors and shapes are as in Figure~\ref{fig:N8-RR-classical}.}
\label{fig:N8-NS-classical}
\end{figure}

For rational $s=\frac{p}{q}$, the tangent line $T_{s}^{\mathrm{NS}}(\textrm{J}_{0}^{\mathrm{NS}})$ is given by the zero locus of the quadratic
\begin{equation}
\textrm{H}^{\textrm{NS}}_{(N,\textrm{m},\textrm{l})}(p,q)=q^{2}\textrm{L}_{0}^{\mathrm{NS}}-q(q-2p)\textrm{J}_{0}^{\mathrm{NS}}-Np(q-p)\,.
\end{equation}
The quadratic defined on the charges \eqref{eq:HQDefR} is mapped to 
\begin{equation}\label{eq:HQNS}
\begin{aligned} 
H^{\textrm{NS}}_{Q}(p,q)&=q^{2}L_{0}^{\mathrm{NS}}-q(q-2p)J_{0}^{\mathrm{NS}}-Np(q-p)\\ &=q^{2}\left(L_{0}^{\mathrm{NS}}-T_{\frac{p}{q}}^{\mathrm{NS}}(J_{0}^{\mathrm{NS}})\right)\,. \end{aligned} 
\end{equation}
Then the tessenlation in the NS frame is given by \eqref{eq:TessellationRboundary} and \eqref{eq:TessellationR}, with $H_Q$ replaced by $H^{\textrm{NS}}_Q$ in \eqref{eq:HQNS}. 
For example, the leading triangle is 
\begin{equation}
\begin{aligned} \frac{0}{1}:&\qquad \textrm{L}_{0}^{\mathrm{NS}}=\textrm{J}_{0}^{\mathrm{NS}}\,,\\ \frac{1}{2}:&\qquad \textrm{L}_{0}^{\mathrm{NS}}=\frac{N}{4}\,,\\ \frac{1}{1}:&\qquad \textrm{L}_{0}^{\mathrm{NS}}=-\textrm{J}_{0}^{\mathrm{NS}}\,. \end{aligned} 
\end{equation}
Figure~\ref{fig:N8-NS-classical} gives the tessellation for $N=8$ in the NS frame.

\subsection{Quasi-tessellation of shifted charge lattice}

Recall that the condition for the terminal chamber is 
\begin{equation}\label{eq:terminality6}
H_{Q'}[\alpha, \gamma]+1 \geq 0
\, \qquad
H_{Q'}[\beta, \delta]+1 \geq 0
\, \qquad
H_{Q'}[\alpha+\beta, \gamma+\delta]+1<0 \,,
\end{equation}
together with constraint \eqref{eq:Xconstraint}. 
Now we show that these lines can be viewed as a deformation of the boundaries of the triangular chambers \eqref{eq:TessellationR} in the (classical) tessellation by the tangent lines. 
The resulting chamber structure will be called the quantum chambers.
In principle, one can examine which (quantum) chamber a charge $Q$ sits in to read off the list of orbifolds that contribute to $d_{\textrm{D}}(Q)$ -- and it would be a precise list in the sense that every orbifold in the list contributes non-trivially, unlike the list one reads off from the classical chamber structure.\footnote{ 
The only caveat is that since the lines given by $H_{Q'}+1=0$ are not tangents, they no longer give a true tessellation, namely, there are cases in which a charge $Q$ can belong to two different quantum chambers.
However, as we will show below, the ambiguity can be easily resolved by also looking at which classical chamber the charge belongs to, and this is equivalent to imposing the constraint 
\eqref{eq:Xconstraint}.
}
Let us first discuss this in the Ramond frame.

\subsubsection{Ramond sector}
For the shifted MCE charge $Q':=(N', m', \ell')$, the shifted polar states are those with 
\begin{equation}
\Delta(Q'):=4N'm'-\ell'^2<0    \,,
\end{equation}
and they sit below the shifted polar parabola:
\begin{equation}
\textrm{m}'=\frac{\textrm{l}'^2 }{4N'}   \,.
\end{equation}
Note that for $N\geq 2$, a polar state is necessarily a shifted polar state:
\begin{equation}
\Delta(Q')<0    \,.
\end{equation}

One could then use the  boundary
\begin{equation}
H_{Q'}(p,q)+1= N'\, p^2+\ell' \, pq +m' \, q^2 +1 =0
\end{equation}
to divide the charge lattice. 
Note that $H_{Q'}(p,q)+1$
no longer gives tangent lines, but instead \textit{shifted tangents}:
\begin{equation}
m'=-\frac pq\ell'-N'\frac{p^2}{q^2}-\frac1{q^2}    \,;
\end{equation}
they are shifted downwards by 
$\frac{1}{q^2}$ in the shifted charge lattice $(\ell',m')$.
(When later drawn in the original charge lattice $(\ell,m)$, they are shifted upwards relative to the classical tangent lines given by $H_Q[p,q]$ for $q>1$.)
One could use these shifted tangents to define triangles as \eqref{eq:terminality6}, however, these shifted tangents do not give a true tessellation since two different triangles can overlap in the 2D space, and some charges $Q$  belong to more than one triangle.
To resolve this ambiguity, we simply impose the constraint \eqref{eq:Xconstraint}, namely, we choose the quantum chamber with $[\frac{\alpha}{\gamma},\frac{\beta}{\delta}]$ that contains $X_*(Q)$.

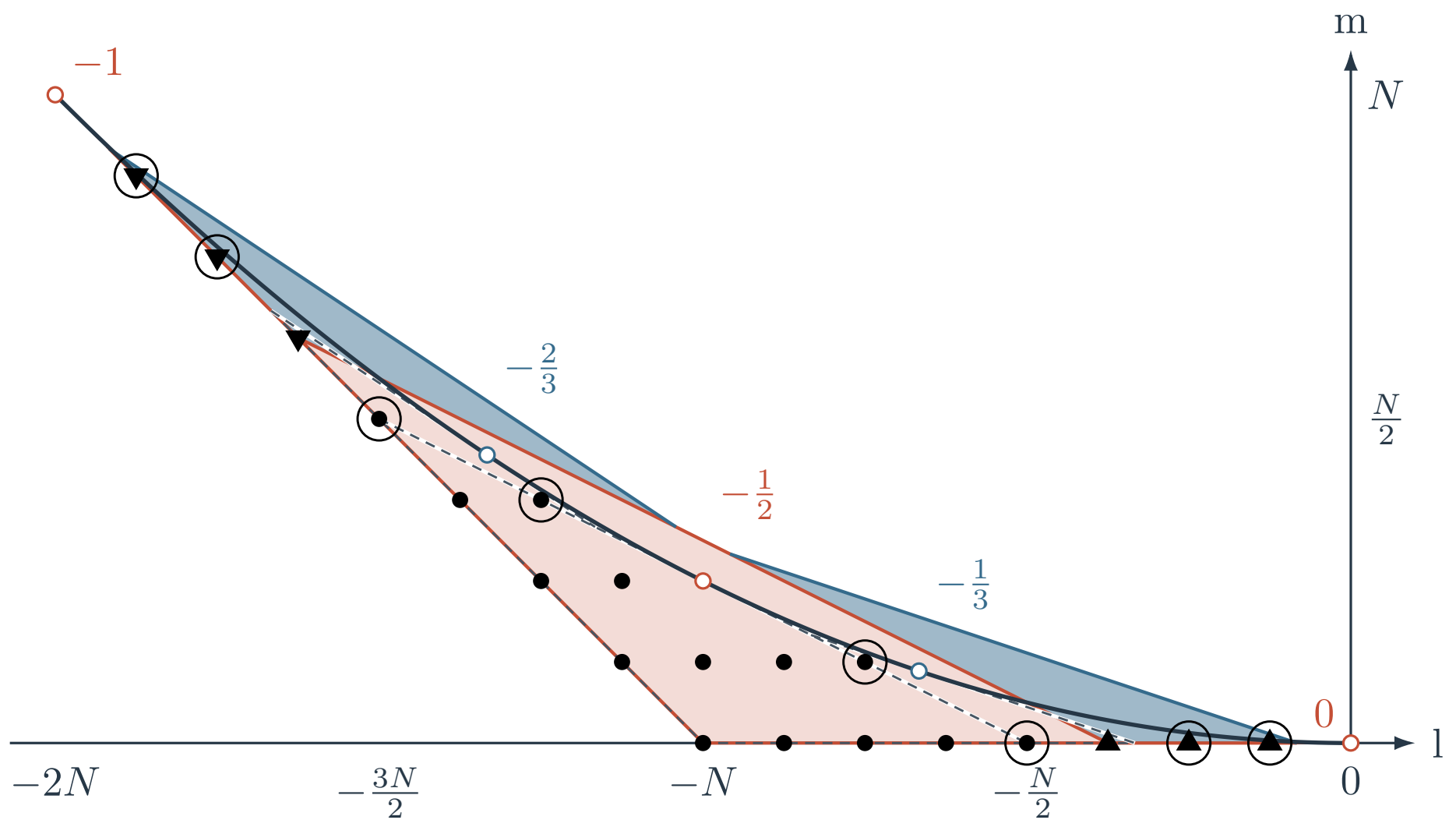
\begin{figure}[H]
\centering
\begin{tikzpicture}[paper figure,x=0.6875cm,y=0.6875cm]
  \definecolor{PaperBlue}{HTML}{356B8C}

  \coordinate (d) at ({-8/3},0);
  \coordinate (e) at ({-20/3},{4/3});
  \coordinate (f) at ({-28/3},{8/3});
  \coordinate (g) at ({-40/3},{16/3});

  \coordinate (D) at ({-2/3},0);
  \coordinate (E) at ({-23/3},{7/3});
  \coordinate (F) at ({-25/3},{8/3});
  \coordinate (G) at ({-46/3},{22/3});

  \fill[PaperPole!20] (-8,0)--(-3,0)--(-13,5)--cycle;
  \fill[PaperBlue!47] (-3,0)--(D)--(E)--cycle
                      (-13,5)--(F)--(G)--cycle;

  \draw[paper axis] (-16.56,0)--(0.8,0) node[right] {$\textrm{l}$};
  \draw[paper axis] (0,0)--(0,8.56) node[above] {$\textrm{m}$};

  \draw[PaperPole,thick] (G)--(-8,0)--(D) (-13,5)--(-3,0);
  \draw[PaperBlue,thick] (G)--(F) (E)--(D);

  \tikzset{classical/.style={
    draw=PaperInk!85,line width=0.55pt,
    dash pattern=on 2.4pt off 1.8pt}}
  \draw[classical] (g)--(-8,0)--(d);
  \draw[classical,preaction={draw=white,solid,line width=0.95pt}]
    (-12,4)--(-4,0) (g)--(f) (e)--(d);

  \draw[paper continuation] (-16,8)--(G) (D)--(0,0);
  \draw[paper boundary] plot[domain=-16:0,samples=120]
    (\x,{\x*\x/32});

  \foreach \x/\y in {-12/4,-10/3,-6/1,-4/0,-2/0,-1/0,-15/7,-14/6}
    \draw[black,line width=0.55pt] (\x,\y) circle[radius=5.2pt];

  \foreach \x/\y in {
    -11/3,-10/2,-9/1,-9/2,-8/0,-8/1,-7/0,-7/1,-6/0,-5/0,
    -6/1,-4/0,-12/4,-10/3}
    \fill[black] (\x,\y) circle[radius=1.95pt];

  \foreach \x/\y/\turn in {
    -3/0/0,-2/0/0,-1/0/0,-13/5/180,-14/6/180,-15/7/180}
    \fill[black,shift={(\x,\y)},rotate=\turn]
      (0,3.5pt)--(-3.1pt,-1.8pt)--(3.1pt,-1.8pt)--cycle;

  \foreach \p/\q/\ink in {1/3/PaperBlue,1/2/PaperPole,2/3/PaperBlue}
    \draw[\ink,fill=white,semithick] ({-16*\p/\q},{8*(\p/\q)^2})
      circle[radius=1.8pt]
      node[above right,text=\ink,yshift=11pt] {$-\tfrac{\p}{\q}$};
  \draw[PaperPole,fill=white,semithick] (-16,8) circle[radius=1.8pt]
    node[above right,text=PaperPole] {$-1$};
  \draw[PaperPole,fill=white,semithick] (0,0) circle[radius=1.8pt]
    node[above left,text=PaperPole] {$0$};

  \foreach \x/\tick in {
    -16/-2N,-12/-\tfrac{3N}{2},-8/-N,-4/-\tfrac{N}{2},0/0}
    \node[below,yshift=-2pt] at (\x,0) {$\tick$};
  \foreach \y/\tick in {4/\tfrac{N}{2},8/N}
    \node[right] at (0,\y) {$\tick$};
\end{tikzpicture}
\caption{The RR sector charge lattice at $N=8$, with classical (gray dashed) and
quantum (solid colored) chamber boundaries. Different shapes correspond to different terminal chambers:
$\bullet\,[0,1]$, $\blacktriangle\,[0,\tfrac{1}{2}]$, and
$\blacktriangledown\,[\tfrac{1}{2},1]$. Black rings mark the eight charges
whose quantum chambers differ from their classical chambers.}
\label{fig:n8-rr-chamber-changes}
\end{figure}

Similar to the procedure that determined the classical chamber,  we apply the iteration explained in Section~\ref{ssec:QuantumAlgorithm} to determine the chamber.
Equivalently, since we have already shown that the constraint \eqref{eq:Xconstraint} is satisfied by the classical terminal chamber and moreover $H_{Q'}[p,q]+1\leq H_Q[p,q]$, this means that the quantum chamber can be selected by starting at the classical terminal chamber of $Q$ and moving towards the root $[0,1]$, stopping at the first $[\frac{\alpha}{\gamma},\frac{\beta}{\delta}]$ that satisfies
\eqref{eq:terminality6}.
In Figure~\ref{fig:n8-rr-chamber-changes} we show the quantum chamber structure for $N=8$.

\noindent 
In this quantum phase diagram, charges in different quantum chambers are denoted with different shapes.
Compare this with the classical chamber structure \ref{fig:N8-RR-classical}, where charges in different classical chambers are denoted with different colors, but with shapes that correspond to their quantum chamber.  
We will list below those charges whose quantum chambers differ from their classical chambers: 
\begin{table}[H]
\centering
\footnotesize
\setlength{\tabcolsep}{3pt}
\renewcommand{\arraystretch}{1.18}
\definecolor{RRBlue}{HTML}{155DA8}
\definecolor{RROrange}{HTML}{C14514}
\definecolor{RRGreen}{HTML}{3F832D}
\definecolor{RRRose}{HTML}{B5386A}
\definecolor{RRGold}{HTML}{A56D07}
\definecolor{RRIndigo}{HTML}{514FA1}
\def\strike#1{%
  \tikz[baseline=(pair.base)]{%
    \node[inner sep=0pt,outer sep=0pt] (pair) {$#1$};
    \draw[line width=0.45pt] (pair.north east)--(pair.south west);}}

\begin{tabular*}{\linewidth}{@{}l@{\extracolsep{\fill}}ccp{0.51\linewidth}@{}}
\hline
RR sector charges  & Classical & Quantum  & $(K,S)$ along the classical path \\
 $(\ell,m)$ & terminal chamber & terminal chamber &  \\
\hline
$\bullet\,(-12,4)$
  & \textcolor{RROrange}{$[\tfrac{1}{2},1]$} & $[0,1]$
  & $(1,-1)$, \strike{(2,-3)} \\
$\bullet\,(-10,3)$
  & \textcolor{RROrange}{$[\tfrac{1}{2},1]$} & $[0,1]$
  & $(1,-1)$, \strike{(2,-3)} \\
\hline
$\bullet\,(-6,1)$
  & \textcolor{RRBlue}{$[0,\tfrac{1}{2}]$} & $[0,1]$
  & $(1,-1)$, \strike{(2,-1)} \\
$\bullet\,(-4,0)$
  & \textcolor{RRBlue}{$[0,\tfrac{1}{2}]$} & $[0,1]$
  & $(1,-1)$, \strike{(2,-1)} \\
\hline
$\blacktriangle\,(-2,0)$
  & \textcolor{RRGreen}{$[0,\tfrac{1}{4}]$} & $[0,\tfrac{1}{2}]$
  & $(1,-1)$, $(2,-1)$, \strike{(3,-1)}, \strike{(4,-1)} \\
\hline
$\blacktriangle\,(-1,0)$
  & \textcolor{RRGold}{$[0,\tfrac{1}{8}]$} & $[0,\tfrac{1}{2}]$
  & $(1,-1)$, $(2,-1)$, \strike{(3,-1)}, \strike{(4,-1)},
    \strike{(5,-1)}, \strike{(6,-1)}, \strike{(7,-1)}, \strike{(8,-1)} \\
\hline
$\blacktriangledown\,(-15,7)$
  & \textcolor{RRIndigo}{$[\tfrac{7}{8},1]$} & $[\tfrac{1}{2},1]$
  & $(1,-1)$, $(2,-3)$, \strike{(3,-5)}, \strike{(4,-7)},
    \strike{(5,-9)}, \strike{(6,-11)}, \strike{(7,-13)}, \strike{(8,-15)} \\
\hline
$\blacktriangledown\,(-14,6)$
  & \textcolor{RRRose}{$[\tfrac{3}{4},1]$} & $[\tfrac{1}{2},1]$
  & $(1,-1)$, $(2,-3)$, \strike{(3,-5)}, \strike{(4,-7)} \\
\hline
\end{tabular*}
\caption{The eight charges whose terminal chamber changes at $N=8$.
The colors follow Figure~\ref{fig:N8-RR-classical} and the shapes follow
Figure~\ref{fig:n8-rr-chamber-changes}. Crossed-out poles lie beyond
the quantum terminal chamber along the classical path.}
\label{tab:n8-chamber-changes-KS}
\end{table}

\subsubsection{Neveu-Schwarz sector}

\begin{figure}[H]
\centering
\begin{tikzpicture}[paper figure,x=1.375cm,y=1.375cm]
  \definecolor{PaperBlue}{HTML}{356B8C}

  \coordinate (A) at (0,0);
  \foreach \side/\bv/\dv/\ev/\BV/\DV/\EV in
    {1/b/d/e/B/D/E,-1/c/g/f/C/G/F} {
    \coordinate (\bv) at ({2*\side},2);
    \coordinate (\dv) at ({8*\side/3},{8/3});
    \coordinate (\ev) at ({2*\side/3},2);
    \coordinate (\BV) at ({5*\side/2},{5/2});
    \coordinate (\DV) at ({11*\side/3},{11/3});
    \coordinate (\EV) at ({\side/6},{5/2});
  }

  \fill[PaperPole!20] (A)--(B)--(C)--cycle;
  \fill[PaperBlue!47] (B)--(D)--(E)--cycle (C)--(F)--(G)--cycle;

  \draw[paper axis] (-4.28,0)--(4.68,0)
    node[right] {$\textrm{J}_{0}^{\mathrm{NS}}$};
  \draw[paper axis] (0,0)--(0,4.44)
    node[above] {$\textrm{L}_{0}^{\mathrm{NS}}$};

  \draw[PaperPole,thick] (G)--(A)--(D) (C)--(B);
  \draw[PaperBlue,thick] (G)--(F) (E)--(D);

  \tikzset{classical/.style={draw=PaperInk!85,line width=0.55pt,
    dash pattern=on 2.4pt off 1.8pt}}
  \draw[classical] (g)--(A)--(d);
  \draw[classical,preaction={draw=white,solid,line width=0.95pt}]
    (c)--(b) (g)--(f) (e)--(d);

  \draw[paper continuation] (-4,4)--(G) (D)--(4,4);
  \draw[paper boundary] plot[domain=-4:4,samples=120] (\x,{2+\x*\x/8});

  \foreach \side in {-1,1}
    \foreach \j/\h in {1/2,2/2,3/3,3.5/3.5}
      \draw[black,line width=0.55pt] ({\side*\j},\h) circle[radius=5.2pt];

  \foreach \k in {0,...,3}
    \foreach \j in {0,...,\k}
      \fill[black] ({\j-\k/2},{\k/2}) circle[radius=1.95pt];
  \foreach \j in {-2,-1,1,2}
    \fill[black] (\j,2) circle[radius=1.95pt];
  \foreach \u in {2.5,3,3.5}
    \foreach \side/\turn in {1/0,-1/180}
      \fill[black,shift={({\side*\u},\u)},rotate=\turn]
        (0,3.5pt)--(-3.1pt,-1.8pt)--(3.1pt,-1.8pt)--cycle;

  \draw[PaperPole,fill=white,semithick] (-4,4) circle[radius=1.8pt]
    node[above right,text=PaperPole] {$-1$};
  \draw[PaperPole,fill=white,semithick] (4,4) circle[radius=1.8pt]
    node[above left,text=PaperPole] {$+1$};
  \foreach \side/\sign in {-1/-,1/+}
    \draw[PaperBlue,fill=white,semithick] ({4*\side/3},{20/9})
      circle[radius=1.8pt] node[above,text=PaperBlue,yshift=30pt]
      {$\sign\tfrac{1}{3}$};
  \draw[PaperPole,fill=white,semithick] (0,2) circle[radius=1.8pt]
    node[above right,text=PaperPole,inner sep=2pt] {$0$};

  \foreach \x/\tick in {-4/-\tfrac{N}{2},-2/-\tfrac{N}{4},0/0,
                          2/\tfrac{N}{4},4/\tfrac{N}{2}}
    \node[below,yshift=-2pt] at (\x,0) {$\tick$};
  \node[above left,inner sep=1.5pt] at (0,2) {$\tfrac{N}{4}$};
  \draw[PaperInk,thin] (0,4)--(-0.07,4) node[left] {$\tfrac{N}{2}$};
\end{tikzpicture}
\caption{NS-NS sector charge lattice at $N=8$, with classical (gray dashed) and
quantum (solid colored) chamber boundaries. Different shapes correspond to different terminal chambers:
$\bullet\,[0,1]$, $\blacktriangle\,[0,\tfrac{1}{2}]$, and
$\blacktriangledown\,[\tfrac{1}{2},1]$. Black rings mark the eight charges
whose quantum chambers differ from their classical chambers.}
\label{fig:n8-ns-chamber-changes}
\end{figure}
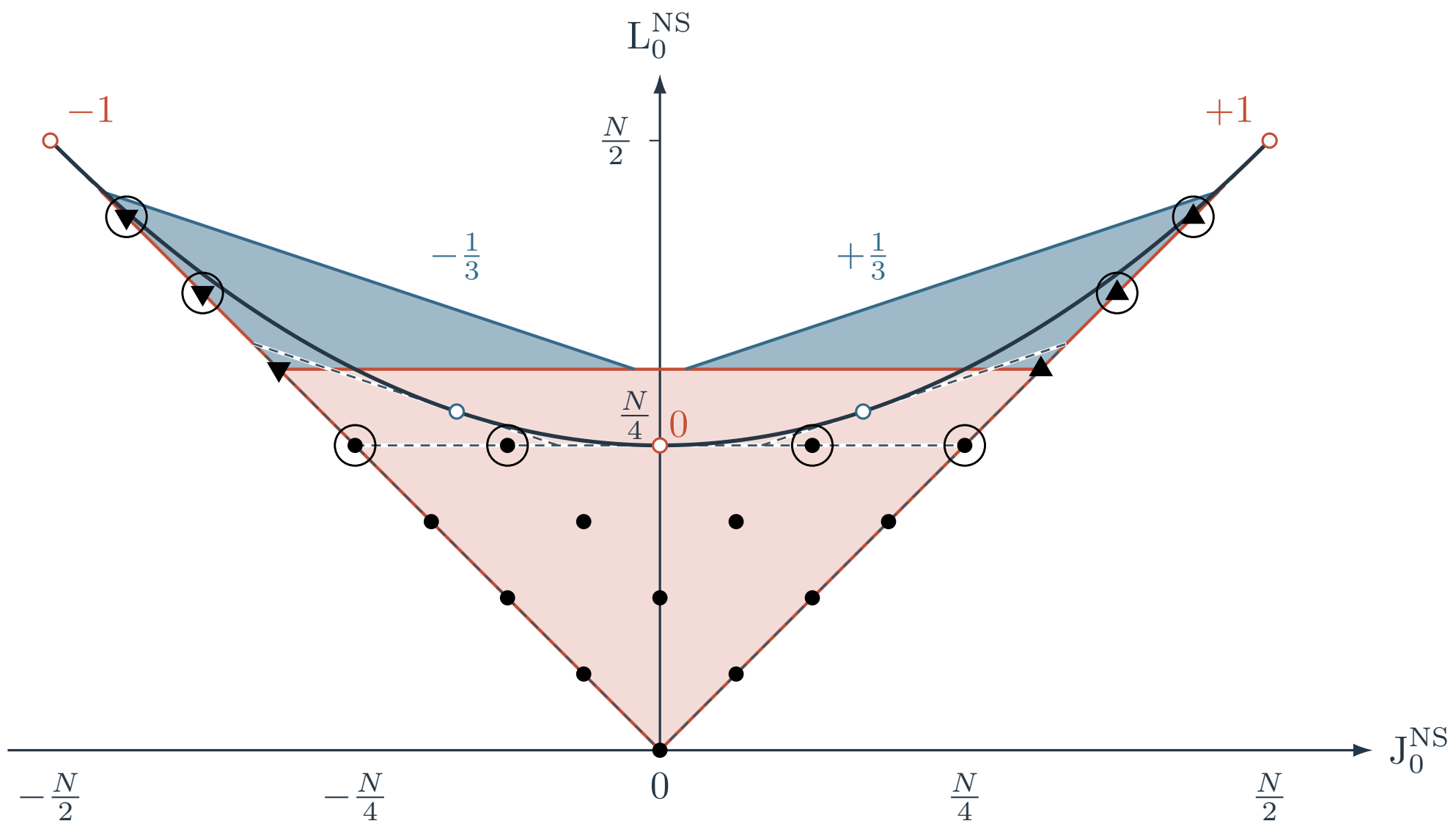

We can also translate this analysis to the NS frame. 
In the NS frame, the shifted boundary is
\begin{equation}
\begin{aligned}
\textrm{L}_{0}^{\mathrm{NS}}
&=\left(1-\frac{2p}{q}\right)\textrm{J}_{0}^{\mathrm{NS}}
  +N\frac{p(q-p)}{q^{2}}
  +\frac{p^{2}-pq+q^{2}-1}{q^{2}}\\
&=T_{\frac{p}{q}}^{\mathrm{NS}}\left(\textrm{J}_{0}^{\mathrm{NS}}\right)
  +\frac{p^{2}-pq+q^{2}-1}{q^{2}}\,.
\end{aligned}
\end{equation}
Figure~\ref{fig:n8-ns-chamber-changes} shows the quantum chamber structure for $N=8$.

\section{Discussion}
\label{sec: discussion}

In this work, we proposed a bulk prescription for computing the degeneracies of polar states at finite $N$ via one-loop contributions from a finite set of asymmetric $\orbifoldK$ backgrounds in type IIB string theory. 
For each polar charge, we provide a contour prescription that specifies the complete finite sum of bulk orbifold contributions to its degeneracy.
We computed the classical charges of these backgrounds using supergravity and their one-loop $\frac{1}{4}$-BPS partition functions using the tensionless worldsheet with a Gauss constraint. 
We matched these results with the residues of the grand-canonical elliptic genus at its linear poles, which determine the polar degeneracies. 
Finally we interpreted the charge-dependent contour choice in terms of the phase structure of the microcanonical charge lattice. 

We conclude with a partial list of open questions:
\begin{enumerate}
    \item It would be important to derive the one-loop supersymmetric partition functions of $\orbifoldK$ via a direct Kaluza-Klein analysis including the massless orbifold twisted sector modes, and see if the results agree with our residues and worldsheet derivation in the elliptic genus limit $\yb \to 1$.
    \item A Euclidean gravitational path integral treatment would also be helpful for elucidating the contour treatment of negative modes and for providing a sum-over-saddles explanation for the Gauss constraint.
    \item It would be nice to have bulk explanation for our charge-dependent contour prescription in Section~\ref{sec:ExpansionMCE}, perhaps along the lines of the dyon-counting analysis of \cite{Cheng:2007ch}.
    \item It would be nice to connect our proposal to fortuity in the D1-D5 system \cite{Chang:2024zqi,Chang:2025rqy,Chang:2025wgo,Zhang:2026jnf}.
    \item In higher-dimensional examples of AdS/CFT, the bulk contributions that account for the finite $N$ BPS spectrum are giant graviton branes \cite{Gaiotto:2021xce,Imamura:2021ytr}. It would be nice to know whether the partition functions on $\orbifold$ can also be understood in terms of branes in $\adsS$.
    \item It would be interesting to use the perspective presented in our work to refine the bulk description of the black hole farey tail \cite{Dijkgraaf:2000fq,Manschot:2007ha}. The bulk interpretation of the farey tail assumes that the polar degeneracies correspond to bulk perturbative spectra, while our work shows that the contributions from non-perturbative saddles are essential for polar data.
\end{enumerate}

\section*{Acknowledgements} 

We thank Alexandre Belin, Nathan Benjamin, Jan Boruch, Alejandra Castro, Chi-Ming Chang, Sean Colin-Ellerin, Matthew Heydeman, Emil Martinec, Hirosi Ooguri, Masaki Shigemori, Tadashi Takayanagi, David Turton, Edward Witten, Xin Wang, Zhenbin Yang, Haoyu Zhang, and especially Matthias Gaberdiel for helpful discussions. 
JHL expresses thanks to the organizers of seminars at CERN, Caltech, UC Berkeley, the University of Southern California, and the University of Chicago, and of the workshops ``Black-Hole Microstructure VIII'' at IPhT, CEA Saclay, ``Pathways to Quantum Black Holes'' at the Galileo Galilei Institute, ``LITP Workshop on Quantum Black Holes'' at the University of Michigan, and ``New Developments in Black Holes and de Sitter Space'' in Xiamen, China, for opportunities to present this work.
WL thanks YITP (Kyoto U.), Kavli-IPMU (Tokyo U.), CUNY, IAS (Princeton), Rutgers U., PCFT (USTC), Quantum Information and Quantum Gravity 2026, and Strings 2026 for the opportunities to present this work. 
WL is particularly grateful to YITP, IAS, and SLMath Berkeley (Program ``Representation Theory Under the Influence of Quantum Field Theory") for the hospitality and stimulating enviroment when part of this work was done.

The work of the group at ETH (JHL) is supported in part by the Simons Foundation grant 994306 (Simons Collaboration on Confinement and QCD Strings) and by the NCCR SwissMAP funded by the Swiss National Science Foundation. 
WL is supported by NSFC No.\ 12447101, 12247103.

\IfFileExists{JHEP.bst}{\bibliographystyle{JHEP}}{\bibliographystyle{utphys}}
\bibliography{references}

\end{document}